\documentclass{article}
\usepackage{booktabs}
\usepackage[T1]{fontenc}
\usepackage{natbib}
\usepackage{helvet}
\DeclareMathAlphabet{\mathsfbi}{T1}{phv}{b}{it}
\usepackage[abs]{overpic}
\usepackage{xcolor,varwidth}
\usepackage{graphicx}
\usepackage{newtxtext}
\usepackage{newtxmath}
\usepackage{natbib}
\usepackage{hyperref}
\definecolor{redNCSU}{rgb}{0.85, 0, 0}
\definecolor{mypink}{rgb}{0.996, 0.414, 0.609}
\definecolor{lightgray}{rgb}{0.5, 0.5, 0.5}
\usepackage{physics}
\usepackage{amsmath}
\usepackage{relsize}
\usepackage{multirow}
\hypersetup{
    colorlinks = true,
    linkcolor  = blue,
    urlcolor   = blue,
    citecolor  = black,
}

\newcommand{\RomanNumeralCaps}[1]
\linenumbers
\newcommand{\volumes}{\rotatebox[origin=c]{180}{\ensuremath{A}}}
\usepackage{mathdots}

\usepackage{arxiv}

\usepackage[utf8]{inputenc} 
\usepackage[T1]{fontenc}    
\usepackage{hyperref}       
\usepackage{url}            
\usepackage{booktabs}       
\usepackage{amsfonts}       
\usepackage{nicefrac}       
\usepackage{microtype}      
\usepackage{lipsum}
\usepackage{graphicx}
\graphicspath{ {./images/} }
\title{Disentangling spanwise asymmetries in unsteady wing wakes: global mode sensitivity and spatio-temporal harmonic resolvent analyses}
\author{
 Maryam Safari \\
  Department of Mechanical and Aerospace Engineering\\
   North Carolina State University\\
  Raleigh, North Carolina, USA. \\
  \texttt{msafari2@ncsu.edu} \\
  \And
 Chi-An Yeh \\
  Department of Mechanical and Aerospace Engineering\\
  North Carolina State University\\
  Raleigh, North Carolina, USA.\\
  \texttt{chian.yeh@ncsu.edu} \\
  }

\begin{document}
\maketitle
\begin{abstract}
We investigate the emergence of long-time spanwise asymmetries in an unsteady wake downstream of a finite-span wing by disentangling flow asymmetries into symmetric and anti-symmetric components using global mode (structural) sensitivity and spatio-temporal harmonic resolvent analysis. The global mode sensitivity analysis shows that asymmetric modes emerge when symmetric and anti-symmetric eigenmodes appear as pairs and exhibit high levels of modal non-normality. The modal non-normality renders the eigenmodes susceptible to asymmetric disturbances, which results in phase interference between the paired symmetric and anti-symmetric modes and unfolds them into highly asymmetric modes. Such interferences further motivate the development of a spatio-temporal harmonic resolvent analysis to examine the cross-frequency phase coupling between modes of different phase velocities. We observe that the flow asymmetries are primarily driven by elliptic vortex instability and its interaction with the wake shear layers. Moreover, we show that, even with a large-amplitude departure in the base flow from the symmetric state, the asymmetric modes obtained from the asymmetric wake can be accurately reconstructed by the symmetric and anti-symmetric modes from the symmetric base flow. This important finding suggests that flow asymmetries can be understood as a superposition of symmetric and anti-symmetric structures that lie under the symmetric base flow, and their phase interference serves as a potential mechanism for the emergence of long-time flow asymmetries. We believe that the present study provides a promising path towards understanding and controlling the emergence of asymmetric flow structures over finite-span wings.
\end{abstract}


\section{Introduction}
\label{Introduction}
Unsteady flows over spanwise-symmetric aerodynamic bodies that are expected to be nominally symmetric can become asymmetric in the spanwise direction \citep{williams2025asymmetries,gursul2004recent,gursul2005review}. Flow asymmetries can manifest in an oscillatory manner, such as vortex shedding in bluff-body wakes \citep{tang1997symmetry,matsumoto1999vortex,song2025symmetry,goswami2022mechanisms}. These oscillatory asymmetries are characterized by energy growth of anti-symmetric instabilities in the flow,  which have been extensively documented in prior studies \citep{mao2014structure,williams1992response}. However, the mechanisms that are responsible for the formation of flow asymmetries over time scales that are significantly longer than those of the oscillatory instabilities remain largely unexplored \citep{williams2025asymmetries, pavia2020salient, keener1977similarity}. 

For aerodynamic flows, such long-time asymmetries are of particular importance for slender bodies and delta wings at high angles of attack, where spanwise asymmetries can occur due to mechanisms such as vortex breakdown \citep{ma2017symmetry,zong2021structural}.  Such long-time asymmetries can generate large side forces, potentially leading to departure from controlled flight \citep{ericsson1990control, hunt1982asymmetric, degani2022development}. The severity of these effects increases dramatically for large-scale flow asymmetries, as the induced forces can rival the body lift \citep{andersson2018instabilities}. Therefore, a deep understanding of the fundamental flow physics that are responsible for the formation of long-time asymmetries is crucial, as it provides insight for designing effective flow control strategies to either reject or promote flow asymmetries to enhance the maneuverability of air vehicles \citep{porter2014closed, shen2016asymmetric, hitzel2018enhanced, williams2008active}.  However, the relevant asymmetric flow physics would be obscured in experimental or numerical configurations that implicitly enforce spanwise symmetry \citep{agrawal1992numerical,visbal1994onset, ghoreyshi2023vortex,rojas2025flow,shah1999turbulent,ribeiro2023triglobal}.

Most studies have tied the formation of these asymmetries to hydrodynamic instabilities in the nominally symmetric flow \citep{degani1991effect, degani1992effect, levy1996systematic, chen2023effect}, as they can amplify small asymmetric disturbances originating from slight geometric imperfections or free-stream fluctuations.  Since these instability waves primarily manifest in the anti-symmetric flow structures \citep{luo2018time, grandemange2013turbulent, rigas2016symmetry}, they can explain the emergence of oscillatory flow asymmetries but fall short of delineating the asymmetries that manifest in a long-time manner.  Clearly, we need to go beyond these anti-symmetric hydrodynamic instabilities, in order to understand, model, and control the formation of long-time asymmetries.  The interactions between flow structures associated with symmetric and anti-symmetric instability modes that evolve on similar time scales can also play a crucial role in the emergence of long-time flow asymmetries.

\subsection{A perspective on long-time flow asymmetries}
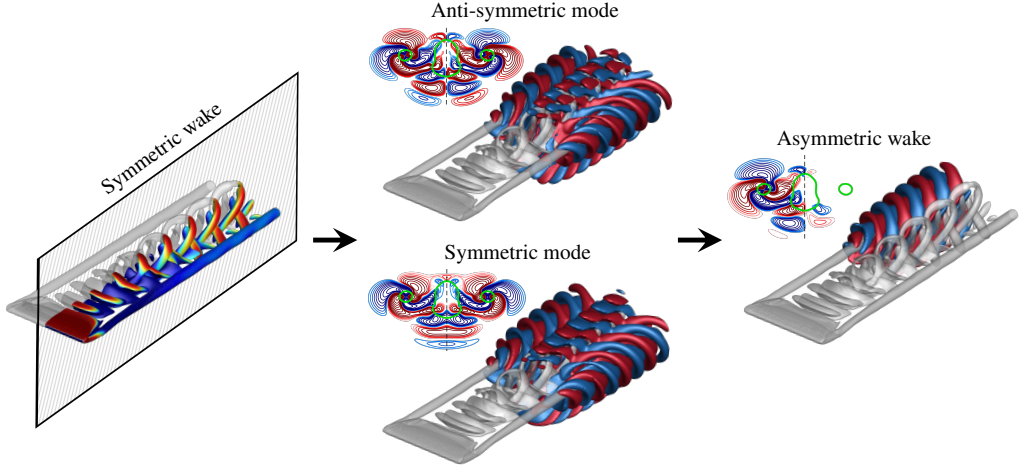
\begin{figure}
\centerline{\begin{overpic}[width=5.32in, trim=1.95cm 0cm 0 0cm]{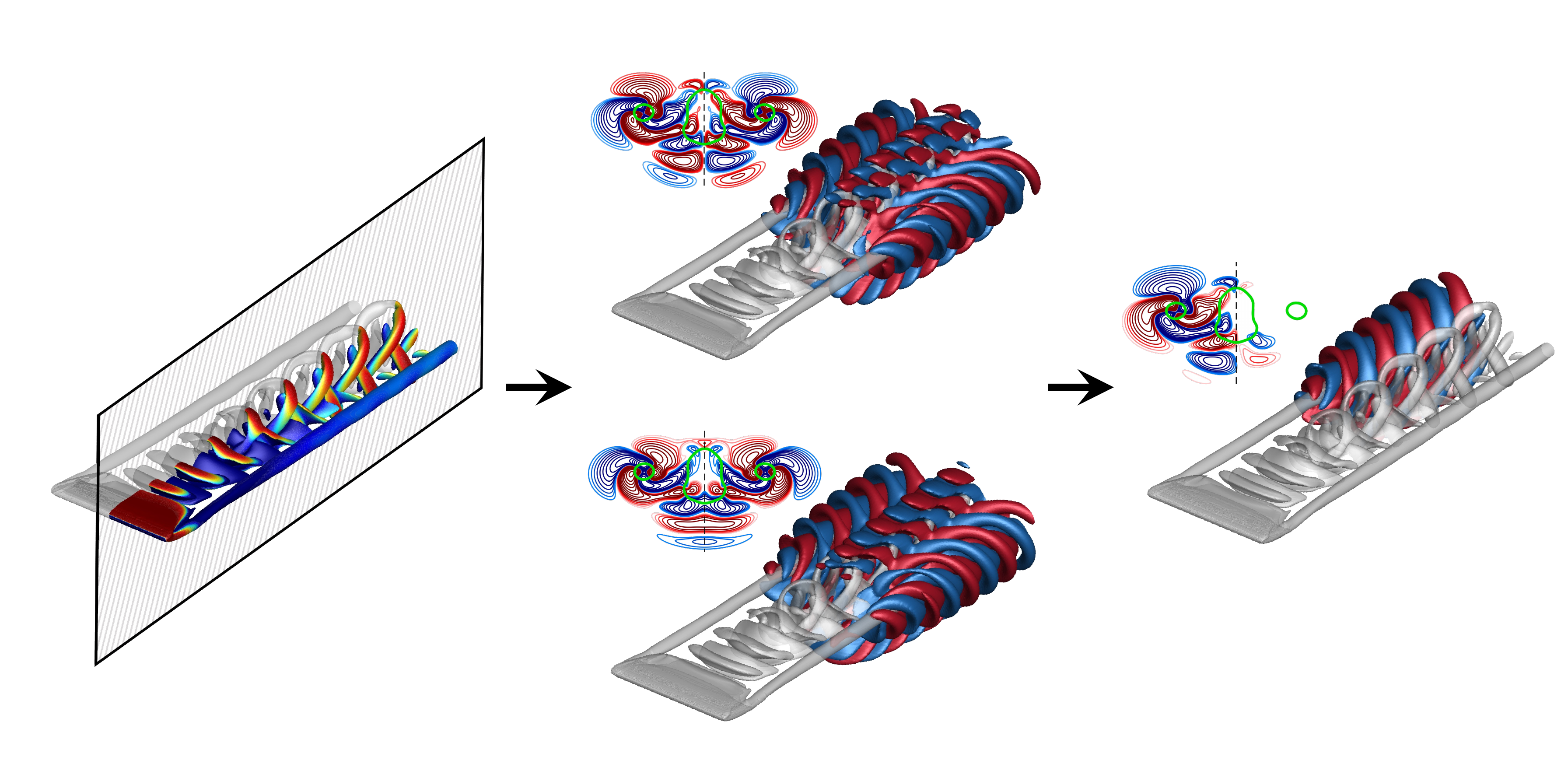} 
\put(35,113){\rotatebox{36}{\fontsize{8}{9}\selectfont Symmetric wake}}
\put(160,180){\rotatebox{0}{\fontsize{8}{9}\selectfont Anti-symmetric mode}}
\put(165,89){\rotatebox{0}{\fontsize{8}{9}\selectfont Symmetric mode}}
\put(290,132){\rotatebox{0}{\fontsize{8}{9}\selectfont Asymmetric wake}}
\end{overpic}}
\caption{The central perspective of the present study: flow asymmetries are viewed as an interference between the symmetric and anti-symmetric structures that reside in the symmetric flow.  Since the resulting flow asymmetry depends on the relative phase between the symmetric and anti-symmetric components, it necessitates the knowledge of the phase dynamics of these components.}
\label{fig.Motivation}
\end{figure}

We argue that the formation of long-time spanwise flow asymmetries over spanwise-symmetric bodies can be understood as an interference between the symmetric and anti-symmetric flow structures that underlie the nominally symmetric flow. This perspective, as summarized in figure~\ref{fig.Motivation}, serves as the starting point of the present study.  It is reminiscent of the interference mechanisms in quantum systems, where a superposition of symmetric and anti-symmetric states can lead to symmetry breaking \citep{griffiths2018introduction}.  Moreover, the relative phase between the oscillatory symmetric and anti-symmetric components determines the asymmetric structure resulting from their superposition, as it governs the spatial pattern of where constructive and destructive interferences would occur on the physical domain.  Therefore, the goal of this study is to disentangle flow asymmetries into symmetric and anti-symmetric components \citep{marasli1989modal} and examine their resemblances to those extracted from the nominally symmetric flows.  If high levels of resemblances are observed, it implies that the symmetric and anti-symmetric unsteady structures lying under a nominally symmetric flow can potentially be used as building blocks to model an asymmetric departure from the symmetric flow state.  We aim to achieve these objectives using modern analytical tools that provide the building-block modal components and their phase information \citep{schmid2014analysis, chavarin2020resolvent, padovan2020analysis, leclercq2023mean}. 

\subsection{An overview of the present study}
In this work, we study the formation of long-time spanwise asymmetries in the wake of a symmetric finite-span wing by disentangling flow asymmetries into symmetric and anti-symmetric components using global mode (structural) sensitivity and spatio-temporal harmonic resolvent (STHR) analyses.  Respectively, these two analyses are extensions of the global linear stability analysis \citep{theofilis2011global} and resolvent analysis \citep{trefethen1993hydrodynamic,mckeon2010critical, jovanovic2005componentwise,rolandi2024invitation}, both allowing us to identify highly amplified perturbations that exhibit high levels of spanwise asymmetries in the wake of the finite wing. 

To understand the origin of these asymmetries, we first perform global linear stability analysis on both a symmetry-enforced and an asymmetrically disturbed wake flow.  This allows us to closely examine how symmetric and anti-symmetric global modes in the symmetry-enforced wake unfold into asymmetric modes in the asymmetrically disturbed wake.  By configuring the asymmetrically disturbed wake as a small-amplitude departure from the corresponding symmetric wake, this becomes a structural sensitivity analysis based on global linear stability \citep{giannetti2007structural, schmid2014analysis, lashgari2014planar, wang2019enhanced, regan2019adjoint}, which we refer to as global mode sensitivity analysis in this work. To further elucidate the role of modal phase interference in the formation of wake asymmetries, we combine two different harmonic resolvent analysis formulations of \citet{chavarin2020resolvent} and \citet{padovan2020analysis} to develop the STHR framework to handle the wake flows that exhibit spatial and temporal periodicities.  The framework not only enables cross-frequency coupling of perturbations propagating at different phase velocities, but it also provides phase information \citep{mckeon2019applications, leclercq2023mean} that has a crucial role in the formation of flow asymmetries.  In the STHR analysis, we further increase the amplitude of the asymmetric departure of the base flow to a finite level and examine how well the asymmetric resolvent modes obtained from the asymmetric base flow can be constructed using the symmetric and anti-symmetric resolvent modes obtained from the symmetry-enforced base flow. By viewing the asymmetric departure from the nominally symmetric base flow as a perturbation mode, we also aim to examine how well the base flow departure can be captured by the resolvent modes of the symmetric base flow.

The structure of the paper is as follows. In~\S\ref{Problemsetup}, we define the model problem considered in this study and document the details on the direct numerical simulations (DNS) of the symmetric and asymmetric wake flows downstream of a finite wing of spanwise symmetry.  The flow fields obtained from the DNS will be treated as the base flows for the global mode sensitivity and STHR analyses, whose domain of analysis is defined in~\S\ref{AnalysisDomain}. In~\S\ref{LinearStability}, we perform the global mode sensitivity analysis to identify the conditions for the eigenmodes from the symmetric base flow to unfold into asymmetric modes in the asymmetric base flow. In~\S\ref{HarmonicResolvent}, we introduce the framework of STHR analysis and use it to disentangle the finite-amplitude flow asymmetries into the symmetric and anti-symmetric resolvent modes of the symmetric wake flow. A summary of our study and conclusions is offered in~\S\ref{Conclusion}.

\section{\label{Problemsetup}Problem setup}
\subsection{\label{Modelprob}Model problem description}
We consider wake flows over a rectangular finite-span wing of spanwise symmetry. The wing has a NACA 0012 cross-section and an aspect ratio of 
$\text{AR} \equiv 2b/L_c = 2.5$, placed at an angle of attack $ \alpha = 10^\circ$. Here, $L_c$ is the chord length of the wing, and $b$ denotes the half-span of the wing. The chosen chord-based Reynolds number is $\text{Re} \equiv u_\infty L_c/\nu_\infty=1,000$, where $u_\infty$ and $\nu_\infty$ are the free-stream velocity and kinematic viscosity, respectively. The choice of the Reynolds number is motivated by prior studies that observed the emergence of flow asymmetries at similar Reynolds numbers \citep{johnson2026flow,grandemange2012reflectional,bury2012transitions,prasad1997instability,haffner2020mechanics, brackston2016stochastic}. A Cartesian coordinate system is used in this study, where the $x$, $y$, and $z$ directions are aligned with the streamwise, transverse, and spanwise directions, respectively. The NACA 0012 airfoil geometry is defined on the $z$-plane, with the leading edge of the mid-span section placed at the origin, as shown in figure~\ref{fig.MeshBC}.
\subsection{\label{DNS}Direct numerical simulation}
\begin{figure}

\centering{\begin{overpic}[width=5.32in]{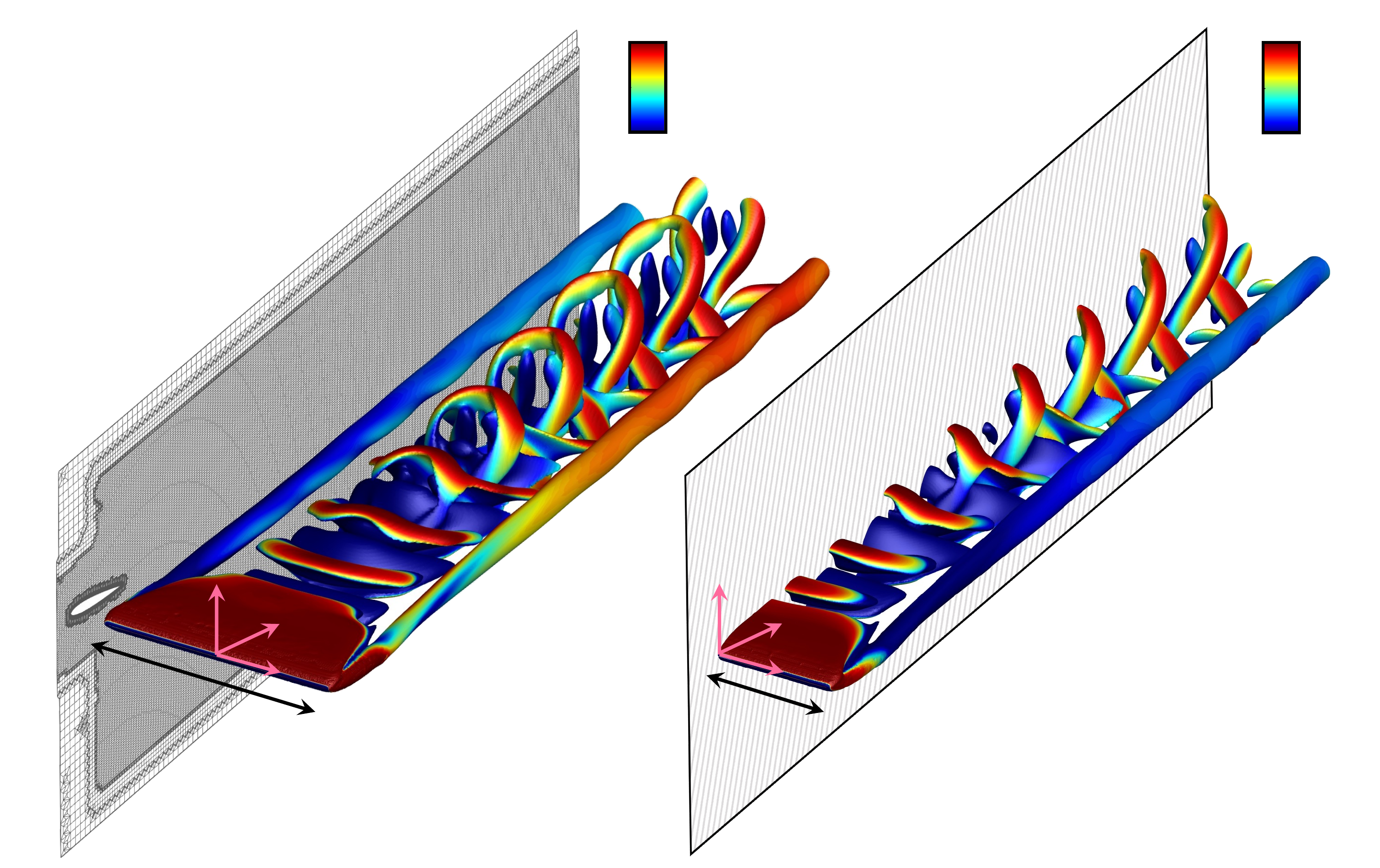} 
\put (15 , 122) {(a)}
\put (188 , 118) {(b)}

\put(178,233){\rotatebox{0}{\fontsize{8}{9}\selectfont $u$}}
\put(188,204){\fontsize{8}{9}\selectfont $0.85$}
\put(188,225){\fontsize{8}{9}\selectfont $1.05$}
\put(354,233){\rotatebox{0}{\fontsize{8}{9}\selectfont $u$}}
\put(364.5,204){\fontsize{8}{9}\selectfont $0.85$}
\put(364.5,225){\fontsize{8}{9}\selectfont $1.05$}
\put(77,58){\fontsize{8}{9}\selectfont \color{mypink}{$z$}}
\put(73,70){\fontsize{8}{9}\selectfont \color{mypink}{$x$}}
\put(53,75){\fontsize{8}{9}\selectfont \color{mypink}{$y$}}
\put(218,58){\fontsize{8}{9}\selectfont \color{mypink}{$z$}}
\put(213,70){\fontsize{8}{9}\selectfont \color{mypink}{$x$}}
\put(193.5,75){\fontsize{8}{9}\selectfont \color{mypink}{$y$}}
\put(240,158){\rotatebox{39}{{{\fontsize{8}{9}\selectfont \color{redNCSU}{\textbf{Far-field}}}}}}
\put(260,54){\rotatebox{39}{{{\fontsize{8}{9}\selectfont \color{redNCSU}{\textbf{Far-field}}}}}}
\put(183,50){\rotatebox{90}{{{\fontsize{8}{9}\selectfont \color{redNCSU}{\textbf{Far-field}}}}}}
\put(338,195){\rotatebox{90}{{{\fontsize{8}{9}\selectfont \color{blue}{\textbf{Outlet}}}}}}
\put(70,162){\rotatebox{39}{{{\fontsize{8}{9}\selectfont \color{redNCSU}{\textbf{Far-field}}}}}}
\put(44,18){\rotatebox{40}{{{\fontsize{8}{9}\selectfont \color{redNCSU}{\textbf{Far-field}}}}}}
\put(8,50){\rotatebox{90}{{{\fontsize{8}{9}\selectfont \color{redNCSU}{\textbf{Far-field}}}}}}
\put(163,195){\rotatebox{90}{{{\fontsize{8}{9}\selectfont \color{blue}{\textbf{Outlet}}}}}}
\put(48,48){\rotatebox{-22}{{{\fontsize{8}{9}\selectfont $2b$}}}}
\put(206,43){\rotatebox{-22}{{{\fontsize{8}{9}\selectfont $b$}}}}
\put(230,125){\rotatebox{40}{{{\fontsize{8}{9}\selectfont Mid-span/root plane}}}}
\put(248,126){\rotatebox{40}{{{\fontsize{8}{9}\selectfont \color{redNCSU}\textbf{Symmetry}}}}}
\end{overpic}}
\caption{Instantaneous isocontours of the $Q$-criterion ($Q L_c^2/u_\infty^2=0.1$) colored by the streamwise velocity $u$: (a) spanwise-asymmetric flow over the full-span wing; (b) symmetry-enforced flow over the half-span wing. The DNS mesh on the root plane ($z = 0$) is also visualized in (a).}
\label{fig.MeshBC}
\end{figure}
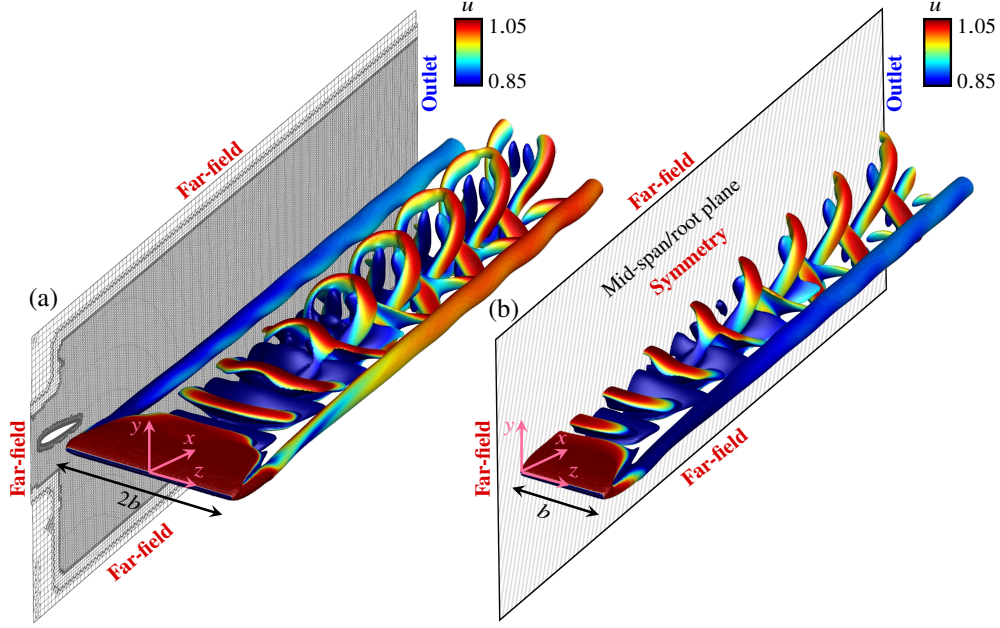
We perform incompressible DNS for the wake flows in both symmetry-enforced and asymmetrically disturbed settings, as shown in figure~\ref{fig.MeshBC}. The symmetry-enforced flow is simulated over a half-span wing with a symmetry boundary condition prescribed over the root plane at the mid-span ($z = 0$); for the asymmetrically disturbed counterpart, the flow is simulated over a full-span wing with a spanwise-asymmetric volumetric forcing introduced to the wing upstream, which will be discussed in detail shortly. The computational domain for the symmetry-enforced flow extends over $x/L_c \in [-15,25]$, $y/L_c \in [-15,15]$, and $z/L_c \in [0,15]$. This domain is mirrored about the root plane for the simulations of asymmetric flows over the full-span wing, such that the domain has a spanwise extent of $z/L_c \in [-15,15]$. For both settings, a Dirichlet boundary condition of $(u,v,w) = (u_\infty,0,0)$ is prescribed for the far-field boundaries, where $u$, $v$, and $w$ are the velocity components in the $x$-, $y$-, and $z$-directions, respectively. A convective boundary condition is employed at the outlet, and the airfoil surface is prescribed with a no-slip boundary condition. 

The incompressible DNS are performed using the CharLES solver package  \citep{ham2006accurate}. The solver uses a second-order accurate finite-volume method and a fractional-step scheme to solve the discretized incompressible Navier--Stokes (NS) equations. A hybrid grid, also shown in figure~\ref{fig.MeshBC}, is generated with mesh refinement near the wing and in the wake region, resulting in approximately 15 million grid points for the full-span wing configuration (for asymmetric flows) and 7.5 million grid points for the half-span wing configuration (for symmetry-enforced flows). The simulations are initialized with a uniform flow and run for $t u_\infty / L_c= 50$ to wash out the transients. We then collect a series of flow field snapshots with a constant time step of $\Delta t u_\infty / L_c = 0.02$ over a convective time of $t u_\infty / L_c = 26.4$, which corresponds to 20 fundamental temporal periods of the flow ($T_0 u_\infty / L_c = 1.32$), to construct the base flows for the analyses.

\subsection{\label{asymforcing}Breaking the symmetry with asymmetric perturbations}
We note that, with the chosen Reynolds number and wing geometry, the flow over the finite wing exhibits numerically perfect spanwise symmetry even in the full-span wing setting. To break the long-time symmetry of the flow, we disturb the flow by introducing a volumetric forcing that is asymmetric about the mid-span/root plane. The constant-in-time forcing follows a spherical Gaussian profile that writes
\begin{equation}
g(\boldsymbol{x}) = a_f  \exp\left[-(\boldsymbol{x}-\boldsymbol{x}_0)^2/\sigma^2\right],
 \label{eq.force}
\end{equation}
and is added to the right-hand side of the streamwise momentum equation. Here, $a_f$ is the forcing amplitude, $\boldsymbol{x}_0 \equiv [x_0, y_0, z_0]$ represents the center of the forcing, and $\sigma$ is the spatial support of the spherical Gaussian with $2\sigma/L_c = 1$. The forcing is placed upstream of the wing at $(x_0, y_0)/L_c = (-1, 0)$, and we will consider different spanwise locations of the forcing that vary from $z_0/b = 0.5$ to $1$ off the root plane $(z = 0)$ to induce different spanwise asymmetries in the wake flows. We will also consider different forcing amplitudes, $a_f$, to trigger different levels of flow asymmetries. These forcing amplitudes are determined in terms of the momentum coefficient,
\begin{equation}
 C_\mu \equiv \frac{a_f \iiint_{\mathbb{R}^3} \exp\left[-(\boldsymbol{x}-\boldsymbol{x}_0)^2/\sigma^2\right]{\rm d}\volumes}{\frac{1}{2}\rho u_\infty^2 \times (2b L_c)},
 \label{eq.Cmu}
\end{equation}
where $C_\mu$ will range from $C_\mu = 10^{-6}$ (for the global mode sensitivity analysis, ~\S\ref{Eigenvalueanalysis}) to $C_\mu = 0.1$ (for the STHR analysis, ~\S\ref{phase_interference}) throughout this study.

\section{Domain of analysis and base flows}
\label{AnalysisDomain}
\begin{figure}
\centerline{\begin{overpic}[width=5.32in]{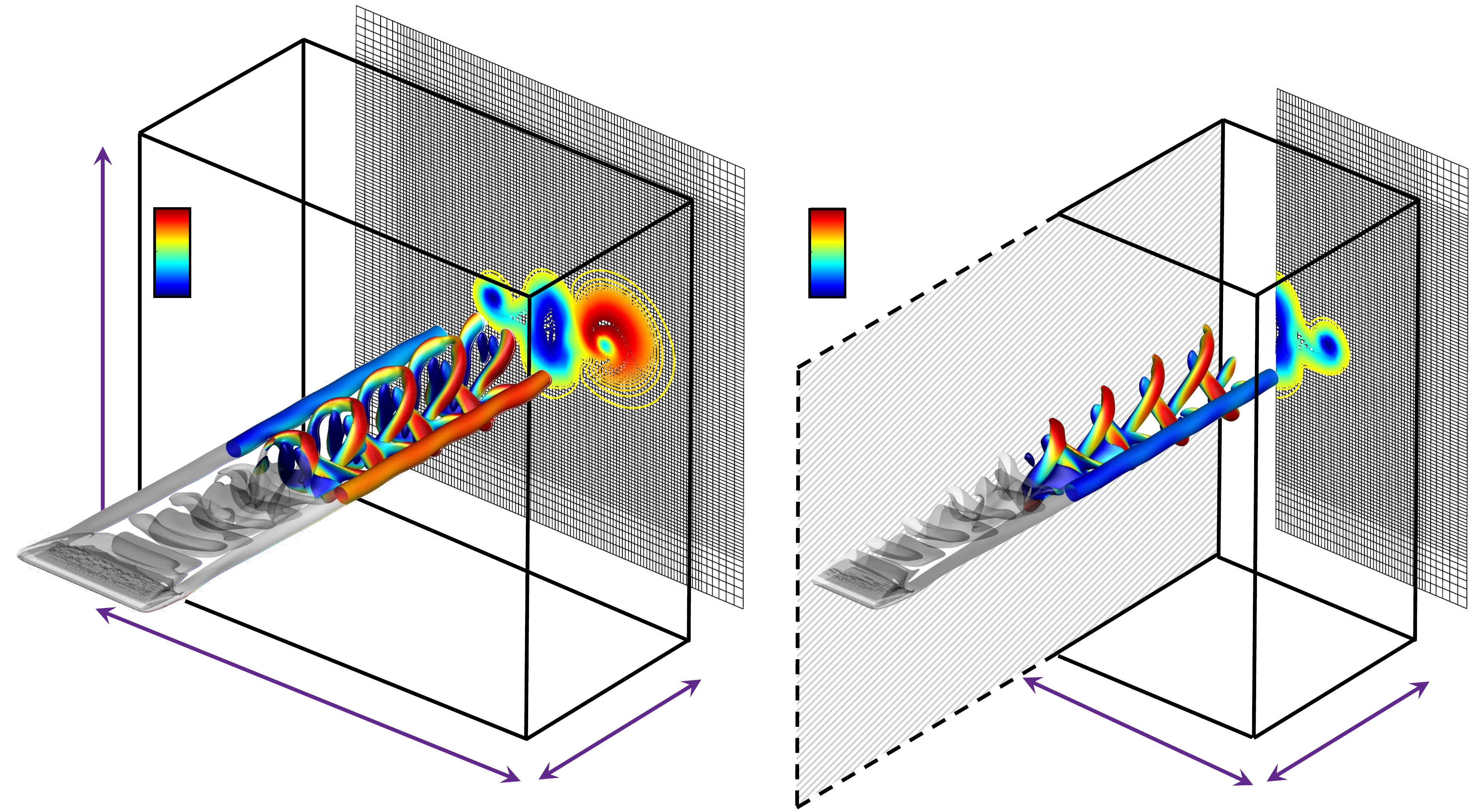} 
\put (20 , 190) {(a)}
\put (203 , 190) {(b)}
\put (65 , 25) {\rotatebox{-20}{{\fontsize{8}{9}\selectfont $16 L_c$ }}}
\put (160 , 12) {\rotatebox{32}{{\fontsize{8}{9}\selectfont$5 L_c$}}}
\put (16 , 112) {\rotatebox{90}{{\fontsize{8}{9}\selectfont$16 L_c$}}}
\put (290 , 14) {\rotatebox{-30}{{\fontsize{8}{9}\selectfont$8 L_c$ }}}
\put (350 , 12) {\rotatebox{32}{{\fontsize{8}{9}\selectfont$5 L_c$ }}}

\put(42,160){\rotatebox{0}{\fontsize{8}{9}\selectfont $\bar{u}$}}
\put(52,134){\fontsize{8}{9}\selectfont $0.85$}
\put(52,152){\fontsize{8}{9}\selectfont $1.05$}

\put(213,160){\rotatebox{0}{\fontsize{8}{9}\selectfont $\bar{u}$}}
\put(222,134){\fontsize{8}{9}\selectfont $0.85$}
\put(222,152){\fontsize{8}{9}\selectfont $1.05$}

\end{overpic}}
\caption{The analysis domain downstream of the wing is defined by $x/L_c \in [5, 10]$ and $y/L_c \in [-8, 8]$, with $z/L_c \in [-8, 8]$ for (a) the asymmetrically disturbed flow (full-span wing) and $z/L_c \in [0, 8]$ for (b) the symmetry-enforced (half-span). The streamwise- time-averaged streamwise velocity, along with the two-dimensional computational grid in the $(y, z)$ plane used for the analyses, are  also visualized for both wake flows.}
\label{fig.BOXDef}
\end{figure}

At the chosen Reynolds number and angle of attack, the flows over the finite-span wing exhibit periodic unsteadiness as the result of vortex shedding accompanied by two counter-rotating wing-tip vortices. Our analyses will focus on a rectangular region over $x/L_c\in [5, 10]$, $y/L_c\in [-8, 8]$, and $z/L_c\in [-8, 8]$ for the asymmetric flows (figure~\ref{fig.BOXDef}(a)), and  $z/L_c\in [0, 8]$ for the symmetry-enforced flow (figure~\ref{fig.BOXDef}(b)).  The flow physics in this region is characterized by strong interactions between the wake deficit and the trailing vortices \citep{edstrand2018parallel}. Such interactions are expected to play an important role in how streamwise convective instabilities amplify small-amplitude disturbances to form wake asymmetries \citep{degani1992effect, degani1992asymmetric}.  

\begin{figure}
\centerline{\begin{overpic}[width=5.32in]{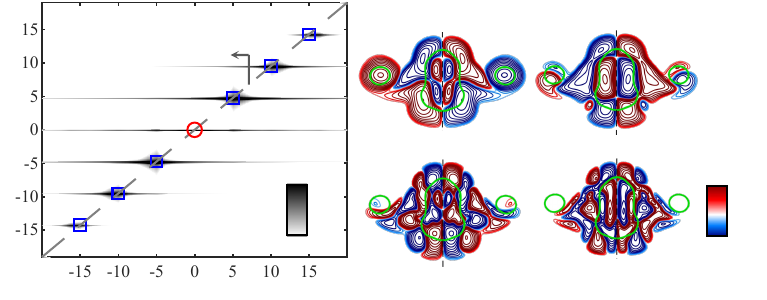} 
\put (22 , 154) {(a)}
\put (185 , 154) {(b)}
\put (354 , 61) {\fontsize{8}{9}\selectfont$\tilde{\zeta}_x$}
\put (367 , 31) {\fontsize{8}{9}\selectfont -0.03}
\put (370 , 52) {\fontsize{8}{9}\selectfont 0.03}
\put (132 , 31) {\rotatebox{90}{\fontsize{8}{9}\selectfont$ \| \tilde{\boldsymbol{q}}_{\omega,k}\|$}}
\put (156, 33) {\fontsize{8}{9}\selectfont $10^{-3}$}
\put (156, 51) {\fontsize{8}{9}\selectfont $10^{-1}$}
\put (35, 121) {\rotatebox{0}{{\fontsize{8}{9}\selectfont $c_0 = \omega_0/k_0 =0.95u_\infty$}}}
\put (88, 2) {\rotatebox{0}{{\fontsize{8}{9}\selectfont $k L_c$}}}
\put (0, 70) {\rotatebox{90}{{\fontsize{8}{9}\selectfont $\omega L_c / u_\infty$}}}

\put (205, 139) {\fontsize{8}{9}\selectfont $n=0$ ~($\bar{\boldsymbol{q}}_0$)}
\put (300, 139) {\fontsize{8}{9}\selectfont $n=1$}
\put (212, 72) {\fontsize{8}{9}\selectfont $n=2$}
\put (300, 72) {\fontsize{8}{9}\selectfont $n=3$}

\put (95, 75) {\fontsize{8}{9}\selectfont $n=0$}
\put (115, 90) {\fontsize{8}{9}\selectfont $n=1$}
\put (132, 107) {\fontsize{8}{9}\selectfont $n=2$}
\put (153, 124) {\fontsize{8}{9}\selectfont $n=3$}
\end{overpic}}
\caption{(a) Space-time Fourier spectrum of the symmetric base flow. The streamwise- and time-averaged component of base flow, $\bar{\boldsymbol{q}}_0(y, z)$, is highlighted by the red circle, while blue markers denote the base flow harmonics. (b) Real part of the streamwise vorticity component of the base flow harmonics of $n\omega_0, nk_0$ with $n=0,~1,~2,~3$. Note that these Fourier modes are visualized beneath a green contour line, which portrays $\bar{\boldsymbol{q}}_0$ by the regions where streamwise velocity exhibits a deficit.}
\label{fig.FFT}
\end{figure}
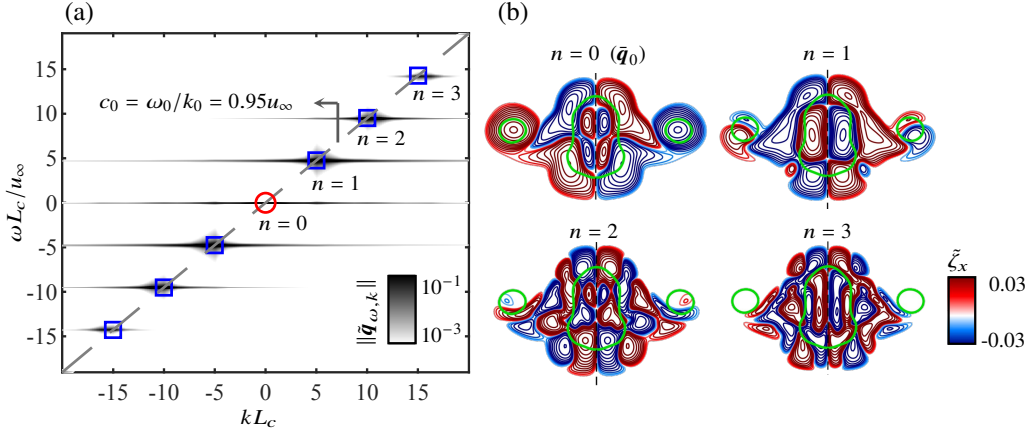

We note that the wake flows in this region exhibit high levels of temporal and spatial periodicity. This is demonstrated in figure~\ref{fig.FFT}(a) by the double Fourier spectrum in frequency ($\omega$) and streamwise wavenumber ($k$) for the symmetric flow.  Here, dominant Fourier components appear as harmonics of fundamental frequency, $\omega_0 L_c/u_\infty = 4.76$, and wavenumber,  $k_0 L_c = 5.00$.  In figure~\ref{fig.FFT}(b), the Fourier modes of the first three spatio-temporal harmonics ($n = 1,~2,~3$) are visualized along with the zeroth harmonic, $\bar{\boldsymbol{q}}_0$ for $n = 0$, by the streamwise vorticity component ($\tilde{\zeta}_x$). These modes show that unsteady oscillations occur mainly in the wake region, with fluctuations around the tip vortices vanishing beyond the first ($n = 1$) harmonic.  Moreover, the appearance of these frequency-wavenumber harmonics indicates that energetic wake structures propagate downstream at a nearly constant phase velocity of $c_0 = \omega_0/k_0 = 0.95 u_\infty$, which will be constantly revisited throughout this paper.  The wake flows in this rectangular domain will be adopted as the base flows in the analyses to follow.  We will also take advantage of the high levels of spatio-temporal periodicity of the wake flow in this region by adopting the parallel-flow assumption in our analyses, which provides spatial homogeneity for the base flows in the streamwise direction.

The flow-field data obtained in the DNS for both symmetric and asymmetric flows are interpolated onto a structured mesh over the rectangular region to construct the base flows for the global mode sensitivity and STHR analyses. The mesh has a $120 \times 120$ non-uniform grid distribution over the $(y,z)$-plane, refined within the region of $(y/L_c, z/L_c) \in [-3,3] \times [-3,3]$ to resolve the instability modes in the wakes and gradually stretched toward the far-field boundaries, as shown in figure~\ref{fig.BOXDef}. The domain size and mesh resolution were found to achieve $3$-digit convergence in the magnitudes of dominant eigenmodes and resolvent gains in the analyses to follow.

\section{\label{LinearStability}Global mode sensitivity analysis: emergence of asymmetric eigenmodes}
In this section, we investigate the sensitivity of the global linear stability modes about the symmetric base flow to symmetry breaking due to the asymmetric departure of the base flow from its intrinsic symmetric state.  To this end, we perform global linear stability analysis \citep{theofilis2003advances, theofilis2011global} on both symmetry-enforced and asymmetrically disturbed wake flows. The asymmetrically disturbed base flow is obtained with a small-amplitude volumetric forcing of $C_\mu = 10^{-6}$, applied at a spanwise location of $z_0/b = 1/2$ upstream of the wing. The base flows considered in the analysis here are the time- and streamwise-averaged wakes in the rectangular domain discussed in figure~\ref{fig.BOXDef}. This results in symmetric and asymmetric two-dimensional base flows, $\bar{\boldsymbol{q}}_0(y,z)$, over the $(y,z)$-plane, as shown by the streamwise velocity fields over the mesh grids in figure~\ref{fig.BOXDef} and the streamwise vorticity component in figure~\ref{fig.FFT}(b).  The linear stability analysis will be performed under a parallel-flow setting by assuming a spatial homogeneity of $\bar{\boldsymbol{q}}_0$ in the streamwise direction.

\subsection{\label{linear Stability fromulation}Formulation}
Consider decomposing a three-dimensional instantaneous flow field into a time- and streamwise-averaged base flow and a small-amplitude unsteady perturbation as
\begin{equation}
    \boldsymbol{q}(x, y, z, t) = \bar{\boldsymbol{q}}_0(y,z) + \boldsymbol{q}^\prime(x, y, z, t).
\end{equation}
Here, $\bar{\boldsymbol{q}}_0(y,z) \equiv [\bar{u}, \bar{v}, \bar{w}, \bar{p}]^T$ denotes the time- and streamwise-averaged flow within the rectangular domain of analysis discussed in~\S\ref{AnalysisDomain}, where the subscript 0 is used to indicate the time- and streamwise-invariance of the base flow. Also, $\boldsymbol{q}^\prime(x, y, z, t) \equiv [u', v', w', p']^T$ represents an infinitesimal perturbation about this base flow. Substituting this decomposition into the incompressible NS equations and neglecting the higher-order terms with respect to $\boldsymbol{q}'$, we arrive at the linearized perturbation equations written in a general form as
\begin{equation}
    \mathsfbi{M}\frac{\partial\boldsymbol{q}^\prime}{\partial t} = \mathsfbi{A}_{\bar{\boldsymbol{q}}_0} \boldsymbol{q}^\prime,
    ~~\text{where}~~
    \mathsfbi{M} = 
    \begin{bmatrix}
        \boldsymbol{I} & \boldsymbol{0} \\
        \boldsymbol{0}  & \boldsymbol{0}
    \end{bmatrix}.
\label{eq.HlinearNS}
\end{equation}
The singular matrix $\mathsfbi{M}$ reflects the continuity equation in the incompressible NS equations, in which pressure appears in the state vector $\boldsymbol{q}^\prime$ to satisfy the incompressibility constraint. Also, $\mathsfbi{A}_{\bar{\boldsymbol{q}}_0}$ denotes the linearized NS operator about the base flow, $\bar{\boldsymbol{q}}_0$.  Due to the spatial homogeneity of the base flows under the parallel-flow assumption, we can consider the perturbation $\boldsymbol{q}^\prime$ to be periodic both in time and the streamwise direction \citep{schmid2001stability}.  This allows for a modal representation of $\boldsymbol{q}^\prime$ that writes
\begin{equation}
    \boldsymbol{q}^\prime(\boldsymbol{x}, t) = \int_{-\infty}^{\infty} \int_{-\infty}^{\infty} \hat{\boldsymbol{q}}_{\lambda, k}(y, z) \mathrm{e}^{\lambda t - \mathrm{i} k x} {\rm d}{\lambda}~{\rm d}{k},
    \label{eq:ansatz}
\end{equation}
where $\lambda$ is the temporal frequency, $k$ is the streamwise wavenumber, and $\hat{\boldsymbol{q}}_{\lambda,k}$ is the modal shape over the $(y, z)$-plane for the $(\lambda, k)$-pair. 

In this study, we are interested in a temporal linear stability problem, where a complex-valued temporal frequency, $\lambda \in \mathbb{C}$, and a real-valued streamwise wavenumber, $k \in \mathbb{R}$, are considered in the modal representation \eqref{eq:ansatz}. Substituting the modal representation into 
equation~\eqref{eq.HlinearNS} results in a generalized eigenvalue problem as
\begin{equation}
    \lambda \mathsfbi{M} \hat{\boldsymbol{q}}_{\lambda,k}
    =
    \mathsfbi{A}_{\bar{\boldsymbol{q}}_0}(k)\hat{\boldsymbol{q}}_{\lambda,k},
\label{eq.NSfreq}
\end{equation}
where $\lambda = \lambda_\mathrm{r} + \mathrm{i}\lambda_\mathrm{i}$ becomes the eigenvalue of the linearized operator $\mathsfbi{A}_{\bar{\boldsymbol{q}}_0}(k)$ for a prescribed real streamwise wavenumber $k$, and $\hat{\boldsymbol{q}}_{\lambda,k} \equiv [\hat{u}, \hat{v}, \hat{w}, \hat{p}]_{\lambda,k}^T$ is the corresponding bi-global eigenmode on the $(y, z)$-plane. The real part of the eigenvalue, $\lambda_\mathrm{r}$, indicates the growth rate, and the imaginary part, $\lambda_\mathrm{i}$, represents the frequency of the eigenmode.

The linear stability analysis discussed above is performed for both the symmetry-enforced and asymmetrically disturbed wake flows, respectively $\bar{\boldsymbol{q}}_0^\text{sym}$ and $\bar{\boldsymbol{q}}_0^\text{asym}$, to examine how the eigenmodes about the symmetric base flow respond to a small-amplitude disturbance that renders the nominally symmetric base flow asymmetric.  This can be understood as a sensitivity analysis of the eigenmode, $(\lambda,~\hat{\boldsymbol{q}}_{\lambda,k})$, to an infinitesimal asymmetric departure of the base flow, $\delta \bar{\boldsymbol{q}}$, from the symmetric state, $\bar{\boldsymbol{q}}_0^\text{sym}$.  That is, we can represent the asymmetric base flow due to the asymmetric disturbance introduced upstream of the wing as $\bar{\boldsymbol{q}}_0^\text{asym} = \bar{\boldsymbol{q}}_0^\text{sym} + \delta \bar{\boldsymbol{q}}$.  This infinitesimal departure $\delta \bar{\boldsymbol{q}}$ translates to a small perturbation to the linear operator constructed about the symmetric wake, allowing the linear operator about the asymmetric wake to be expressed as
\begin{equation}
\mathsfbi{A}_{\bar{\boldsymbol{q}}_0^\text{asym}} = \mathsfbi{A}_{\bar{\boldsymbol{q}}_0^\text{sym}+\delta \bar{\boldsymbol{q}}} = \mathsfbi{A}_{\bar{\boldsymbol{q}}_0^\text{sym}} + \delta \mathsfbi{A}_{\delta\bar{\boldsymbol{q}}}.
\label{eq.perturbedA}
\end{equation}
With the equation above, the linear stability analysis of the asymmetric wake
\begin{equation*}
    \lambda \mathsfbi{M} \hat{\boldsymbol{q}}_{\lambda,k}
    =
    \mathsfbi{A}_{\bar{\boldsymbol{q}}_0^\text{asym}}(k)\hat{\boldsymbol{q}}_{\lambda,k},
\end{equation*}
becomes a perturbed eigenvalue problem \citep{schmid2014analysis, marquet2008sensitivity,giannetti2007structural} for the symmetric wake, which writes
\begin{equation}
(\lambda + \delta \lambda)
\mathsfbi{M}\left(\hat{\boldsymbol{q}}_{\lambda,k} + \delta \hat{\boldsymbol{q}}_{\lambda,k} \right)
=
\left[ \mathsfbi{A}_{\bar{\boldsymbol{q}}_0^\text{sym}}(k) + \delta \mathsfbi{A}_{\delta\bar{\boldsymbol{q}}}(k) \right]
\left( \hat{\boldsymbol{q}}_{\lambda,k} + \delta \hat{\boldsymbol{q}}_{\lambda,k} \right).
\label{eq.NSfreq_pert}
\end{equation}
Here, $\delta \lambda$ is the eigenvalue shift due to the asymmetric departure of the base flow, representing the resulting change in modal growth rate and frequency. The term $\delta \hat{\boldsymbol{q}}_{\lambda,k}$ denotes the associated change in the eigenmode.  

Our goal in the present global mode sensitivity analysis is to identify a set of dominant eigenmodes that are particularly susceptible to an asymmetric departure of the base flow, such that the perturbed eigenmodes exhibit high levels of asymmetry. The corresponding frequencies and eigenmode structures will also shed light on the flow physics that are responsible for breaking the spanwise symmetry in a long-time sense.  Moreover, since the sensitivity of the eigenmodes to the perturbations of $\delta \mathsfbi{A}_{\delta\bar{\boldsymbol{q}}}$ is closely related to the non-normality of $\mathsfbi{A}_{\bar{\boldsymbol{q}}_0^\text{sym}}$ \citep{trefethen2005spectra, schmid2007nonmodal}, we will also closely examine the level of non-normality of each eigenmode about the symmetric wake.  This non-normality will be quantified on a mode-by-mode basis via an inner product between the direct and adjoint eigenmodes for each eigenvalue.  In this study, this modal non-normality is quantified by

\begin{equation}
    \kappa_{\lambda, k} = 1 - {\langle \hat{\boldsymbol{q}}_{\lambda, k},~\hat{\boldsymbol{q}}_{\lambda, k}^\dagger \rangle},
\end{equation}
and we will examine the role of modal non-normality in the emergence of asymmetric eigenmodes.

When the linear stability analysis is performed for the symmetry-enforced wake downstream of the half-span wing, it is important to note that the oscillatory global modes can exhibit either symmetric or anti-symmetric structures about the root plane for the symmetric base flow \citep{marasli1989modal, sipp2003widnall}.  As pointed out earlier, the interference between the symmetric and anti-symmetric modes can result in asymmetric modes, which could in turn lead to a long-time asymmetry of the base flow due to nonlinear effects. Therefore, for the linear stability analysis of the symmetric base flow, we seek separately the symmetric and anti-symmetric eigenmodes by prescribing two sets of boundary conditions at the root plane. These two boundary conditions are written as
\begin{subequations}
\begin{align}
    \left.\left[
        \frac{\partial \hat{u}}{\partial z},~
        \frac{\partial \hat{v}}{\partial z},~
        \hat{w},~
        \frac{\partial \hat{p}}{\partial z}
    \right]
    \right|_{z=0} = [0,0,0,0] & \quad\text{for symmetric modes,}\\
    \left.\left[
        \hat{u},~
        \hat{v},~
        \frac{\partial \hat{w}}{\partial z},~
        \hat{p}
    \right]
    \right|_{z=0} = [0,0,0,0] & \quad\text{for anti-symmetric modes,}
    \label{eq.bcs}
\end{align}
\end{subequations}
which result in two linear operators, $\mathsfbi{A}_{\bar{\boldsymbol{q}_0}^\text{sym}}^\text{sym}$ and $\mathsfbi{A}_{\bar{\boldsymbol{q}_0}^\text{sym}}^\text{anti-sym}$, for the symmetric base flow residing on the half-domain, $z/L_c \in [0, 8]$. For the asymmetric base flows, the linear operator, $\mathsfbi{A}_{\bar{\boldsymbol{q}_0}^\text{asym}}$, is constructed for the full-domain, $z/L_c \in [-8, 8]$, with a homogeneous boundary condition $[\hat{u},\hat{v},\hat{w},\partial \hat{p}/\partial n] = [0, 0, 0, 0]$ prescribed at the far-field.  The eigenvalue problems in equation~\eqref{eq.NSfreq} for $\mathsfbi{A}_{\bar{\boldsymbol{q}_0}^\text{sym}}^\text{sym}$,
$\mathsfbi{A}_{\bar{\boldsymbol{q}_0}^\text{sym}}^\text{anti-sym}$, and
$\mathsfbi{A}_{\bar{\boldsymbol{q}_0}^\text{asym}}$ are solved using an implicitly restarted Arnoldi method with a tolerance of $10^{-6}$.

\subsection{Results}
\label{Eigenvalueanalysis}

\begin{figure}
 \centerline{{\begin{overpic}[width=5.32in]{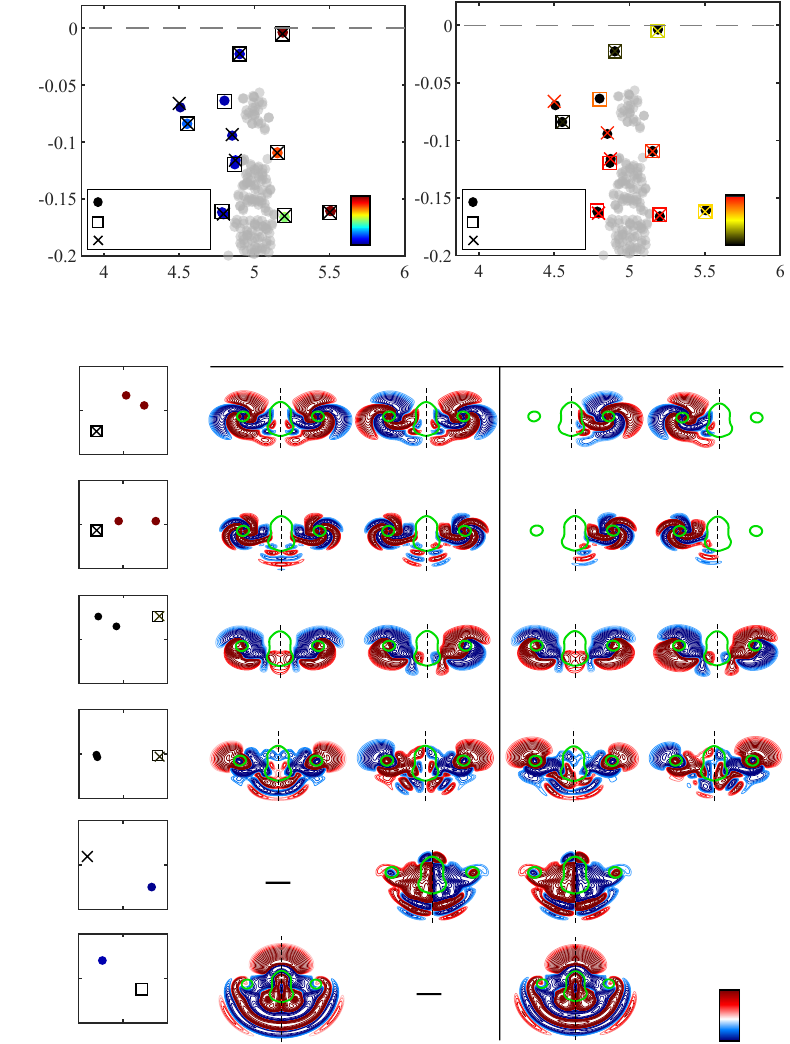} 
   \put (24 , 0) {\fontsize{8}{9}\selectfont$\lambda_i^c-\Delta$}
   \put (55 , 0) {\fontsize{8}{9}\selectfont$\lambda_i^c$}
   \put (70 , 0) {\fontsize{8}{9}\selectfont$\lambda_i^c+\Delta$}
   
   \put (12 , 286) {\fontsize{8}{9}\selectfont$-0.006$}
   \put (12 , 305) {\fontsize{8}{9}\selectfont$-0.004$}
   \put (12 , 324) {\fontsize{8}{9}\selectfont$-0.002$}
   
   \put (12 , 231) {\fontsize{8}{9}\selectfont$-0.166$}
   \put (12 , 251) {\fontsize{8}{9}\selectfont$-0.161$}
   \put (12 , 269) {\fontsize{8}{9}\selectfont$-0.156$}  

   \put (12 , 175) {\fontsize{8}{9}\selectfont$-0.024$}
   \put (12 , 194) {\fontsize{8}{9}\selectfont$-0.023$}
   \put (12 , 213) {\fontsize{8}{9}\selectfont$-0.022$}  

   \put (12 , 119) {\fontsize{8}{9}\selectfont$-0.087$}
   \put (12 , 138) {\fontsize{8}{9}\selectfont$-0.084$}
   \put (12 , 157) {\fontsize{8}{9}\selectfont$-0.081$}  

   \put (12 , 66) {\fontsize{8}{9}\selectfont$-0.072$}
   \put (12 , 84) {\fontsize{8}{9}\selectfont$-0.067$}
   \put (12 , 103) {\fontsize{8}{9}\selectfont$-0.062$}  

    \put (12 , 10) {\fontsize{8}{9}\selectfont$-0.065$}
   \put (12 , 28) {\fontsize{8}{9}\selectfont$-0.064$}
   \put (12 , 48) {\fontsize{8}{9}\selectfont$-0.063$} 
   
   \put (40 , 510) {(a)}
   \put (222, 511) {(b)}
   \put (39, 340) {(c)}
   \put (136 , 498) {{\fontsize{8}{9}\selectfont \color{violet}(1)}}
   \put (150 , 412) {{\fontsize{8}{9}\selectfont \color{violet}(2)}}
   \put (103 , 486) {{\fontsize{8}{9}\selectfont \color{violet}(3)}}
   \put (77 , 441) {{\fontsize{8}{9}\selectfont \color{violet}(4)}}
   \put (73 , 460) {{\fontsize{8}{9}\selectfont \color{violet}(5)}}
   \put (101 , 466) {{\fontsize{8}{9}\selectfont \color{violet}(6)}}

   \put (319 , 498) {{\fontsize{8}{9}\selectfont \color{violet}(1)}}
   \put (332 , 412) {{\fontsize{8}{9}\selectfont \color{violet}(2)}}
   \put (285 , 486) {{\fontsize{8}{9}\selectfont \color{violet}(3)}}
   \put (259 , 441) {{\fontsize{8}{9}\selectfont \color{violet}(4)}}
   \put (255 , 460) {{\fontsize{8}{9}\selectfont \color{violet}(5)}}
   \put (283 , 466) {{\fontsize{8}{9}\selectfont \color{violet}(6)}}
   
   \put (55 , 319.5) {{\fontsize{8}{9}\selectfont \color{violet}(1)}}
   \put (55 , 264.5) {{\fontsize{8}{9}\selectfont \color{violet}(2)}}
   \put (55 , 209) {{\fontsize{8}{9}\selectfont \color{violet}(3)}}
   \put (55 , 153.5) {{\fontsize{8}{9}\selectfont \color{violet}(4)}}
   \put (55 , 100) {{\fontsize{8}{9}\selectfont \color{violet}(5)}}
   \put (55 , 45) {{\fontsize{8}{9}\selectfont \color{violet}(6)}}

   \put (84 , 287){\rotatebox{90} {\fontsize{7}{9}\selectfont$\lambda_i^c=5.1875$}}
   \put (93 , 290) {\rotatebox{90}{\fontsize{7}{9}\selectfont$\Delta = 0.002$}}

   \put (84 , 232){\rotatebox{90} {\fontsize{7}{9}\selectfont$\lambda_i^c=5.5045$}}
   \put (93 , 235) {\rotatebox{90}{\fontsize{7}{9}\selectfont$\Delta = 0.006$}}

   \put (84 , 177){\rotatebox{90} {\fontsize{7}{9}\selectfont$\lambda_i^c=4.9020$}}
   \put (93 , 180) {\rotatebox{90}{\fontsize{7}{9}\selectfont$\Delta = 0.001$}}

   \put (84 , 122){\rotatebox{90} {\fontsize{7}{9}\selectfont$\lambda_i^c=4.5540$}}
   \put (93 , 125) {\rotatebox{90}{\fontsize{7}{9}\selectfont$\Delta = 0.003$}}

   \put (84 , 68){\rotatebox{90} {\fontsize{7}{9}\selectfont$\lambda_i^c=4.5040$}}
   \put (93 , 71) {\rotatebox{90}{\fontsize{7}{9}\selectfont$\Delta = 0.005$}}

   \put (84 , 13){\rotatebox{90} {\fontsize{7}{9}\selectfont$\lambda_i^c=4.8012$}}
   \put (93 , 16) {\rotatebox{90}{\fontsize{7}{9}\selectfont$\Delta = 0.001$}}

   \put (52 , 407) {\fontsize{7.5}{9}\selectfont Asymmetric}
   \put (52 , 397) {\fontsize{7.5}{9}\selectfont Symmetric}
   \put (52 , 389) {\fontsize{7.5}{9}\selectfont Anti-symmetric}

   \put (234 , 407) {\fontsize{7.5}{9}\selectfont Asymmetric}
   \put (234 , 397) {\fontsize{7.5}{9}\selectfont Symmetric}
   \put (234 , 389) {\fontsize{7.5}{9}\selectfont Anti-symmetric}
   
   \put (100 , 334) {\fontsize{8}{9}\selectfont Symmetric modes}
   \put (169 , 334) {\fontsize{8}{9}\selectfont Anti-symmetric modes}
   \put (278 , 334) {\fontsize{8}{9}\selectfont Asymmetric modes}
   
   \put (352 , 29) {\fontsize{8}{9}\selectfont $\hat{u}$}
   \put (362 , 1) {\fontsize{8}{9}\selectfont -0.03}
   \put (365 , 21.5) {\fontsize{8}{9}\selectfont 0.03}

   \put (173 , 416.5) {\fontsize{8}{9}\selectfont $\psi$}
   \put (183 , 388) {\fontsize{8}{9}\selectfont $0$}
   \put (183 , 407) {\fontsize{8}{9}\selectfont $0.7$}
   
   \put (355 , 415.5) {\fontsize{8}{9}\selectfont $\kappa$}
   \put (365 , 388) {\fontsize{8}{9}\selectfont $0.5$}
   \put (365 , 407) {\fontsize{8}{9}\selectfont $1$}   
   \put (2 , 14) {\rotatebox{90}{\fontsize{8}{9}\selectfont$\lambda_r L_c/u_\infty$}}
   \put (2 , 70) {\rotatebox{90}{\fontsize{8}{9}\selectfont$\lambda_r L_c/u_\infty$}}
   \put (2 , 123) {\rotatebox{90}{\fontsize{8}{9}\selectfont$\lambda_r L_c/u_\infty$}}
   \put (2 , 178) {\rotatebox{90}{\fontsize{8}{9}\selectfont$\lambda_r L_c/u_\infty$}}
   \put (2 , 234) {\rotatebox{90}{\fontsize{8}{9}\selectfont$\lambda_r L_c/u_\infty$}}
   \put (2 , 289) {\rotatebox{90}{\fontsize{8}{9}\selectfont$\lambda_r L_c/u_\infty$}}
   \put (6 , 428) {\rotatebox{90}{\fontsize{8}{9}\selectfont$\lambda_r L_c/u_\infty$}}
   \put (98 , 362) {\rotatebox{0}{\fontsize{8}{9}\selectfont$\lambda_i L_c/u_\infty$}}
   \put (281 , 362) {\rotatebox{0}{\fontsize{8}{9}\selectfont$\lambda_i L_c/u_\infty$}}
\end{overpic}}}
\caption{The overlaid eigenvalue spectra of the symmetric and asymmetric base flows: (a) asymmetric eigenmodes colored by their levels of modal asymmetry, $\psi$; (b) eigenmodes of the symmetric base flow colored by their modal non-normality levels, $\kappa$; (c) symmetric and anti-symmetric eigenmodes obtained from the symmetric base flow 
and asymmetric eigenmodes from the asymmetric base flow. The corresponding eigenvalues are indicated in table~\ref{tbl.EV}.
 }  
\label{fig.LinStabilitySpectrum}
\end{figure}
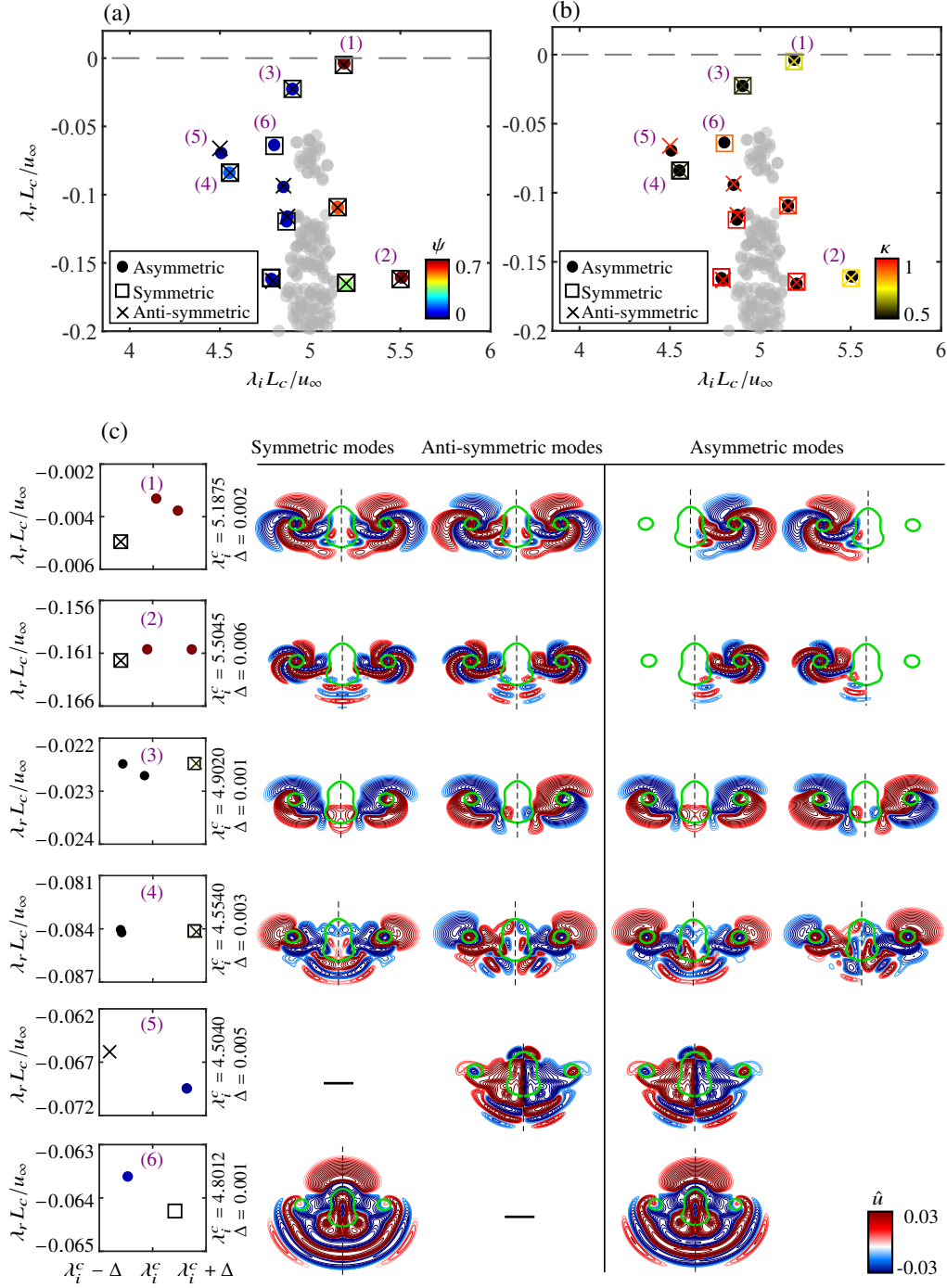

We begin by examining the eigenvalue spectra for both the symmetric and asymmetric base flows. For the analysis here, we focus on the streamwise wavenumber $k L_c = 5$, which corresponds to the dominant streamwise wavelength of the wake structures within the rectangular domain of analysis.
The spectra for both base flows are shown in figure~\ref{fig.LinStabilitySpectrum}. For the eigenmodes from the symmetric base flow in figure~\ref{fig.LinStabilitySpectrum}, we mark the symmetric and anti-symmetric eigenmodes by $\square$ and $\cross$, respectively. The eigenmodes obtained from the asymmetric base flow are marked by $\bullet$, overlaid on the spectrum of the symmetric base flow for a direct comparison. In figure~\ref{fig.LinStabilitySpectrum}(a), we highlight the physical eigenmodes from the asymmetric base flow by the levels of spanwise asymmetry of their modal structure. This asymmetry level is quantified using
\begin{equation}
\psi =  \frac{  \lVert~ |\hat{\boldsymbol{q}}_{\lambda,k}(y, z^+)| 
                      - |\hat{\boldsymbol{q}}_{\lambda,k}(y, z^-)|~
                \rVert
        }{\lVert\hat{\boldsymbol{q}}_{\lambda,k}\rVert} ,
\label{eq.asym}
\end{equation}
where $z^+$ denotes a spanwise coordinate on the $z>0$ side of the domain and $z^-$ is its mirror location about the plane of symmetry, $z=0$. We note that, since equation~\eqref{eq.asym} dismisses the oscillatory nature of the eigenmodes by considering their absolute values, the asymmetry level quantified by \eqref{eq.asym} represents a time-invariant spanwise asymmetry exhibited by an eigenmode. A perfectly symmetric or anti-symmetric eigenmode yields $\psi = 0$, while large values of $\psi$ indicate high levels of asymmetry in the eigenmodes. The normalization ensures that $\psi \in [0, 1]$.  Similar to figure~\ref{fig.LinStabilitySpectrum}(a), in figure~\ref{fig.LinStabilitySpectrum}(b) we color the eigenmodes from the symmetric base flow by their levels of modal non-normality, $\kappa = 1 - \langle \hat{\boldsymbol{q}},~\hat{\boldsymbol{q}}^\dagger \rangle$.  An eigenmode is said to exhibit high levels of modal non-normality if the value of $\kappa$ is far from zero.  Spurious eigenmodes are shown in gray in figure~\ref{fig.LinStabilitySpectrum}(a) and (b). We also choose six representative groups of eigenmodes, marked as clusters (1) to (6). For each cluster of eigenmodes, we show a zoom-in spectrum in figure~\ref{fig.LinStabilitySpectrum}(c) to examine the local scattering of eigenvalues before and after the asymmetric disturbance is introduced to the base flow.  The eigenmodes from both symmetric and asymmetric base flows are also visualized in figure~\ref{fig.LinStabilitySpectrum}(c) by the real part of the streamwise velocity components.  Note that these eigenmodes are visualized on top of a green contour line, which portrays the base flow by the regions where streamwise velocity exhibits a deficit.

With the results in figure~\ref{fig.LinStabilitySpectrum}(a) and (b), we make an important observation that highly asymmetric eigenmodes from the asymmetric base flow emerge specifically in the spectral regions where symmetric and anti-symmetric eigenmodes from the symmetric base flow appear as pairs and exhibit pronounced modal non-normality. This finding is particularly evident for the eigenmode pairs in clusters (1) and (2) in the spectra, where the highest levels of asymmetry exhibited by the eigenmodes are closely tied to the high modal non-normality of the unperturbed symmetric and anti-symmetric eigenmode pairs. The modal non-normality in these regions increases the sensitivity of the eigenmodes to asymmetric disturbances, resulting in interference between the symmetric and anti-symmetric mode pair to form two highly asymmetric modes from the asymmetric base flow. As opposed to the eigenmodes in clusters (1) and (2), those in clusters (3) and (4) inversely support the condition for the emergence of highly asymmetric modes: although the eigenmodes in clusters (3) and (4) appear as pairs of symmetric and anti-symmetric modes, these modes from the symmetric base flow exhibit low levels of non-normality and are hence robust against the asymmetric departure of the base flow.  As a result, the perturbed eigenmodes from the asymmetric base flow exhibit low levels of asymmetry due to their low sensitivity to asymmetric disturbances. Lastly, the eigenmodes in clusters (5) and (6) are examples where symmetric and anti-symmetric modes do not come in pairs. For the eigenmodes in these two clusters, since the perturbed eigenmodes are not accompanied by a symmetric or anti-symmetric counterpart to interfere with, the perturbed eigenmodes from the asymmetric base flow exhibit low levels of asymmetry.  We also note that, while the above observations were made for the asymmetric base flow obtained with a spherical Gaussian forcing discussed in~\S\ref{asymforcing}, they sustain even for an asymmetrically disturbed wake flow with random volumetric forcing, as concluded in appendix \ref{apndx.randforcing}.  

\begin{table}
  \begin{center}
  \def~{\hphantom{0}}

  \begin{tabular}{lcccc}

           & Symmetric mode& Anti-symmetric mode& Asymmetric mode 1 & Asymmetric mode 2 \\
    \specialrule{1pt}{3pt}{3pt}
       (1) &
       \scalebox{0.92}{$-0.004951 + 5.186271\mathrm{i}$} &
       \scalebox{0.92}{$-0.004953 + 5.186270 \mathrm{i}$} &
       \scalebox{0.92}{$-0.003312 + 5.187614\mathrm{i}$} &
       \scalebox{0.92}{$-0.003770 + 5.188433 \mathrm{i}$} \\[4pt]

       (2) & \scalebox{0.92}{$-0.161820 + 5.500886\mathrm{i}$} & \scalebox{0.92}{$-0.161820 + 5.500886 \mathrm{i}$} & \scalebox{0.92}{$-0.160555 + 5.503884\mathrm{i}$} & \scalebox{0.92}{$-0.160563 + 5.508942\mathrm{i}$}\\[4pt]
       (3) & \scalebox{0.92}{$-0.022477 + 4.902835 \mathrm{i}$} & \scalebox{0.92}{$-0.022477 + 4.902871 \mathrm{i}$} & \scalebox{0.92}{$-0.022484 + 4.901487\mathrm{i}$} & \scalebox{0.92}{$-0.022707 + 4.901897 \mathrm{i}$}\\[4pt]
       (4) & \scalebox{0.92}{$-0.083968 + 4.556451\mathrm{i}$} & \scalebox{0.92}{$-0.083918 + 4.556331\mathrm{i}$} & \scalebox{0.92}{$-0.083844 + 4.552176\mathrm{i}$} & \scalebox{0.92}{$-0.084010 + 4.552241\mathrm{i}$} \\[4pt]
       (5) & - & \scalebox{0.92}{$-0.066025 +
       4.499970\mathrm{i}$} & \scalebox{0.92}{$-0.069462 + 4.507199\mathrm{i}$} &  -\\[4pt]
       (6) & \scalebox{0.92}{$-0.064246 + 4.801617\mathrm{i}$} & - & \scalebox{0.92}{$-0.063598 + 4.800732\mathrm{i}$} & - \\[4pt]
  \end{tabular}

\caption{The eigenvalues associated with clusters (1)-(6) in figure~\ref{fig.LinStabilitySpectrum}. The symmetric and anti-symmetric modes are obtained from the symmetric base flow, while asymmetric modes 1 and 2 correspond to the eigenmodes computed from the asymmetrically disturbed base flow.}
  \label{tbl.EV}
  \end{center}
\end{table}

To investigate the physical mechanisms for the manifestation of the flow asymmetries, we next examine the structures of the eigenmodes in figure~\ref{fig.LinStabilitySpectrum}(c) for each eigenmode cluster. We begin with the modal structures for the eigenvalues in clusters (1) and (2). Notably, the anti-symmetric mode in cluster (1) corresponds to the eigenvalue with the highest growth rate in the spectrum. In figure~\ref{fig.LinStabilitySpectrum}(c), we observe that the eigenmodes in these two clusters demonstrate elliptic vortex instabilities \citep{benton2021crow,leweke2016dynamics,tsai1976stability}, with both symmetric and anti-symmetric structures. Upon introducing small-amplitude asymmetric disturbances, each eigenmode pair undergoes modal interference between the symmetric and anti-symmetric structures, producing two distinct asymmetric eigenmodes from the asymmetric base flow. 
Moreover, the asymmetric eigenmodes exhibit higher growth rates than those of the symmetric and anti-symmetric mode pairs from the symmetric base flow in clusters (1) and (2). This is evident in the locally zoom-in spectra, with corresponding eigenvalues reported in table~\ref{tbl.EV}. This emphasizes the role of modal interference between the symmetric and anti-symmetric global modes in the formation of spanwise-asymmetric modes. We note that similar observations were made in the symmetry-breaking problems in quantum mechanics, where asymmetric disturbances trigger interference between paired modes \citep{magri2023linear}.

Eigenmodes in clusters (3) and (4) are examples where paired symmetric and anti-symmetric modes from the symmetric base flow do not result in highly asymmetric modes from the asymmetric base flow, due to low modal non-normality. The low non-normality makes the eigenmodes less susceptible to the asymmetric departure of the base flow, preventing the interference between the pairs of symmetric and anti-symmetric modes. Consequently, the eigenmodes from the asymmetric base flow preserve almost identical structures to those from the symmetric base flow and thus exhibit very low levels of asymmetry. Also, the eigenvalues in clusters (3) and (4) undergo much smaller displacement due to the asymmetric departure of the base flow than they do in clusters (1) and (2), as is clearly seen in the eigenvalues reported in table~\ref{tbl.EV}.  These findings suggest that, while the appearance of paired symmetric and anti-symmetric modes from the symmetric base flow is a prerequisite for highly asymmetric modes to emerge from the asymmetric base flow, the resulting modal asymmetry also depends on the modal non-normality of the eigenmode pair, which dictates the sensitivity of the eigenmodes to disturbances.  For the single eigenmodes in clusters (5) and (6), eigenmodes from the asymmetric base flow inherit the same symmetric or anti-symmetric structures as those from the symmetric base flow and exhibit low levels of modal asymmetry.  This once again underscores the role of modal interference in the formation of asymmetric wake structures.  Interestingly, the structures of these single eigenmodes are characterized by strong fluctuations in the wake region, as opposed to those in clusters (1)-(4) where modal structures show strong fluctuations around the tip vortices. Moreover, the wake-dominated single modes exhibit phase velocities closer to that of the base flow.

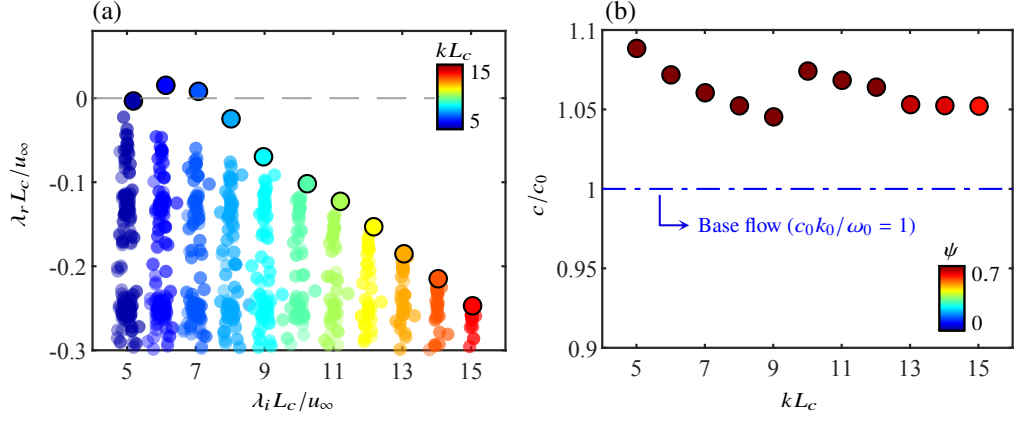
\begin{figure}
  \centerline{\begin{overpic}[width=5.32in]{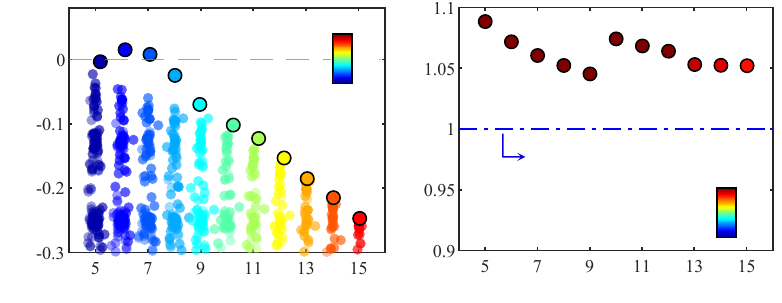} 
   \put (34, 149) {(a)}
   \put (227 , 149) {(b)}
   \put (163 , 135) {\fontsize{8}{9}\selectfont$k L_c$}
   \put (176 , 108) {\fontsize{8}{9}\selectfont 5}
   \put (176 , 126.5) {\fontsize{8}{9}\selectfont 15}
    \put (3 , 71) {\rotatebox{90}{\fontsize{8}{9}\selectfont$\lambda_r L_c/u_\infty$}}
    \put (198 , 75) {\rotatebox{90}{\fontsize{8}{9}\selectfont$c/c_0$}}
   \put (293 , 3) {\rotatebox{0}{\fontsize{8}{9}\selectfont$k L_c$}}
   \put (94 , 3) {\rotatebox{0}{\fontsize{8}{9}\selectfont$\lambda_i L_c/u_\infty$}}
   \put (262, 69) {\rotatebox{0}{{\fontsize{8}{9}\selectfont {\color{blue}Base flow ($c_0 k_0/ \omega_0 =1$)}}}}
   \put (354 , 60) {\rotatebox{0}{\fontsize{8}{9}\selectfont$\psi$}}
   \put (365, 50.5) {\rotatebox{0}{\fontsize{8}{9}\selectfont$0.7$}}
   \put (365, 32) {\rotatebox{0}{\fontsize{8}{9}\selectfont$0$}}
\end{overpic}}
\caption{(a) Eigenvalue spectrum of the asymmetrically disturbed base flow for different streamwise wavenumbers, $k$. The eigenmode with the highest growth rate at each $k$ is highlighted with a black-edged marker. (b) Phase velocity of the dominant eigenmodes, $c=\lambda_i/k$, plotted against the corresponding streamwise wavenumber, with the markers colored by the asymmetry level of the corresponding eigenmodes. The blue dashed line denotes the phase velocity of the base flow, $c_0 = \omega_0/k_0 $.}  
\label{fig.Cntsphase}
\end{figure}

The linear stability analysis performed this far was concerned with a fixed streamwise wavenumber of $kL_c = 5$. In figure~\ref{fig.Cntsphase}, we extend the range of streamwise wavenumbers to $kL_c \in [5, 15]$ with an increment of $\Delta k L_c= 1$ for the analyses.  The resulting spectra for different values of $k$ are shown in figure~\ref{fig.Cntsphase}(a).  Here, the spectra obtained from different values of $k$ are overlaid with different colors, and the eigenmode with the highest growth rate at each $k$ is highlighted by a black-edged marker.  Recall that these dominant eigenmodes, such as those appearing in cluster (1) in figure~\ref{fig.LinStabilitySpectrum}, are also those that exhibit the highest levels of asymmetry, which are of particular interest for the present study.  For these dominant eigenmodes, we compute their phase velocities as $c = \lambda_i/k$ and plot them against the corresponding wavenumbers in figure~\ref{fig.Cntsphase}(b).  Each data point of these phase velocities is colored by the asymmetry level of the corresponding eigenmodes.  In the same figure, we also show a blue dashed line to mark the phase velocity of the dominant unsteady wake structures from the base flow, as shown in figure~\ref{fig.FFT}. This allows us to examine the variation of the phase velocities of these highly asymmetric dominant eigenmodes over the chosen range of streamwise wavenumbers and compare them with that of the base flow. 

From the analysis in figure~\ref{fig.Cntsphase}, we observe that, over the chosen range of wavenumbers, the phase velocities of these dominant eigenmodes exhibit small fluctuations about their mean (less than $3\%$). This indicates that the development of these eigenmodes leverages the convective physics of the base flow and propagates downstream with a nearly constant phase velocity across different values of $k$. Moreover, we make an important observation that the phase velocities of these dominant highly asymmetric eigenmodes are all slightly higher than that of the base flow. This observation suggests that the interactions between the wake structures of different phase velocities may play a crucial role in the formation of flow asymmetries in the wake flow.  This potential mechanism motivates the STHR analysis in the next section to investigate the interaction dynamics among perturbations of different phase velocities.

\section{Spatio-temporal harmonic resolvent analysis}
\label{HarmonicResolvent}
The global mode sensitivity analysis in the previous section suggests that the formation of the wake asymmetry can be attributed to the modal phase interference between the symmetric and anti-symmetric modes of the intrinsically symmetric wake flow.  Also, the dominant eigenmodes that exhibit high levels of modal asymmetry all propagate at similar phase velocities that are slightly higher than that of the base flow.  These findings necessitate an analysis that provides insights into the phase dynamics of perturbations while accounting for interactions between perturbations that travel at different phase velocities.  Below, we introduce a spatio-temporal harmonic resolvent (STHR) analysis that is capable of uncovering non-arbitrary phase dynamics while allowing for interactions among perturbations, in order to investigate the physical mechanisms that are responsible for the emergence of long-time flow asymmetries in the wake. 

\subsection{Formulation}
\label{HR_Mathematicalformulation}
To motivate the need for STHR analysis for the present study, let us first briefly discuss the classical resolvent analysis \citep{trefethen1993hydrodynamic, mckeon2010critical, jovanovic2005componentwise, yeh2019resolvent} to remark on its limitations.   The classical resolvent analysis considers the linearized NS equations about a time-invariant base flow, $\bar{\boldsymbol{q}}_0$.  For incompressible flows, this yields a system of equations that can be expressed as 
\begin{equation}
\mathsfbi{M}\frac{\mathrm{d}\boldsymbol{q}^\prime}{\mathrm{d} t} = \mathsfbi{A}_{\bar{\boldsymbol{q}}_0} \boldsymbol{q}^\prime + 
\mathsfbi{M}
\boldsymbol{f^\prime}.
\label{eq.linearNS}
\end{equation}
Note that this is the same equation as \eqref{eq.HlinearNS} with an additional forcing term, $\boldsymbol{f}^\prime$, on the right-hand side. With the spatial homogeneity in the base flow in the streamwise direction, we can express the flow response $\boldsymbol{q}^\prime$ and forcing $\boldsymbol{f}^\prime$ as the sum of spatio-temporal Fourier modes as,
\begin{equation}
    \left[\boldsymbol{q}^\prime(\boldsymbol{x}, t),~\boldsymbol{f}^\prime(\boldsymbol{x}, t)\right] 
    = \int_{-\infty}^{\infty} \int_{-\infty}^{\infty} 
    \left[
        \hat{\boldsymbol{q}}(y, z),~
        \hat{\boldsymbol{f}}(y, z)
    \right]_{\omega, k} 
    \mathrm{e}^{\mathrm{i}(\omega t - k x)} {\rm d}{\omega}~{\rm d}{k},
\end{equation}
where $[\hat{\boldsymbol{q}},~\hat{\boldsymbol{f}}]_{\omega, k}$ are the bi-global modes for response and forcing, respectively, at the streamwise wavenumber $k$ and frequency $\omega$. By substituting the Fourier representations into equation~\eqref{eq.linearNS},  we arrive at linearized NS equations in the Fourier space given by
\begin{equation}
    \hat{\boldsymbol{q}}_{\omega,k}
    =
    \mathsfbi{R}_{\bar{\boldsymbol{q}}_0}(\omega,k) \hat{\boldsymbol{f}}_{\omega,k},
    \quad\text{where}~~
    \mathsfbi{R}_{\bar{\boldsymbol{q}}_0}(\omega,k) = \left[\mathrm{i} \omega  
\mathsfbi{M}
- \mathsfbi{A}_{\bar{\boldsymbol{q}}_0}(k)\right]^{-1}
\mathsfbi{M}
\label{eq.classicresolvent}
\end{equation}
is the resolvent operator for the linearized NS equations. It can be understood as a transfer function between the forcing, $\hat{\boldsymbol{f}}_{\omega,k}$, and response, $\hat{\boldsymbol{q}}_{\omega,k}$ \citep{schmid2001stability, jovanovic2005componentwise}. Resolvent analysis generally seeks a singular value decomposition (SVD) of $\mathsfbi{R}_{\bar{\boldsymbol{q}}_0}(\omega,k)$, where the extracted left singular vectors are referred to as the response modes, the right singular vectors as the forcing modes, and the corresponding singular value is the amplification of the forcing-response pair. 

We note that, since the resolvent operator, $\mathsfbi{R}_{\bar{\boldsymbol{q}}_0}$, is constructed about a time-invariant base state ($\bar{\boldsymbol{q}}_0$), the phase dynamics represented in equation~\eqref{eq.classicresolvent} involve only relative phases between the forcing and response modes at the same frequency. In other words, the phase of the response mode depends solely on that of the forcing, and hence the phase dynamics of resolvent modes at different frequencies are completely decoupled \citep{mckeon2019applications}. Therefore, to study the role of phase in the emergence of flow asymmetries, we extend the classical resolvent analysis to the STHR framework that provides non-arbitrary phase information for the resolvent modes while enabling interactions between modes of different phase velocities.  The STHR analysis allows us to handle base flows that exhibit periodicity in both space and time.  Such a periodic behavior well characterizes the wake flow in the rectangular domain, as discussed in figure~\ref{fig.FFT}. The STHR framework  combines the resolvent formulations of \citet{chavarin2020resolvent}, which is concerned with spatially periodic base flows, and 
\citet{padovan2020analysis}, which is concerned with temporally periodic base flows.  Below, we discuss the formulation of STHR analysis in detail while focusing on how it unravels interactions between perturbations of different phase velocities and provides the non-arbitrary phase dynamics among them.

Consider the same forced linearized NS equation in \eqref{eq.linearNS} but about a time-varying base flow, $\bar{\boldsymbol{q}}(\boldsymbol{x}, t)$.  equation~\eqref{eq.linearNS} becomes 
\begin{equation}
    \mathsfbi{M}\frac{\mathrm{d}\boldsymbol{q}^\prime}{\mathrm{d} t} 
    = 
    \mathsfbi{A}_{\bar{\boldsymbol{q}}(\boldsymbol{x}, t)} \boldsymbol{q}^\prime 
    + 
    \mathsfbi{M}\boldsymbol{f^\prime}.
\label{eq.PerdclinearNS}
\end{equation}
Here, we limit the analysis to base flows of spatio-temporal periodicity, and $\mathsfbi{A}_{\bar{\boldsymbol{q}}(\boldsymbol{x}, t)}$ is the linearized NS operator about such a base flow.  Assuming the spatial periodicity of the base flow in the streamwise direction, a Fourier representation of $\bar{\boldsymbol{q}}(\boldsymbol{x}, t)$ can be found as
\begin{equation}
\bar{\boldsymbol{q}}(\boldsymbol{x},t) = 
\sum_{n = -\infty}^{+\infty} \tilde{\boldsymbol{q}}_{n \omega_0, n k_0}(y, z) \mathrm{e}^{\mathrm{i}n(\omega_0 t - k_0 x)}, 
\label{eq.periodicBFFourier}
\end{equation}
where $\omega_0$ and $k_{0}$ represent the fundamental frequency and  streamwise wavenumber of the base flow, respectively, and $n \in \mathbb{Z}$ is the index of the spatio-temporal harmonics.  Note that such a representation of the base flow is conceptually analogous to the parallel-flow assumption, as the Fourier coefficients $\tilde{\boldsymbol{q}}_{n \omega_0, n k_0}(y, z)$ are constant in the streamwise direction. Moreover, the phase information of the base flow is embedded in its complex-valued Fourier coefficients, $\tilde{\boldsymbol{q}}_{n \omega_0, n k_0}$.  Similar to the Fourier representation of the base flow, we seek response and forcing in the form of 
\begin{equation}
\label{eq.FourierRep}
    \left[\boldsymbol{q}^\prime(\boldsymbol{x},t),~\boldsymbol{f}^\prime(\boldsymbol{x},t)\right]
    = \mathrm{e}^{\mathrm{i}\gamma t}
        \sum_{m=-\infty}^{+\infty} \sum_{r=-\infty}^{+\infty}
        \left[  \hat{\boldsymbol{q}}(y,z),~
                \hat{\boldsymbol{f}}(y,z)
        \right]_{m \omega_0+\gamma,\,r k_0}
        \mathrm{e}^{\mathrm{i}(m \omega_0 t - r k_0 x)},
\end{equation}
where $\gamma \in \mathbb{R}$ and $(m,r) \in \mathbb{Z}$. Note that the exponential prefactor, $\mathrm{e}^{\mathrm{i}\gamma t}$, shifts the frequency of the perturbations away from those of the base flow ($m\omega_0$), and we can choose different values of $\gamma$ to prescribe such a frequency shift \citep{padovan2022analysis}.

Let us make some important remarks on the modal representations of the base flow \eqref{eq.periodicBFFourier} and perturbations \eqref{eq.FourierRep} in regard to their phase velocities. For the base flow, its Fourier representation in equation~\eqref{eq.periodicBFFourier} implies that all Fourier modes of $\bar{\boldsymbol{q}}(\boldsymbol{x}, t)$ propagate at a constant phase velocity, which is given by
\[
    c_0 = \frac{n\omega_0}{nk_0} = \frac{\omega_0}{k_0}.
\]
As opposed to the base flow, however, the modal form of the perturbations in equation~\eqref{eq.FourierRep} permits the perturbations to travel at different phase velocities from that of the base flow, $c_0$.  Observing \eqref{eq.FourierRep}, the phase velocity of a forcing-response pair, $[\hat{\boldsymbol{q}},~\hat{\boldsymbol{f}}]_{m \omega_0+\gamma,\,r k_0}$, is given by 
\begin{equation}
    c = \frac{m\omega_0+\gamma}{rk_0} = \frac{m}{r} \times c_0 + \frac{\gamma}{rk_0}.
    \label{eq.modalphasevel}
\end{equation}
When $\gamma = 0$, the phase velocity becomes $mc_0/r$, which can be different from $c_0$ if $m \neq r$.  Also, when $m$ and $r$ hold opposite signs, the resulting negative phase velocity propagates the perturbation upstream along the wake, providing insights into mechanisms associated with absolute instabilities \citep{huerre1985absolute, monkewitz1988absolute}. Moreover, when $\gamma \neq 0$ and the perturbation frequency is shifted from a harmonic of the base flow, different values of $\gamma$ allow us to fine-tune the phase velocity of the perturbation around $m c_0/r$. This capability of fine tuning the phase velocities for the perturbations is crucial in the present study, since the results in figure~\ref{fig.Cntsphase}(b) indicate that the dominant eigenmodes of high levels of spanwise asymmetry propagate at a phase velocity slightly higher than that of the base flow. 

Since the base flow has a spatio-temporal periodicity, the linearized NS operator in equation~\eqref{eq.linearNS}, $\mathsfbi{A}_{\bar{\boldsymbol{q}}(\boldsymbol{x}, t)}$, inherits such a periodicity of the base flow.  For a linearized NS operator acting on perturbations of a wavenumber $k = rk_0$, the inherited periodicity of the operator admits a Fourier representation for $\mathsfbi{A}_{\bar{\boldsymbol{q}}(\boldsymbol{x}, t)}(rk_0)$, given by 
\begin{align}
    \mathsfbi{A}_{\bar{\boldsymbol{q}}(\boldsymbol{x}, t)}(rk_0)
    &= \sum_{n = -\infty}^{+\infty} \mathsfbi{\hat{A}}_{n\omega_0, n k_0} (rk_0) \mathrm{e}^{\mathrm{i}n( \omega_0 t -  k_0 x)}.  
\label{eq.LinOptA}
\end{align}
Note that the Fourier component for $n = 0$, or $\mathsfbi{\hat{A}}_{0, 0}(rk_0)$, is equivalent to the linear NS operator about the time- and streamwise-averaged base flow that appeared in equations \eqref{eq.linearNS} and \eqref{eq.classicresolvent}, i.e. $
    \mathsfbi{\hat{A}}_{0, 0}(rk_0) = \mathsfbi{A}_{\bar{\boldsymbol{q}}_0}(rk_0).$ Also, the Fourier components for $n \neq 0$, $\mathsfbi{\hat{A}}_{n\omega_0, nk_0}(rk_0)$ is associated with the linearized convective terms about $\tilde{\boldsymbol{q}}_{n\omega_0, nk_0}$, the Fourier components of the base flow.  Hence, the phase information of $\tilde{\boldsymbol{q}}_{n \omega_0, n k_0}$ is thereby encoded into $\mathsfbi{\hat{A}}_{n\omega_0, n k_0}$ with $n \neq 0$ \citep{leclercq2023mean}.  Explicit forms of $\mathsfbi{\hat{A}}_{n\omega_0, n k_0}$ are provided in appendix \ref{apndx.Ahats}. 

Substituting the Fourier representations in equations \eqref{eq.LinOptA} and \eqref{eq.FourierRep} into equation~\eqref{eq.PerdclinearNS}, the linearized NS equation about the spatio-temporal periodic base flow becomes
\begin{equation}
    \mathrm{i}(m \omega_0+\gamma) \mathsfbi{M} \hat{\boldsymbol{q}}_{m \omega_0+\gamma, r k_0} 
    = 
    \sum_{n + \check{m} = m} \sum_{n + \check{r} = r}
    \mathsfbi{\hat{A}}_{n\omega_0, nk_0} (\check{r}k_0) \hat{\boldsymbol{q}}_{\check{m}\omega_0+\gamma, \check{r}k_0} 
    +
    \mathsfbi{M}\hat{\boldsymbol{f}}_{m \omega_0+\gamma, r k_0},
\label{eq.harmonic}
\end{equation}
with $(n, m, r, \check{m}, \check{r}) \in \mathbb{Z}$ and $\gamma \in \mathbb{R}$.  The equation above represents a system of equations that governs the evolution of coupled flow response $\hat{\boldsymbol{q}}$ to forcing $\hat{\boldsymbol{f}}$, between all frequencies and wavenumbers, $(m \omega_0+\gamma, r k_0)$. The first term on the right-hand side couples the response at the frequency-wavenumber of ($m \omega_0+\gamma, r k_0$) with that of ($\check{m} \omega_0+\gamma, \check{r} k_0$), through the base flow harmonic of ($n \omega_0, n k_0$). It is important to recall that the phase of the base flow is encoded into the linear operators $\mathsfbi{\hat{A}}_{n\omega_0, nk_0}$ with $n \ne 0$.  Therefore, as opposed to the classical resolvent, the phases of $\hat{\boldsymbol{q}}$ and $\hat{\boldsymbol{f}}$ of different frequencies and wavenumbers are all related to that of the base flow and are no longer arbitrary.

Following the system of equations in \eqref{eq.harmonic}, we can find an explicit expression for the coupled response modes in a matrix form,
\begin{equation}
    \label{eq.TQ}
    \hat{\boldsymbol{\mathcal{Q}}} 
    = 
    \mathsfbi{H}_{\tilde{\boldsymbol{q}}}
    \left(\gamma\right)
        \hat{\boldsymbol{\mathcal{F}}},
    \quad\text{where}~~
    \mathsfbi{H}_{\tilde{\boldsymbol{q}}} \left(\gamma\right)
    =
    \left[\mathrm{i} \gamma \mathsfbi{I}_\mathsfbi{M} 
        - \mathsfbi{T}_{\tilde{\boldsymbol{q}}} 
    \right]^{-1} \mathsfbi{I}_\mathsfbi{M}    
\end{equation}
is referred to as the STHR operator about the Fourier-transformed base flow, $\tilde{\boldsymbol{q}}$.  In equation~\eqref{eq.TQ}, the vectors $\hat{\boldsymbol{\mathcal{Q}}}$ and $\hat{\boldsymbol{\mathcal{F}}}$ respectively denote the stacked response and forcing modes of different frequencies and wavenumbers. The matrix $\mathsfbi{I}_{\mathsfbi{M}}$ is a block-diagonal matrix with each diagonal block given by $\mathsfbi{M}$.  The matrix $\mathsfbi{T}_{\tilde{\boldsymbol{q}}}$ is referred to as the Hill matrix, whose block structure is associated with the Fourier components of the linear NS operators, $\mathsfbi{\hat{A}}_{n\omega_0, n k_0}$. We note the resemblance of equation~\eqref{eq.TQ} to equation~\eqref{eq.classicresolvent} from the classical resolvent analysis, but emphasize again that equation~\eqref{eq.TQ} provides non-arbitrary phase information across different frequency–wavenumber components contained in $\hat{\boldsymbol{\mathcal{Q}}}$ and $\hat{\boldsymbol{\mathcal{F}}}$, since these components are dynamically coupled through the base flow.  

Theoretically, the matrices and stacked vectors in equation~\eqref{eq.TQ} are all infinite in size, given the infinite Fourier series representations of the base flow and perturbations. For numerical practice, we need to truncate these Fourier series such that $n$, $m$, and $r$ in \eqref{eq.periodicBFFourier} and \eqref{eq.FourierRep} are all finite integers and do not go to infinity.  That is, the indices for the truncated Fourier modes satisfy
$n \in \{-\tilde{n},\dots,0,\dots,\tilde{n}\}$,
$m \in \{-\tilde{m},\dots,0,\dots,\tilde{m}\}$, and
$r \in \{-\tilde{r},\dots,0,\dots,\tilde{r}\}$, 
where $\tilde{n}$, $\tilde{m}$, and $\tilde{r}$ are finite integers, and the corresponding frequencies and wavenumbers belong to the sets of
\begin{equation}
    \begin{cases}
        \boldsymbol{\Omega}_b
            &=\{-\tilde{n}, \dots, 0, \dots, \tilde{n} \}\omega_0 \\
        \boldsymbol{K}_b
            &=\{-\tilde{n}, \dots, 0, \dots, \tilde{n} \}k_0
    \end{cases}
    ~~\text{and}\quad
    \begin{cases}
        \boldsymbol{\Omega} 
        &=\{ -\tilde{m}, \ldots, 0, \ldots, \tilde{m} \} \omega_0 + \gamma\\ 
    \boldsymbol{K}
        &=\{ -\tilde{r}, \ldots, 0, \ldots, \tilde{r} \} k_0
    \end{cases}
   ~~\text{,}
    \label{eq.fwsets}
\end{equation}
respectively for the base flow and perturbations.  Since the Hill matrix and STHR operator are both related to these finite sets of frequencies and wavenumbers, we have
\begin{equation}
    \mathsfbi{T}_{\tilde{\boldsymbol{q}}} 
    = 
    \mathsfbi{T}_{\tilde{\boldsymbol{q}}}
    \left(
        \boldsymbol{\Omega}, \boldsymbol{K}; ~\boldsymbol{\Omega}_b, \boldsymbol{K}_b
    \right) 
    \quad\text{and}\quad
    \mathsfbi{H}_{\tilde{\boldsymbol{q}}} 
    = 
    \mathsfbi{H}_{\tilde{\boldsymbol{q}}}
    \left(
        \gamma,\boldsymbol{\Omega}, \boldsymbol{K}; ~\boldsymbol{\Omega}_b, \boldsymbol{K}_b
    \right).
\end{equation}
More details on the construction of these matrices for chosen sets of frequencies and wavenumbers are provided in appendix \ref{apndx.HRMath}.  Once the finite-size Hill matrix $\mathsfbi{T}_{\tilde{\boldsymbol{q}}}$ is formed, our STHR analysis then seeks an SVD of the corresponding STHR operator as 
\begin{equation}
    \mathsfbi{H}_{\tilde{\boldsymbol{q}}} 
    \left[\mathrm{i} \gamma \mathsfbi{I}_\mathsfbi{M} 
        - \mathsfbi{T}_{\tilde{\boldsymbol{q}}}
    \left(\boldsymbol{\Omega}, \boldsymbol{K}; ~\boldsymbol{\Omega}_b, \boldsymbol{K}_b\right) 
    \right]^{-1} \mathsfbi{I}_\mathsfbi{M}
   = \mathsfbi{Q} \boldsymbol{\Sigma}  \mathsfbi{F}^*.
\end{equation}
The stacked forcing modes $\hat{\boldsymbol{\mathcal{F}}}$ of different frequencies and wavenumbers in $\boldsymbol{\Omega}$ and $\boldsymbol{K}$ correspond to the columns of $\mathsfbi{F}$.  Similarly, the response modes are the columns of $\mathsfbi{Q}$. The amplifications between the response and forcing modes are encapsulated in the diagonal matrix $\boldsymbol{\Sigma}$ as the ranked singular values. Note that for non-uniform grids, a diagonal matrix that contains quadrature weights is needed to similarity-transform $\mathsfbi{H}_{\tilde{\boldsymbol{q}}}$ before the SVD.  We omit the details here for brevity. 

Unlike classical resolvent analysis, where the singular values account for the energy amplification of perturbations at a single frequency, the singular values in $\boldsymbol{\Sigma}$ capture the energy amplification over the sets of frequencies and wavenumbers in $\boldsymbol{\Omega}$ and $\boldsymbol{K}$, respectively. Therefore, we use the leading 
singular value, $\sigma_1$, and each frequency-wavenumber component, $\hat{\boldsymbol{q}}_{m\omega_0+\gamma, rk_0}$, in the leading left singular vector $\hat{\boldsymbol{\mathcal{Q}}}_1$ of $\mathsfbi{Q}$, to quantify the component-wise amplification.  This is given by
\begin{equation}
    \sigma_{m\omega_0+\gamma, rk_0} = \sigma_1 \lVert \hat{\boldsymbol{q}}_{m\omega_0+\gamma, rk_0} \rVert,
    \label{eq.compamp}
\end{equation}
and will be examined throughout the STHR analysis.

\subsection{Base flows and numerical settings}
\label{HR_BaseFlow}
We adopt the dominant spatio-temporal harmonics of the wake flow in the rectangular domain (see figure~\ref{fig.FFT}(b)) as the base flow for the present STHR analysis. We recall that the adoption of these spatio-temporal harmonics of the wake flow implies that the base flow has a constant phase velocity ($c_0/u_\infty = 0.95$), echoing with the Fourier representation for the base flow in equation~\eqref{eq.periodicBFFourier}. In the STHR analysis, we choose $n \in \{0,\pm1,\pm2\}$ for the base flow and $(m, r) \in \{0,\pm1,\pm2,\pm3\}$ for the perturbations, or $\tilde{n} = 2$ and $\tilde{m} = \tilde{r} = 3$ in equation~\eqref{eq.fwsets}.  This results in 49 components of $\hat{\boldsymbol{q}}_{m\omega_0 + \gamma, rk_0}$ with different frequency-wavenumber combinations stacked into the STHR state vector, $\hat{\boldsymbol{\mathcal{Q}}}$.

\begin{figure}
\centering{\begin{overpic}[trim=0 0 0 -0.4in, clip, width=5.32in]{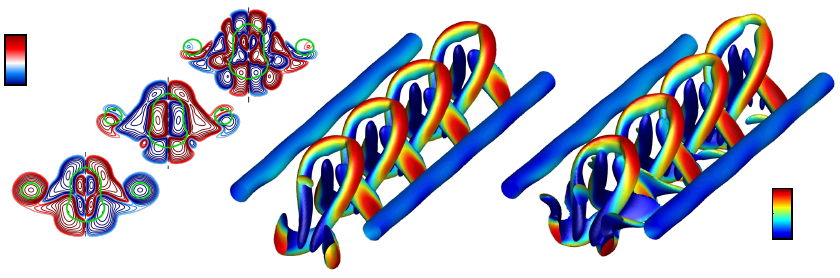} 
\put (2 , 135) {(a)}
\put (160, 135) {(b)}
\put (277 , 135) {(c)}

\put (117 , 60) {\rotatebox{42}{\fontsize{8}{9}\selectfont Base flow for STHR ($\tilde{n} =2$)}}
\put (240 , 60) {\rotatebox{40}{\fontsize{8}{9}\selectfont DNS simulation}}
\put (3 , 115) {\fontsize{8}{9}\selectfont$\tilde{\zeta}_x$}
\put (14, 87.5) {\fontsize{8}{9}\selectfont -0.03}
\put (17, 105.5) {\fontsize{8}{9}\selectfont 0.03}

\put (355 , 43) {\fontsize{8}{9}\selectfont$u$}
\put (365 , 17) {\fontsize{8}{9}\selectfont $0.85$}
\put (365 , 35) {\fontsize{8}{9}\selectfont $1.05$}
\put (30, 58) {\fontsize{8}{9}\selectfont $n=0$}
\put (68, 93) {\fontsize{8}{9}\selectfont $n=1$}
\put (105, 125) {\fontsize{8}{9}\selectfont $n=2$}

\end{overpic}}
\caption{The base flow considered in the STHR analysis: (a) visualizations of $\tilde{\boldsymbol{q}}_{n\omega_0,nk_0}$ with $n \in \{0,~1,~2\}$, as adopted from figure~\ref{fig.FFT}; (b) the spatio-temporal periodic base flow constructed using the harmonic components in (a), visualized by $Q$-criterion isosurface ($QL_c^2/u_\infty^2=0.1$); (c) an instantaneous (non-truncated) flow field obtained from DNS for direction comparison with (b).  The symmetry-enforced wake flow is considered as an example here. All visualizations are mirrored about the symmetry plane to obtain full-span representations.}
\label{fig.HRBaseflow}
\end{figure}
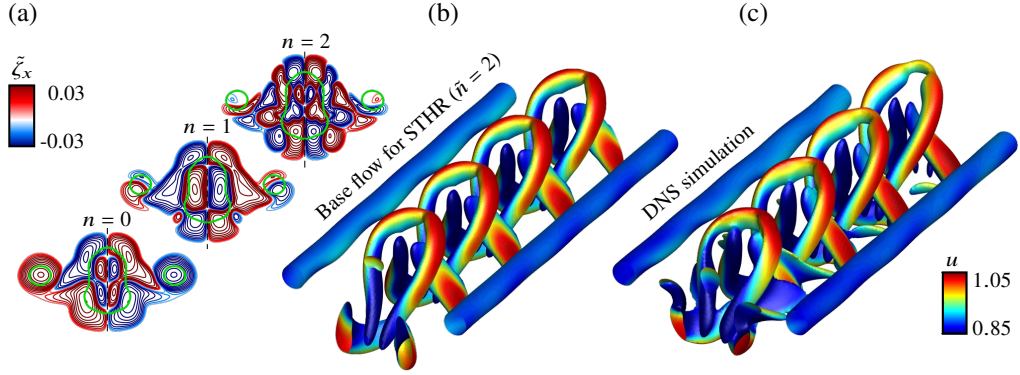

These adopted harmonics that define the spatio-temporal periodic base flow are shown in figure~\ref{fig.HRBaseflow}, along with three-dimensional visualizations of such a (truncated) base flow and the original (non-truncated) flow field obtained from the DNS of symmetry-enforced wake flow for direct comparison.  Note that in this figure, the flow fields are mirrored about the symmetry plane to obtain full-span visualizations. The figure shows that the base flow constructed by these harmonics captures the dominant unsteady wake structures, including those in the wake shear layers and the tip vortices.  The same truncation is considered for the asymmetric base flows, and we note that the applications of asymmetric forcings at different locations and amplitudes do not shift the dominant frequencies and wavenumbers and their harmonics from those of the symmetry-enforced wake flow.  Convergence in the leading singular values of the STHR operator against the number of harmonics for both the base flow and the perturbations is examined in the appendix \ref{apndx.HRTruncation}.

We also note that, for the base flows considered in the present study, the associated Hill matrix, $\mathsfbi{T}_{\tilde{\boldsymbol{q}}}$, holds positive Floquet exponents.  This is observed in the Floquet--Hill analysis discussed in appendix \ref{apndx.Floquet}. Therefore, we introduce a small discounting parameter, $\gamma_d$, to the STHR formulation \citep{jovanovic2004modeling,yeh2020resolvent,leclercq2023mean}. The resulting harmonic resolvent operator becomes
\begin{equation}
    \mathsfbi{H}_{\bar{\boldsymbol{q}}} = [(\mathrm{i} \gamma + \gamma_d) \mathsfbi{I}_\mathsfbi{M} - \mathsfbi{T}_{\bar{\boldsymbol{q}}}]^{-1}.
    \label{eq.discHR}
\end{equation}
The SVD of the STHR operator defined in equation~\eqref{eq.discHR} is performed using an Arnoldi iteration with a Krylov space dimension of $64$ and a residual tolerance of $10^{-6}$.  

\subsection{Emergence of asymmetric modes in asymmetric base flows}
\label{subharmonic_HR}
The STHR analysis in this section starts with a slightly asymmetric wake flow and compares the results to those from the symmetry-enforced wake, in a similar manner to the global mode sensitivity analysis in~\S\ref{LinearStability}. To further examine the emergence of long-time asymmetries in the wake flow, we then compare these results with those obtained from a finite-amplitude asymmetrically disturbed wake flow with $C_\mu = 0.01$. These analyses not only allow us to identify perturbations that are responsible for breaking the long-time spanwise symmetry in the wing wake, but also reveal whether the underlying mechanism responsible for infinitesimal departures from the symmetric state remains relevant as the wake develops finite-amplitude departures from its nominally symmetric configuration.  We consider the frequency shift, $\gamma$, in equation~\eqref{eq.TQ}, over the range $\gamma L_c/u_\infty\in[-0.64,\,0.64]$, to enable cross-frequency interactions between perturbations propagating with different phase velocities. The need to account for perturbation dynamics at phase velocities different from that of the base flow is motivated by the observations in figure~\ref{fig.Cntsphase}(b), which shows that the dominant asymmetric structures propagate at phase velocities slightly higher than that of the base flow. The range of $\gamma$ considered here is also based on the scattering of Floquet exponents of the Hill matrix, $\mathsfbi{T}_{\tilde{\boldsymbol{q}}}$,  obtained in~\S\ref{apndx.Floquet}. The results indicate that the Floquet exponents of $\mathsfbi{T}_{\tilde{\boldsymbol{q}}}$ are confined to narrow frequency bands centered around the dominant harmonics, beyond which the corresponding perturbations are strongly damped and exhibit negligible amplification.

\begin{figure}
\centerline{\begin{overpic}[width=5.32in]{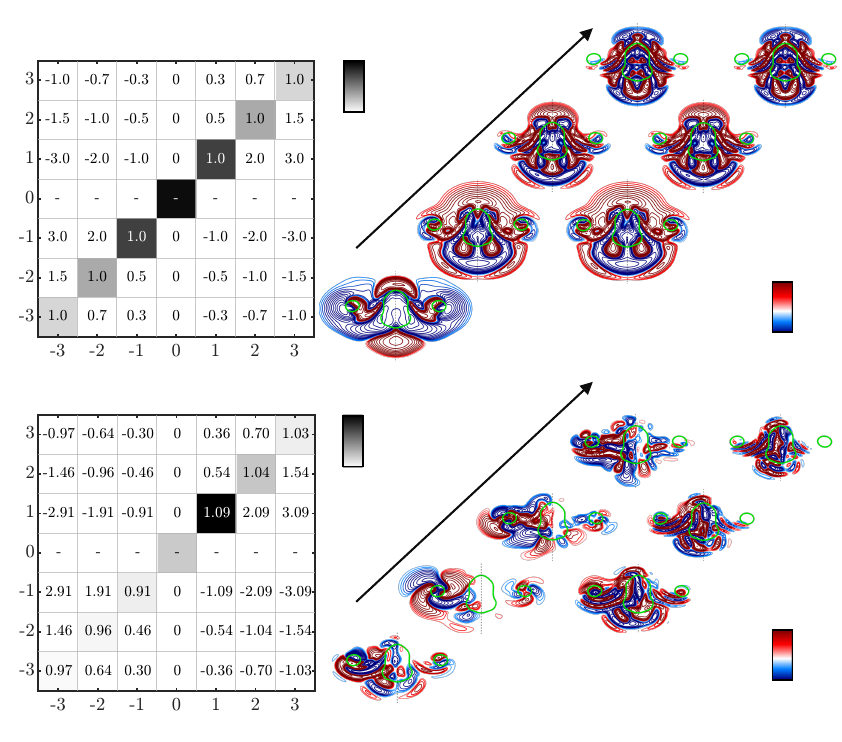}
 \put (0, 325) {(a)}
 \put (235, 325) {(b)}
 \put (0, 163) {(c)}
 \put (235, 163) {(d)}
 \put (60, 317) {\fontsize{8}{9}\selectfont $\gamma L_c/u_\infty=0$}
 \put (52, 155) {\fontsize{8}{9}\selectfont $\gamma L_c/u_\infty=0.43$}
 \put (355, 54) {\fontsize{8}{9}\selectfont $\hat{u}$}
 \put (363.5, 28) {\fontsize{8}{9}\selectfont -0.03}
 \put (366, 46) {\fontsize{8}{9}\selectfont 0.03}
 \put (355, 212) {\fontsize{8}{9}\selectfont $\hat{u}$}
 \put (363.5, 186.5) {\fontsize{8}{9}\selectfont -0.03}
 \put (366, 204) {\fontsize{8}{9}\selectfont 0.03}
 \put (148, 113) {\rotatebox{90}{\fontsize{8}{9}\selectfont $\sigma_{m\omega_0+\gamma, rk_0}$}}
 \put (168, 125.5) {\fontsize{8}{9}\selectfont 0}
 \put (168, 144) {\fontsize{8}{9}\selectfont 30}
 \put (148, 276) {\rotatebox{90}{\fontsize{8}{9}\selectfont $\sigma_{m\omega_0+\gamma, rk_0}$}}
 \put (168, 287) {\fontsize{8}{9}\selectfont 0}
 \put (168, 305.5) {\fontsize{8}{9}\selectfont 60}
\put (155 ,65) {\rotatebox{43}{\fontsize{8}{9}\selectfont $(m\omega_0+\gamma,rk_0),\;\; m=r=0,\pm1,\pm2,\pm3$}} 
\put (155, 228) {\rotatebox{43}{\fontsize{8}{9}\selectfont $(m\omega_0,rk_0),\;\; m=r=0,\pm1,\pm2,\pm3$}} 
\put (77, 168) {\fontsize{8}{9}\selectfont $\omega$}
\put (77, 5) {\fontsize{8}{9}\selectfont $\omega$}
\put (0 , 82) {\rotatebox{90}{\fontsize{8}{9}\selectfont $k$}}
\put (0 , 245) {\rotatebox{90}{\fontsize{8}{9}\selectfont $k$}}
\put (138, 14.5) {\rotatebox{0}{\fontsize{8}{9}\selectfont $\times \omega_0 + \gamma$}}
\put (138, 176) {\rotatebox{0}{\fontsize{8}{9}\selectfont $\times \omega_0 + \gamma$}}
 \put (0, 311) {\rotatebox{0}{\fontsize{8}{9}\selectfont $\times k_0$}}
 \put (0 , 149) {\rotatebox{0}{\fontsize{8}{9}\selectfont $\times k_0$}}
 \put (171, 51) {\rotatebox{0}{\fontsize{7}{9}\selectfont $m=0$}}
 \put (171, 216) {\rotatebox{0}{\fontsize{7}{9}\selectfont $m=0$}}
 \put (209, 82) {\rotatebox{0}{\fontsize{7}{9}\selectfont $m=1$}}
 \put (208, 258) {\rotatebox{0}{\fontsize{7}{9}\selectfont $m=1$}}
 \put (276, 258) {\rotatebox{0}{\fontsize{7}{9}\selectfont $m=-1$}}
 \put (280, 82) {\rotatebox{0}{\fontsize{7}{9}\selectfont $m=-1$}}
 \put (242, 294.5) {\rotatebox{0}{\fontsize{7}{9}\selectfont $m=2$}}
 \put (311, 294.5) {\rotatebox{0}{\fontsize{7}{9}\selectfont $m=-2$}}
 \put (242, 116) {\rotatebox{0}{\fontsize{7}{9}\selectfont $m=2$}}
 \put (311, 116) {\rotatebox{0}{\fontsize{7}{9}\selectfont $m=-2$}}
 \put (280, 151) {\rotatebox{0}{\fontsize{7}{9}\selectfont $m=3$}}
 \put (282, 329) {\rotatebox{0}{\fontsize{7}{9}\selectfont $m=3$}}
 \put (346, 151) {\rotatebox{0}{\fontsize{7}{9}\selectfont $m=-3$}}
 \put (349, 329) {\rotatebox{0}{\fontsize{7}{9}\selectfont $m=-3$}}
\end{overpic}}
\caption{Amplifications of all frequency–wavenumber components, $\sigma_{m\omega_0+\gamma, r k_0}$, about the slightly asymmetric base flow for $\gamma L_c/u_\infty=0$ and $0.43$ in (a) and (c), respectively. The text values shown over each panel of frequency–wavenumber component denote the corresponding phase velocity normalized by the base-flow phase velocity, i.e.~$c/c_0$. The real part of
the streamwise velocity of the corresponding response modes, $\hat{\boldsymbol{q}}_{m\omega_0+\gamma,rk_0}$ with $m=r$, are shown in (b) and (d) respectively for $\gamma L_c/u_\infty=0$ and $0.43$ .}
\label{fig.ampl_block}
\end{figure}
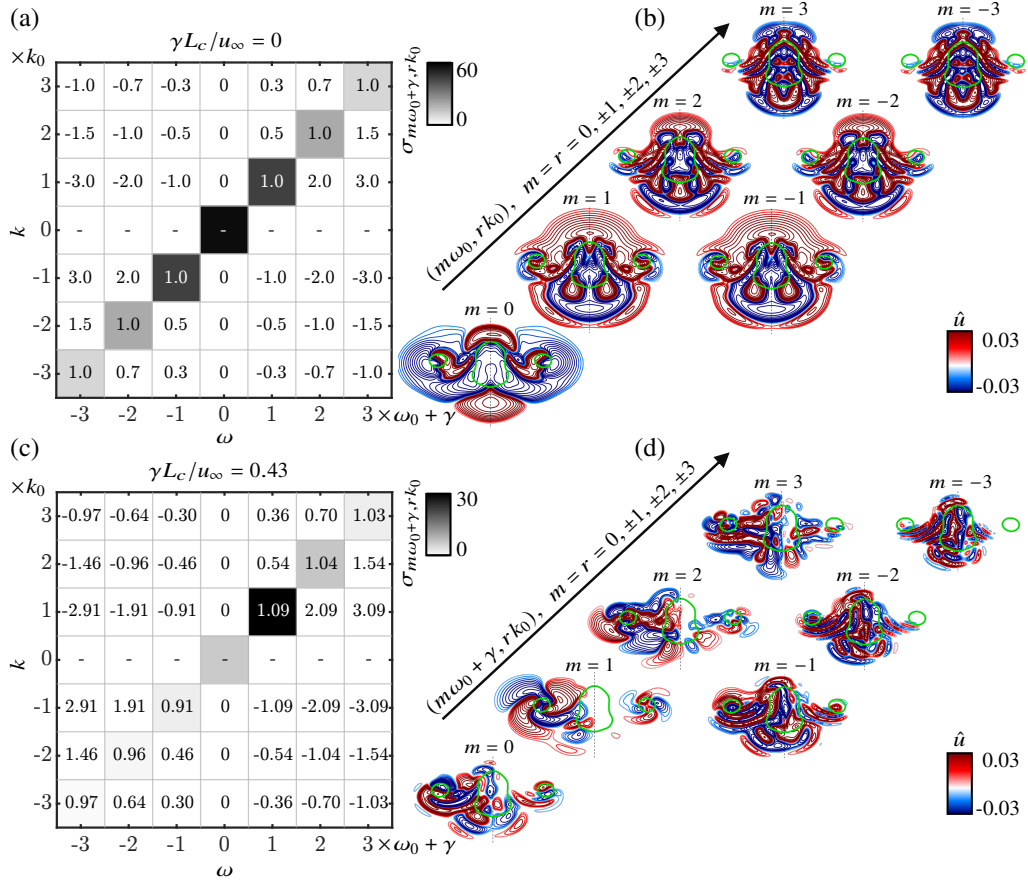

We start our discussion with the STHR analysis about the slightly disturbed asymmetric base flow, as shown in figure~\ref{fig.ampl_block}.  Here, we consider two different values of frequency shift, $\gamma L_c/u_\infty = 0$ and $0.43$, in this figure.  Note that with $\gamma = 0$ and $m=r$, the corresponding $\hat{\boldsymbol{q}}_{m\omega_0+\gamma, rk_0}$ travels at the same phase velocity as the base flow ($c/c_0 = 1$). On the other hand, the choice of $\gamma L_c/u_\infty = 0.43$ is motivated by the phase velocity of the dominant asymmetric eigenmode at $k = k_0$ observed in figure~\ref{fig.Cntsphase}(b), which travels at a phase velocity of $c/c_0 = 1.09$. With $\gamma L_c/u_\infty = 0.43$ and $m = r = 1$, the resulting phase velocity of the corresponding $\hat{\boldsymbol{q}}_{m\omega_0+\gamma, rk_0}$, as given by equation~\eqref{eq.modalphasevel}, has the same value of $c/c_0 = 1.09$ as that of the dominant eigenmode at $k = k_0$ observed in figure~\ref{fig.Cntsphase}(b).  In figure~\ref{fig.ampl_block}, we focus on the leading component-wise amplification of perturbations, $\sigma_{m\omega_0+\gamma, rk_0}$ as defined in \eqref{eq.compamp}, and the corresponding response modes, $\hat{\boldsymbol{q}}_{m\omega_0+\gamma, rk_0}$. The component-wise amplification is shown in figures \ref{fig.ampl_block}(a) and (c) by the shaded panel for each frequency-wavenumber component, and the text value on top of each panel indicates the corresponding phase velocity normalized by that of the base-flow phase velocity, $c/c_0$.  We remind the readers that, for the present STHR analysis, these amplifications of 49 frequency-wavenumber components are obtained from a single SVD of the STHR operator. This is opposed to the classical resolvent analysis in equation~\eqref{eq.classicresolvent}, where a series of SVDs are performed for different wavenumbers prescribed in the resolvent operator. 

The results in figure~\ref{fig.ampl_block} show that strong amplification occurs primarily for the downstream-propagating perturbations whose phase velocities are close to that of the base flow. For both values of $\gamma$ in figure~\ref{fig.ampl_block}(a) and (c), only the components along the diagonal blocks with $m=r$ exhibit significant amplification. For these components, their phase velocities, $c=c_0+\gamma/(rk_0)$, lie within a narrow band around that of the base flow, $c_0=\omega_0/k_0$. As shown in figure~\ref{fig.ampl_block}(b), perturbations that propagate with the same phase velocity as the base flow contribute to both wake shear-layer and tip-vortex instabilities for lower harmonics, while for higher harmonics they remain largely confined in the wake region. Despite the slight departure of the base flow from the symmetric state, these perturbations maintain their symmetric structures, resulting in low levels of modal asymmetry. This is consistent with the findings in~\S\ref{Eigenvalueanalysis}, which show that the dominant wake modes in this frequency range appear as single modes without their symmetric or anti-symmetric counterparts. Note that for $\gamma L_c/u_\infty=0$, perturbations form complex-conjugate pairs between $ (m \omega_0, rk_0)$ and $(-m \omega_0, -rk_0)$, corresponding to perturbations with the same phase velocities. However, introducing a nonzero frequency shift, $\gamma$, allows for interactions between perturbations of different phase velocities, i.e.~$c=c_0+\gamma/(rk_0)$ and $c=c_0-\gamma/(rk_0)$. In the case where $\gamma > 0$, the phase velocities of the modes along the diagonal blocks with $(m, r) \in \{0, +1, +2, +3\}$ are higher than that of the base flow, as reflected by the corresponding values of $c/c_0>1$. In figure~\ref{fig.ampl_block}(c), we observe that these perturbations with higher phase velocities than that of the base flow also exhibit high amplifications, compared to those with lower phase velocities. This echoes with the observation made in figure~\ref{fig.Cntsphase}, where the dominant instabilities in the wake propagate downstream at phase velocities slightly higher than that of the base flow. The corresponding modal structures in figure~\ref{fig.ampl_block}(d) show interactions between asymmetric elliptic instabilities and the wake shear layers for perturbations with different phase velocities. 

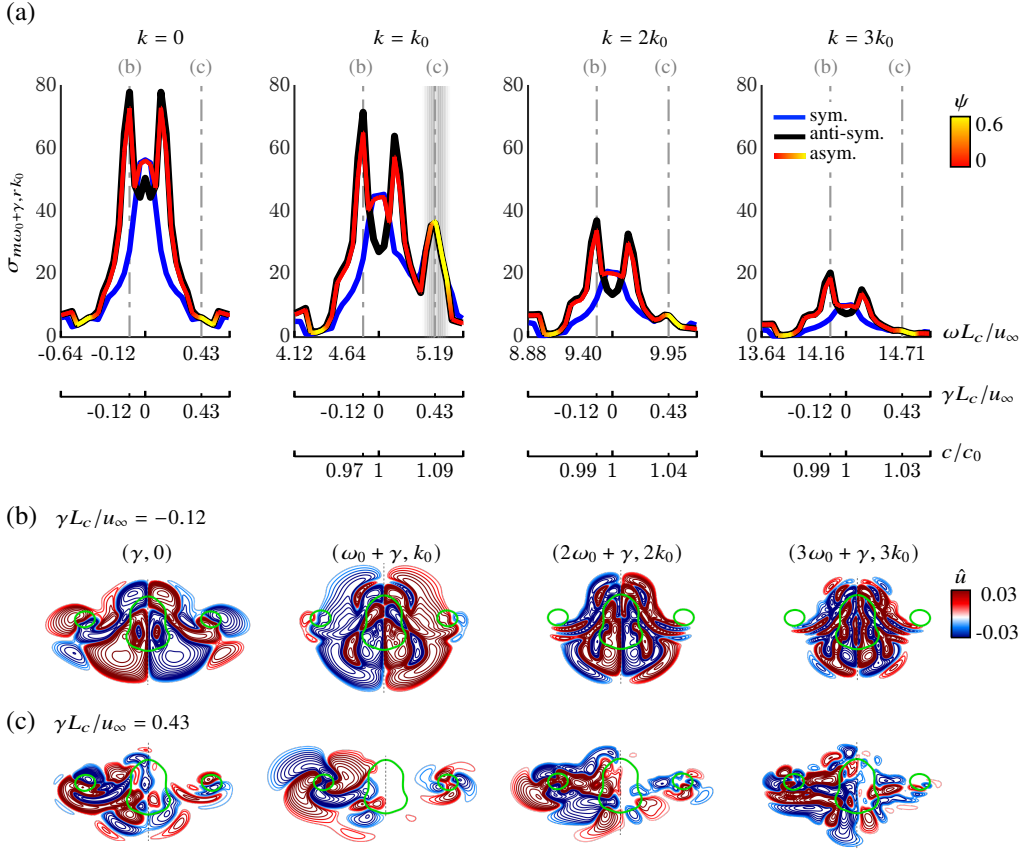
\begin{figure}
\centerline{\begin{overpic}[width=5.32in]{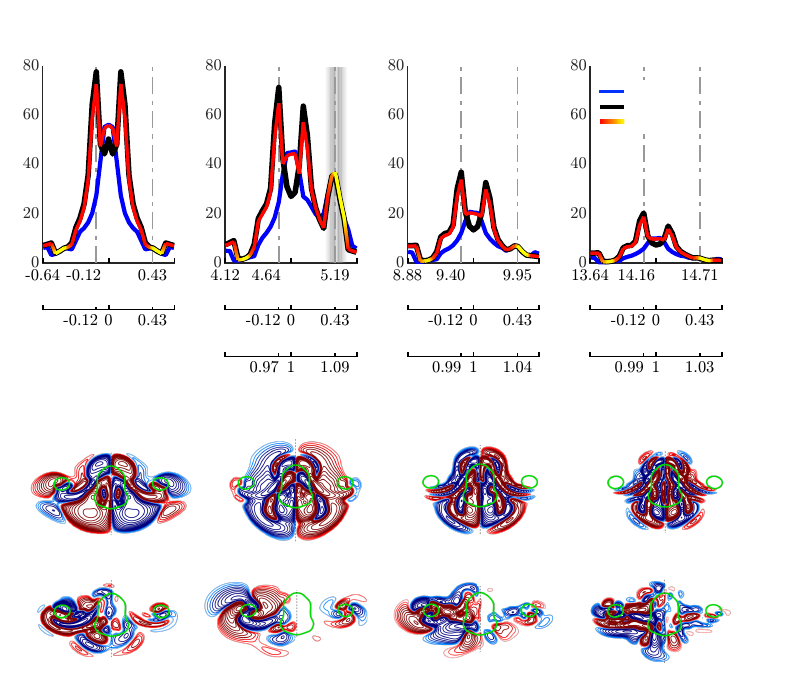} 
    \put (355 , 262) {\includegraphics[scale=0.8]{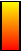}}
    \put (355 , 85) {\includegraphics[scale=0.8]{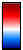}}
    \put (0, 318) {\rotatebox{0}{(a)}}
    \put (0, 128) {\rotatebox{0}{(b) \fontsize{7.8}{9}\selectfont\;\;$\gamma L_c/u_\infty=-0.12$}}
    \put (0, 51) {\rotatebox{0}{(c) \fontsize{7.8}{9}\selectfont \;\;$\gamma L_c/u_\infty=0.43$}}

    \put (302.5, 280) {\fontsize{7.5}{9}\selectfont sym.}
    \put (302.5, 273) {\fontsize{7.5}{9}\selectfont anti-sym.}
    \put (302.5, 265.5) {\fontsize{7.5}{9}\selectfont asym.}
    \put (357, 106) {\fontsize{7.8}{9}\selectfont $\hat{u}$}
    \put (365, 85) {\fontsize{7.8}{9}\selectfont -0.03}
    \put (367, 99) {\fontsize{7.8}{9}\selectfont 0.03}
    \put (356.5, 285) {\fontsize{7.8}{9}\selectfont $\psi$}
    \put (365, 262.5) {\fontsize{7.8}{9}\selectfont $0$}
    \put (365, 276) {\fontsize{7.8}{9}\selectfont $0.6$}
    \put (1, 222) {\rotatebox{90}{\fontsize{7.8}{9}\selectfont $\sigma_{m\omega_0+\gamma, rk_0}$}}
    
    \put (352, 198) {\rotatebox{0}{\fontsize{7.8}{9}\selectfont $\omega L_c/u_\infty$}}
    \put (352, 175) {\rotatebox{0}{\fontsize{7.8}{9}\selectfont $\gamma L_c/u_\infty$}}
    \put (352, 153) {\rotatebox{0}{\fontsize{7.8}{9}\selectfont $c/c_0$}}
    
    \put (49, 309) {\rotatebox{0}{\fontsize{7.8}{9}\selectfont $k=0$}}
    \put (138, 309) {\rotatebox{0}{\fontsize{7.8}{9}\selectfont $k=k_0$}}
    \put (224, 309) {\rotatebox{0}{\fontsize{7.8}{9}\selectfont $k=2k_0$}}
    \put (309, 309) {\rotatebox{0}{\fontsize{7.8}{9}\selectfont $k=3k_0$}}
    \put (42, 298) {\rotatebox{0}{\fontsize{7.8}{9}\selectfont {\color{lightgray}{(b)}}}} 
    \put (69, 298) {\rotatebox{0}{\fontsize{7.8}{9}\selectfont {\color{lightgray}{(c)}}}}
    \put (129, 298) {\rotatebox{0}{\fontsize{7.8}{9}\selectfont {\color{lightgray}{(b)}}}}
    \put (158, 298) {\rotatebox{0}{\fontsize{7.8}{9}\selectfont {\color{lightgray}{(c)}}}}
    \put (217, 298) {\rotatebox{0}{\fontsize{7.8}{9}\selectfont {\color{lightgray}{(b)}}}}
    \put (244, 298) {\rotatebox{0}{\fontsize{7.8}{9}\selectfont {\color{lightgray}{(c)}}}}
    \put (304, 298) {\rotatebox{0}{\fontsize{7.8}{9}\selectfont {\color{lightgray}{(b)}}}}
    \put (332, 298) {\rotatebox{0}{\fontsize{7.8}{9}\selectfont {\color{lightgray}{(c)}}}}
    
    \put (43, 115) {\rotatebox{0}{\fontsize{7.8}{9}\selectfont $(\gamma,0)$}}
    \put (122, 115) {\rotatebox{0}{\fontsize{7.8}{9}\selectfont $(\omega_0+\gamma,k_0)$}}
    \put (205, 115) {\rotatebox{0}{\fontsize{7.8}{9}\selectfont $(2\omega_0+\gamma,2k_0)$}}
    \put (294, 115) {\rotatebox{0}{\fontsize{7.8}{9}\selectfont $(3\omega_0+\gamma,3k_0)$}}
\end{overpic}}
\caption{(a) Distribution of leading harmonic resolvent amplification across frequency-wavenumber pairs for symmetric and asymmetric base flows. Energy amplifications for the symmetric base flow are shown by symmetric (blue line) and anti-symmetric (black line) components, and that for the asymmetric base flow is colored according to the degree of asymmetry ($\psi$) of the corresponding response mode. The real part of the streamwise velocity response modes corresponding to the perturbations at frequency-wavenumber pairs $(m\omega_0+\gamma,rk_0)$ with (b) $\gamma L_c/u_\infty=-0.12$ and (c) $\gamma L_c/u_\infty=0.43$, where $m=r=0,1,2,$ and $3$.}
\label{fig.subharmonics}
\end{figure}

In figure~\ref{fig.subharmonics}, we show the distribution of the component-wise amplifications,  $\sigma_{m\omega_0+\gamma, rk_0}$, across different $(\omega, k)$-components for both the symmetric and slightly asymmetric base flows. Here, we focus on the modes with $(m, r) \in \{0, 1, 2, 3\}$. The component-wise amplifications associated with the symmetric base flow are obtained for both symmetric (blue line) and anti-symmetric (black line) modal structures, and those associated with the asymmetric base flow are colored by the level of asymmetry, $\psi$, of the corresponding response modes. We observe that the symmetric base flow exhibits three dominant amplification peaks that are particularly prominent for $k = k_0$ modes: two dominant peaks near the phase velocity of the base flow ($c = c_0\pm 0.12/k_0$) and another peak at a higher phase velocity ($c = c_0+0.43/k_0$).  The first two dominant peaks are associated with anti-symmetric perturbations, and the third peak at $c/c_0 = 1.09$ appears when symmetric and anti-symmetric modes from the symmetric base flow exhibit comparable levels of amplification.  Compared with the symmetric base flow, introducing slight asymmetry to the base flow reduces the amplifications associated with the first two peaks, while the amplification associated with perturbations of higher phase velocity ($c/c_0 = 1.09$) remains nearly unchanged. More importantly, the color variation along the component-wise amplification profile of the asymmetric flow indicates that the modal structures near this peak are associated with the highest level of asymmetry in the response mode, as highlighted by the gray-shaded window in figure~\ref{fig.subharmonics}. These findings indicate that perturbations propagating with phase velocities within the highlighted interval are responsive to asymmetric forcing. These observations are closely related to the findings in~\S\ref{Eigenvalueanalysis}, where the eigenvalues with the highest growth rates correspond to perturbations whose phase velocity is slightly higher than that of the base flow. In~\S\ref{Eigenvalueanalysis}, we concluded that these dominant eigenvalues are prone to symmetry-breaking due to the appearance of the paired symmetric and anti-symmetric eigenmodes and the high sensitivity due to their modal non-normality. This conclusion is consistent with our findings of the harmonic resolvent analysis in regard to the emergence of highly amplified asymmetric modes in this frequency range.

To further elucidate the physical mechanisms of symmetry-breaking in the wake flow, we examine the corresponding response modes associated with perturbations within the gray-shaded window. The most amplified response mode within the highlighted interval and other perturbations that interacts with this mode via the harmonic resolvent operator are highlighted by gray dashed lines in figure~\ref{fig.subharmonics}. The response modes suggest that symmetry-breaking of these perturbations is closely tied to the unsteady structures of the wing-tip vortices and their interactions with the wake shear layers. At a frequency slightly shifted from the fundamental harmonic, $(\omega_0 + \gamma, k_0)$ with $\gamma L_c/u_\infty=0.43$, the response is dominated by a highly amplified vortex instability that manifests as an asymmetric helical structure, indicating that elliptic vortex instability is the primary mechanism that leads to the flow asymmetry in the wing wake. As the order of harmonics increases, the response modes show stronger interactions between the tip-vortex structures and the surrounding wake shear layers, with progressively smaller scales across both the vortex core and the shear region. This enhanced wake-vortex interaction allows the asymmetry associated with the helical vortex structure to also manifest within the wake shear layers, extending the flow asymmetry from the wing tip to the root plane.

\begin{figure}
\centerline{\begin{overpic}[width=5.32in]{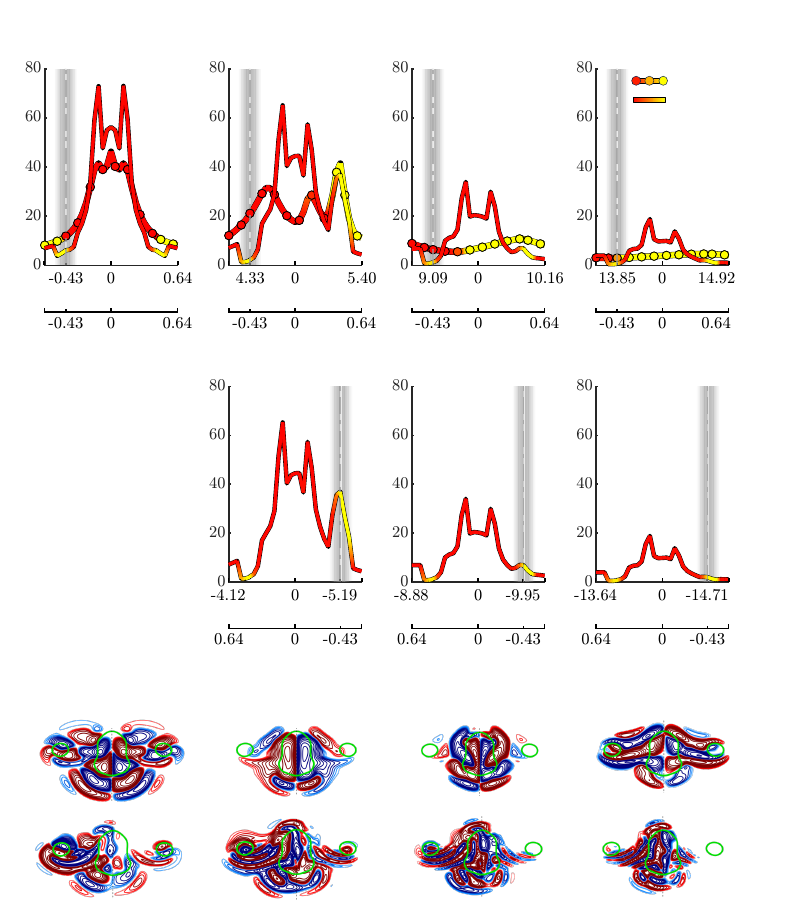} 
 \put (355 , 380) {\includegraphics[scale=0.8]{figs/autumn_cb.pdf}}
 \put (355 , 70) {\includegraphics[scale=0.8]{figs/bluered_cb.pdf}}
 \put (0, 425) {\rotatebox{0}{(a)}}
 \put (0, 113) {\rotatebox{0}{(b) \fontsize{7.8}{9}\selectfont\;\;$\gamma L_c/u_\infty=-0.43$}}

\put (1, 340) {\rotatebox{90}{\fontsize{7.8}{9}\selectfont $\sigma_{m\omega_0+\gamma, rk_0}$}}
\put (90, 186) {\rotatebox{90}{\fontsize{7.8}{9}\selectfont $\sigma_{m\omega_0+\gamma, rk_0}$}}
\put (354, 310) {\rotatebox{0}{\fontsize{7.8}{9}\selectfont $\omega L_c/u_\infty$}}
\put (354, 288) {\rotatebox{0}{\fontsize{7.8}{9}\selectfont $\gamma L_c/u_\infty$}}
\put (354, 158) {\rotatebox{0}{\fontsize{7.8}{9}\selectfont $\omega L_c/u_\infty$}}
\put (354, 137) {\rotatebox{0}{\fontsize{7.8}{9}\selectfont $\gamma L_c/u_\infty$}}
\put (322, 397.5) {\fontsize{7.5}{9}\selectfont CR}
 \put (322, 388.5) {\fontsize{7.5}{9}\selectfont STHR}
 \put (357, 92) {\fontsize{7.8}{9}\selectfont $\hat{u}$}
 \put (366, 70) {\fontsize{7.8}{9}\selectfont -0.03}
 \put (368.5, 83) {\fontsize{7.8}{9}\selectfont 0.03}
 \put (356, 403) {\fontsize{7.8}{9}\selectfont $\psi$}
 \put (365, 380) {\fontsize{7.8}{9}\selectfont 0}
 \put (365, 395) {\fontsize{7.8}{9}\selectfont 0.6}

  \put (27, 410) {\rotatebox{0}{\fontsize{7.8}{9}\selectfont {\color{lightgray}{(b)}}}}
 \put (115, 410) {\rotatebox{0}{\fontsize{7.8}{9}\selectfont {\color{lightgray}{(b)}}}}
 \put (202, 410) {\rotatebox{0}{\fontsize{7.8}{9}\selectfont {\color{lightgray}{(b)}}}}
 \put (290, 410) {\rotatebox{0}{\fontsize{7.8}{9}\selectfont {\color{lightgray}{(b)}}}}
 \put (49, 419) {\rotatebox{0}{\fontsize{7.8}{9}\selectfont $k=0$}}
 \put (138, 419) {\rotatebox{0}{\fontsize{7.8}{9}\selectfont $k=k_0$}}
 \put (224, 419) {\rotatebox{0}{\fontsize{7.8}{9}\selectfont $k=2k_0$}}
 \put (309, 419) {\rotatebox{0}{\fontsize{7.8}{9}\selectfont $k=3k_0$}}
 \put (138, 262) {\rotatebox{0}{\fontsize{7.8}{9}\selectfont $k=-k_0$}}
 \put (224, 262) {\rotatebox{0}{\fontsize{7.8}{9}\selectfont $k=-2k_0$}}
 \put (309, 262) {\rotatebox{0}{\fontsize{7.8}{9}\selectfont $k=-3k_0$}}
 
   \put (43, 97) {\rotatebox{0}{\fontsize{7.8}{9}\selectfont $(\gamma,0)$}}
  \put (122, 97) {\rotatebox{0}{\fontsize{7.8}{9}\selectfont $(\omega_0+\gamma,k_0)$}}
  \put (205, 97) {\rotatebox{0}{\fontsize{7.8}{9}\selectfont $(2\omega_0+\gamma,2k_0)$}}
  \put (294, 97) {\rotatebox{0}{\fontsize{7.8}{9}\selectfont $(3\omega_0+\gamma,3k_0)$}}
  \put (5, 75) {\rotatebox{90}{\fontsize{8}{9}\selectfont CR}}
  \put (5, 20) {\rotatebox{90}{\fontsize{8}{9}\selectfont STHR}}
\end{overpic}}
\caption{(a) Leading amplification across frequency–wavenumber pairs from classical resolvent and STHR analyses of the infinitesimal-asymmetry base flow. (c) Comparison of the real part of the streamwise velocity response modes, $\hat{u}$, obtained from the classical resolvent and STHR analyses at $(m\omega_0+\gamma,rk_0)$ with $\gamma L_c/u_\infty=0.43$ and $(m,r)\in\{0,1,2,3\}$. }
\label{fig.ClasResHRes}
\end{figure}

Beyond the cross-frequency interactions highlighted in figure~\ref{fig.subharmonics}, symmetry-breaking can also be attributed to the interactions between the modes that are coupled as a complex-conjugate mode pair by the STHR operator.  This can be seen by comparing the STHR results to those from the classical resolvent analysis, as shown in figure~\ref{fig.ClasResHRes}. 
In this figure, we show the gain profiles of the asymmetric base flow obtained from both the STHR and classical resolvent analyses, in which perturbation dynamics at different frequency-wavenumber components are decoupled. Color variations along the gain profiles indicate the level of asymmetry of the corresponding response modes. Observing the gain profiles and modal asymmetries from the STHR analysis in figure~\ref{fig.ClasResHRes}, we find that the response modes within the gray-shaded bands exhibit high levels of modal asymmetry, even if these modes are associated with low amplifications. This is particularly evident for the modes at ($k=k_0$), which exhibit pronounced modal asymmetries despite their weak amplification. In contrast, the response modes obtained from classical resolvent analysis within the gray-shaded bands retain their symmetric or anti-symmetric structures and exhibit low levels of modal asymmetries. This indicates that cross-frequency interactions play a crucial role in the emergence of modal asymmetries. These asymmetric modes emerge due to their interactions with another highly amplified mode of high levels of asymmetry, instead of the intrinsic behavior of their individual dynamics. Note that the gain profile of the harmonic resolvent exhibits complex-conjugate symmetry about the zero frequency.  This means that the harmonic resolvent modes of $(m\omega_0+\gamma, r k_0)$ with $m=r=-1$ and near $\gamma L_c/u_\infty = -0.43$ exhibit the same levels of high gain and modal asymmetries as they do for $m=r=1$ and near $\gamma L_c/u_\infty =0.43$. The complex conjugacy and the resulting symmetric gain profile about zero frequency provide an explanation for the emergence of highly asymmetric, yet weakly amplified, modes within the gray-shaded band near $\gamma L_c/u_\infty = -0.43$ and $m = r = 1$, where such high levels of modal asymmetry are absent in the classical resolvent analysis. These highly asymmetric, yet low-gain, modes are coupled with the modes within the gray-shaded band near $\gamma L_c/u_\infty = -0.43$ and $m = r = -1$, which exhibit both high gain and high levels of modal asymmetry. The mechanism of the resulting modal asymmetries can be elucidated by examining the corresponding response modes within the shaded windows in the gain profile. In the classical resolvent analysis, response modes remain largely confined to the wake shear region with anti-symmetric structures. Conversely, when cross-frequency interactions are enabled in the STHR analysis, the response modes exhibit asymmetric structures in the wake and develop pronounced asymmetric helical structures in the tip-vortex region. The signature of asymmetric vortex structures in the response modes confirms that the emergence of asymmetries in these perturbations is associated with cross-frequency interaction with highly amplified asymmetric unsteadiness near the tip vortices.

\begin{figure}
\centerline{\begin{overpic}[width=5.32in]{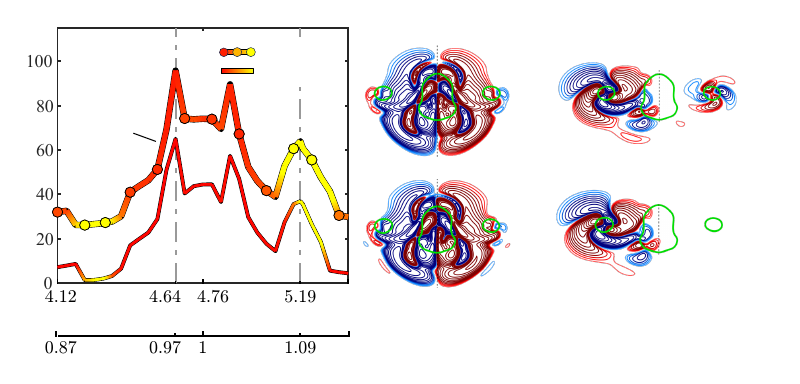}
 \put (35 , 140) {\includegraphics[scale=0.8]{figs/autumn_cb.pdf}}
 \put (355 , 50) {\includegraphics[scale=0.8]{figs/bluered_cb.pdf}}
\put (29 , 176) {(a)}
\put (179 , 176) {(b)}
\put (270 , 176) {(c)}
\put (78 , 176) {{\color{lightgray}{(b)}}}
\put (138 , 176) {{\color{lightgray}{(c)}}}
\put (5, 85) {\rotatebox{90}{\fontsize{8}{9}\selectfont $\sigma_{\omega_0+\gamma, k_0}$}}
\put (84, 29) {\rotatebox{0}{\fontsize{8}{9}\selectfont $\omega L_c/u\infty$}}
\put (94, 4) {\rotatebox{0}{\fontsize{8}{9}\selectfont $c/c_0$}}
\put (35, 125) {\rotatebox{0}{\fontsize{7.8}{9}\selectfont shifted by 25}}
\put (227, 165) {\rotatebox{0}{\fontsize{8}{9}\selectfont Infinitesimal-asymmetry}}
\put (235, 100){\rotatebox{0}{\fontsize{8}{9}\selectfont Finite-asymmetry}}
 \put (357, 71) {\fontsize{8}{9}\selectfont $\hat{u}$}
 \put (365, 50) {\fontsize{8}{9}\selectfont -0.03}
 \put (367.5, 63) {\fontsize{8}{9}\selectfont 0.03}
 \put (36, 163) {\fontsize{8}{9}\selectfont $\psi$}
 \put (45, 140) {\fontsize{8}{9}\selectfont 0}
 \put (45, 154) {\fontsize{8}{9}\selectfont 0.6}
 \put (125, 157) {\rotatebox{0}{\fontsize{7.8}{9}\selectfont finite-asym}}
\put (124, 148){\rotatebox{0}{\fontsize{7.8}{9}\selectfont $\epsilon$-asym}}
\end{overpic}}
\caption{(a) Leading amplification of perturbations at ($\omega_0+\gamma, k_0$) for base flows of the infinitesimal- and finite-level asymmetry. The amplification profiles are colored by the asymmetry metric, $\psi$, of the corresponding response modes. For visual clarity, the amplification profile corresponding to the asymmetric base flow obtained from the finite-level disturbance is shifted vertically by 25. The real part of streamwise velocity component of the response modes, $\hat{u}$, associated with the perturbations at the phase velocity of (b) $c/c_0=0.97$, and (c) $c/c_0=1.09$, are compared for both asymmetric flows.}
\label{fig.Finite_Asym}
\end{figure}

To further examine whether these symmetry-breaking mechanisms persist as the wake asymmetries increase to a finite level, we also compare the STHR amplification profiles of asymmetric base flows that are obtained from both infinitesimal and finite-level disturbances in figure~\ref{fig.Finite_Asym}. Here, the amplification profiles are colored by the asymmetry level of the corresponding response modes. We observe that introducing finite-amplitude asymmetry to the base flow primarily enhances the amplification of dominant perturbations, while leaving the overall trend of modal asymmetry largely unchanged compared to the slightly asymmetric wake. The low levels of modal asymmetry associated with the first two dominant peaks indicate that the corresponding response modes remain predominantly anti-symmetric, similar to those of the symmetric base flow. Similarly, the perturbations responsible for breaking the long-time spanwise symmetry in the slightly asymmetric base flow continue to exhibit strong modal asymmetry in the base flow of finite-level asymmetry. The corresponding response modes in figure~\ref{fig.Finite_Asym} closely resemble those observed in the infinitesimal-asymmetry flow, retaining the pronounced asymmetric helical structures in the wing-tip vortices and their interactions with the wake shear layers. These observations suggest that the helical vortex instability mechanism identified in the infinitesimal-asymmetry wake remains the dominant mechanism responsible for symmetry-breaking even after the wake develops finite-amplitude asymmetry.

\subsection{Base flows of finite-level asymmetries} \label{phase_interference}
The asymmetric base flows we have considered up to this point were obtained by introducing asymmetric forcing upstream of the wing, ranging from infinitesimal amplitudes to finite-amplitude forcing levels up to $C_\mu=0.01$.  In this section, we consider the amplitude of the asymmetric forcing over a range of $C_\mu \in [0.01,~0.1]$. Our goal is to examine if the asymmetric modes arising from the so-obtained asymmetric base flows with stronger finite-amplitude forcing can still be understood as a result of phase interference between the symmetric and anti-symmetric modes obtained from the symmetry-enforced wake flow. 

To this end, we first form a basis that contains the symmetric and anti-symmetric harmonic resolvent modes from the symmetric wake flow and project the asymmetric modes obtained from the asymmetric base flows onto this basis of symmetric-flow modes. That is, we reconstruct these asymmetric modes using the symmetric and anti-symmetric harmonic resolvent modes obtained from the symmetric base flow via 
\begin{equation}
    \tilde{\boldsymbol{\mathcal{Q}}}_i^\text{asym} 
    = \sum\limits_{j=1}^{n_s} 
        \langle
            \hat{\boldsymbol{\mathcal{Q}}}_i^\text{asym},~\hat{\boldsymbol{\mathcal{Q}}}_j^\text{sym}
        \rangle~
        \hat{\boldsymbol{\mathcal{Q}}}_j^\text{sym}.
\label{eq.reconst_asym}
\end{equation}
Here, $\hat{\boldsymbol{\mathcal{Q}}}_i^\text{asym}$ denotes the $i$-th harmonic resolvent mode from the asymmetric flow, $\hat{\boldsymbol{\mathcal{Q}}}_j^\text{sym}$ denotes the $j$-th symmetric-flow mode of either symmetric or anti-symmetric structures, 
$\langle
    \hat{\boldsymbol{\mathcal{Q}}}_i^\text{asym},~\hat{\boldsymbol{\mathcal{Q}}}_j^\text{sym}
\rangle$ 
performs a projection (inner product) of $\hat{\boldsymbol{\mathcal{Q}}}_i^\text{asym}$ onto $\hat{\boldsymbol{\mathcal{Q}}}_j^\text{sym}$, and $\tilde{\boldsymbol{\mathcal{Q}}}_i^\text{asym}$ denotes the reconstructed asymmetric mode using the basis of symmetric-flow modes, and $n_s$ is the number of symmetric-flow modes contained in the basis. Here, we focus on the modes that are primarily responsible for symmetry-breaking in the wake flow by considering the frequency set of $\boldsymbol{\Omega} = \gamma +  \{-3,\ldots,0,\ldots,3\}\,\omega_0$, $\gamma L_c/u_\infty=0.43$, and streamwise wavenumber \(\boldsymbol{K} = \{-3,\ldots,0,\ldots,3\}\,k_0\) (see~\S\ref{subharmonic_HR}).

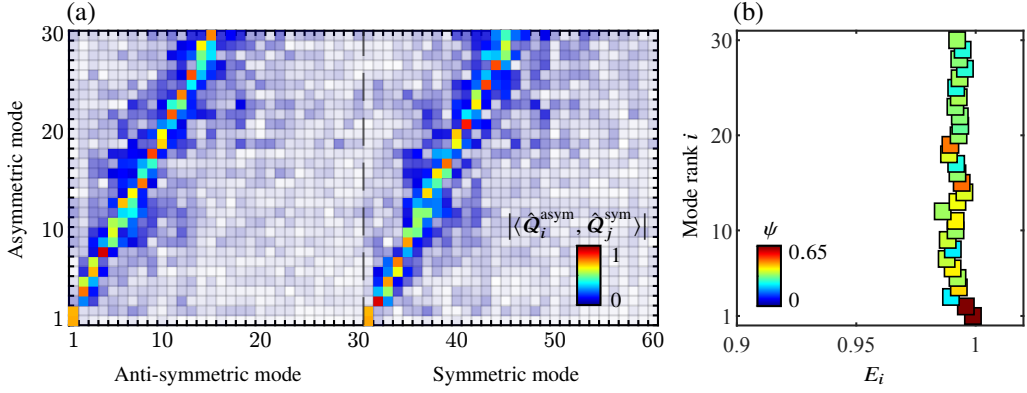
\begin{figure}
  \centering{\begin{overpic}[width=5.32in]{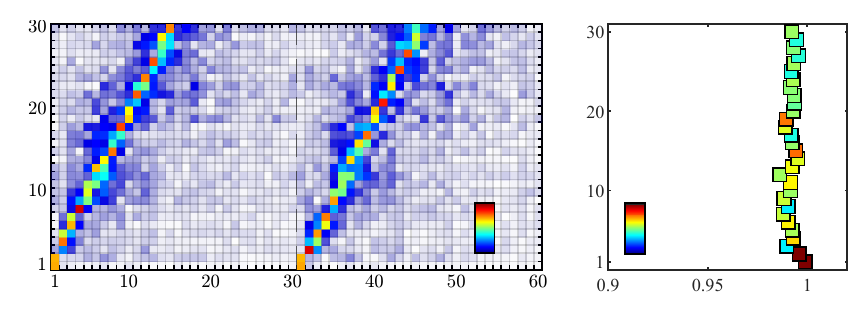}
  \put (22 , 138) {(a)}
  \put (273 , 138) {(b)}
  \put (188 , 58){\rotatebox{0} {\fontsize{8}{9}\selectfont$\big| \langle \hat{\boldsymbol{\mathcal{Q}}}_i^\text{asym},\hat{\boldsymbol{\mathcal{Q}}}_j^\text{sym}\rangle\big|$}}
   \put (284 , 57){\rotatebox{0} {\fontsize{8}{9}\selectfont$\psi$}}
   \put (294 , 48){\rotatebox{0} {\fontsize{8}{9}\selectfont$0.65$}}
   \put (294 , 30){\rotatebox{0} {\fontsize{8}{9}\selectfont$0$}}
  \put (227,30){\fontsize{8}{9}\selectfont 0}
   \put (227,47){\fontsize{8}{9}\selectfont 1} 
  \put (254 , 56){\rotatebox{90} {\fontsize{8}{9}\selectfont Mode rank $i$}}
  \put (322, 2){\rotatebox{0} {\fontsize{8}{9}\selectfont $E_i$}}
  \put (0 , 48){\rotatebox{90} {\fontsize{8}{9}\selectfont Asymmetric mode}}
  \put (40 , 2) {\fontsize{8}{9}\selectfont Anti-symmetric mode}
  \put (160 , 2) {\fontsize{8}{9}\selectfont Symmetric mode}
\end{overpic}}
\caption{(a) The projection coefficient matrix obtained from the projection of asymmetric modes, $\hat{\boldsymbol{\mathcal{Q}}}_i^\text{asym}$, onto the symmetric/anti-symmetric modes obtained from symmetric base flow, $\hat{\boldsymbol{\mathcal{Q}}}_j^\text{sym}$. (b) Energy of the 30 leading asymmetric response modes that can be captured through the phase interference of symmetric and anti-symmetric response modes, color-coded according to the degree of asymmetry of the corresponding modes.} 
\label{fig.E_matrix}
\end{figure}

Let us examine this reconstruction in equation~\eqref{eq.reconst_asym} by first considering the asymmetric wake obtained with an asymmetric forcing of $C_\mu = 0.01$.  Following the harmonic resolvent analysis of the symmetric base flow, we construct a basis consisting of $30$ symmetric and $30$ anti-symmetric response modes, denoted by $\hat{\boldsymbol{\mathcal{Q}}}_j^\text{sym}$. The asymmetric response modes, $\hat{\boldsymbol{\mathcal{Q}}}_i^\text{asym}$, are then projected onto the subspace spanned by this basis. In figure~\ref{fig.E_matrix}(a), we present the magnitudes of these projection coefficients, or
$
    \big|\langle \hat{\boldsymbol{\mathcal{Q}}}_i^\text{asym}, \hat{\boldsymbol{\mathcal{Q}}}_j^\text{sym}\rangle\big|
$
for each $(i, j)$-pair, and summarize them in a matrix form. This allows us to quantify the reconstruction accuracy of $\tilde{\boldsymbol{\mathcal{Q}}}_i^\text{asym}$ by calculating the amount of energy of the asymmetric mode captured by the basis of symmetric-flow modes.  That is
\begin{equation}
E_i = \sum\limits_{j = 1}^{n_s}\big|\langle \hat{\boldsymbol{\mathcal{Q}}}_i^\text{asym}, \hat{\boldsymbol{\mathcal{Q}}}_j^\text{sym}\rangle\big|^2,
\label{eq.Ei}
\end{equation}
with $n_s = 60$ for this analysis. Figure~\ref{fig.E_matrix}(b) shows the energy of the first 30 asymmetric response modes captured through the superposition, with the markers colored by the level of asymmetry of the corresponding modes. We observe that more than $98\%$ of the energy of the first 30 asymmetric modes is captured through the interference among the symmetric-flow modes in the basis, despite high levels of asymmetry in some of the modes.  Considering that the asymmetric base flow here is characterized by a finite-amplitude departure from its nominally symmetric state, we find these levels of reconstruction accuracy highly remarkable.

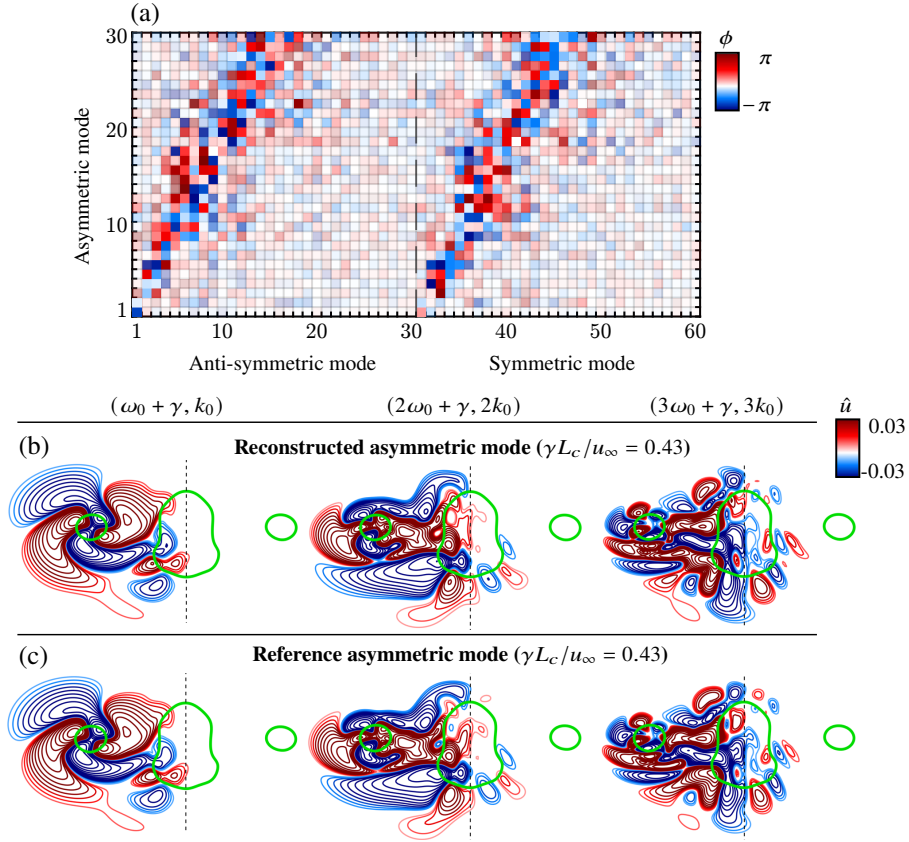
\begin{figure}
  \centerline{\begin{overpic}[width=5.32in]{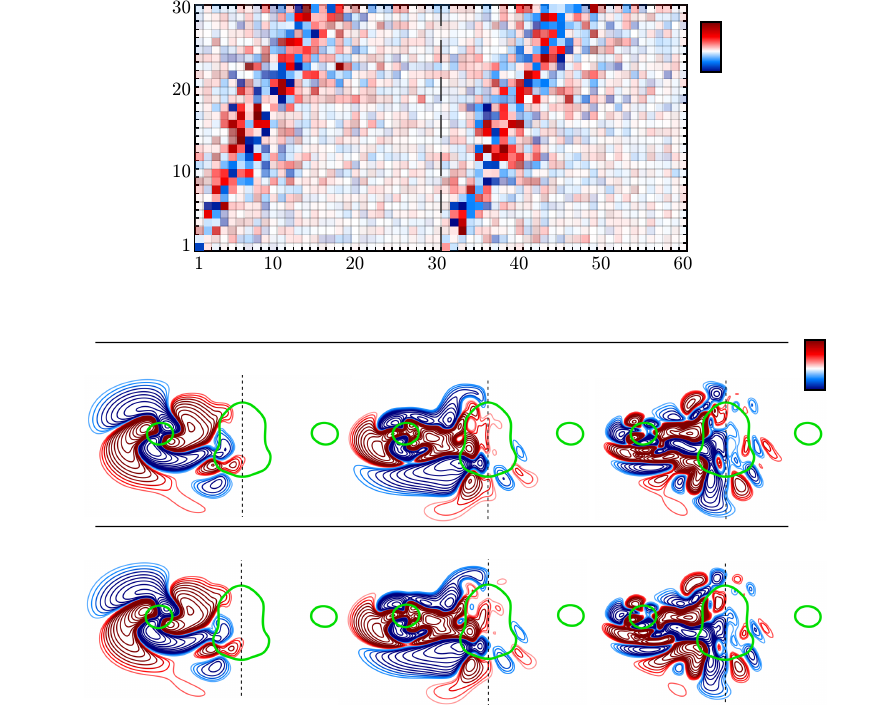} 
   \put (84, 309) {(a)}
   \put (42, 146) {(b)}
   \put (42, 67) {(c)}

   \put (63 , 217){\rotatebox{90} {\fontsize{8}{9}\selectfont Asymmetric mode}}
   \put (106 , 177.5) {\fontsize{8}{9}\selectfont Anti-symmetric mode}
   \put (219 , 177.5) {\fontsize{8}{9}\selectfont Symmetric mode}
   \put (76, 162){\rotatebox{0}{\fontsize{8}{9}\selectfont $(\omega_0+\gamma, k_0)$}}
  \put (180, 162) {\fontsize{8}{9}\selectfont $(2\omega_0+\gamma, 2k_0)$}
  \put (279, 162) {\fontsize{8}{9}\selectfont $(3\omega_0+\gamma, 3k_0)$}
   \put (351, 162) {\fontsize{8}{9}\selectfont $\hat{u}$}
   \put (362, 154) {\fontsize{8}{9}\selectfont 0.03}
   \put (359, 137) {\fontsize{8}{9}\selectfont -0.03}
   \put (305, 300.5) {\fontsize{8}{9}\selectfont $\phi$}
   \put (320, 293) {\fontsize{8}{9}\selectfont $\pi$}
   \put (314, 275) {\fontsize{8}{9}\selectfont $-\pi$}
    \put (123 , 146) {\fontsize{8}{9}\selectfont \textbf{Reconstructed asymmetric mode ($\gamma L_c /u_\infty=0.43$)}}
   \put (130 , 68) {\fontsize{8}{9}\selectfont \textbf{Reference asymmetric mode ($\gamma L_c /u_\infty=0.43$)}}
\end{overpic}}
\caption{(a) Computed phase angles, $\phi$, from the projection of the asymmetric modes $\hat{\boldsymbol{\mathcal{Q}}}_i^\text{asym}$ onto symmetric and anti-symmetric modes $\hat{\boldsymbol{\mathcal{Q}}}_j^\text{sym}$. Reconstructed asymmetric response modes in (b) at the frequency-wavenumber pairs  $(m \omega_0 + \gamma, rk_0)$, with $\gamma L_c/u_\infty=0.43$ and $m=r=1,2,3$, compared with the corresponding reference asymmetric modes obtained directly from the asymmetric base flow in (c).}  
\label{fig.Phase}
\end{figure}

Note that the harmonic resolvent modes are complex-valued vectors.  Therefore, the projection coefficient, 
$\langle
    \hat{\boldsymbol{\mathcal{Q}}}_i^\text{asym},~\hat{\boldsymbol{\mathcal{Q}}}_j^\text{sym}
\rangle$, provides not only the magnitude, but also the phase for the symmetric and anti-symmetric modes to interfere and result in an asymmetric mode. This phase can be determined by 
\begin{equation}
\phi_{ij}=\arg\!
    \left(
    \left\langle
        \hat{\boldsymbol{\mathcal{Q}}}_i^\text{asym},~\hat{\boldsymbol{\mathcal{Q}}}_j^\text{sym}
    \right\rangle
    \right).
\label{eq:phaseinf}
\end{equation}
These phases from the projection coefficients are shown in figure~\ref{fig.Phase}(a). Following equation \ref{eq.reconst_asym}, we reconstruct the asymmetric modes by superposing the symmetric and anti-symmetric modes, scaled by the magnitude of their projection coefficients and rotated according to the corresponding phases. For demonstration purposes, we reconstruct the leading response modes of the asymmetric flow associated with the symmetry-breaking perturbations at the frequency-wavenumber pairs $(m \omega_0 + \gamma, rk_0)$, with $\gamma L_c/u_\infty=0.43$ and $m=r=1,2,3$. Figure~\ref{fig.Phase}(b) presents the reconstructed response modes, which are compared with the corresponding asymmetric modes directly obtained from the asymmetric base flow in figure~\ref{fig.Phase}(c). This comparison demonstrates that the phase interference between the symmetric and anti-symmetric modes effectively captures the evolution of the asymmetric helical vortex instability and its interaction with the wake shear layer in the asymmetric wake.

\begin{figure}
\centerline{\raggedleft{\begin{overpic}[width=5.32in]{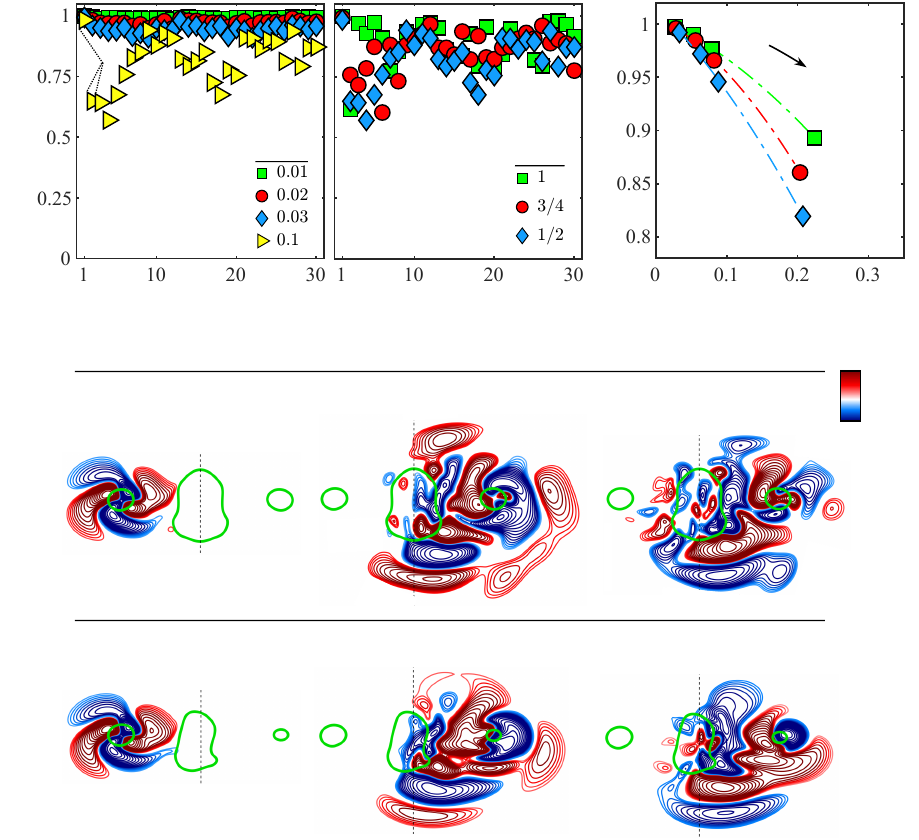}
  \put (32 , 354) {(a)}
  \put (140 , 354) {(b)}
  \put (277 , 354) {(c)}
  \put (32 , 200) {(d)}

  \put (110 , 288) {\fontsize{8}{9}\selectfont $C_\mu$}
  \put (218 , 286) {\fontsize{8}{9}\selectfont $z_0/b$}
  \put (35 , 249) {\fontsize{8}{9}\selectfont $z_0/b=1/2$}
  \put (143 , 249) {\fontsize{8}{9}\selectfont $C_\mu = 0.1$}
  \put (339, 255) {\fontsize{7}{9}\selectfont $z_0/b=1/2$}
  \put (339, 275) {\fontsize{7}{9}\selectfont $z_0/b=3/4$}
  \put (346 , 290) {\fontsize{7}{9}\selectfont $z_0/b=1$}
  \put (325 , 334) {\rotatebox{-22}{\fontsize{7}{9}\selectfont $C_\mu\uparrow$}}
  \put (43 , 326.5) {\fontsize{8}{9}\selectfont \color{violet}(d)}

  \put (5 , 290){\rotatebox{90} {\fontsize{8}{9}\selectfont $E_i$}}
  \put (60 , 221){\rotatebox{0} {\fontsize{8}{9}\selectfont Mode rank $i$}}
  \put (170 , 221){\rotatebox{0} {\fontsize{8}{9}\selectfont Mode rank $i$}}
   \put (322 , 221){\rotatebox{0} {\fontsize{8}{9}\selectfont $\psi_b$}}
  \put (248 , 290){\rotatebox{90} {\fontsize{8}{9}\selectfont $\bar{E}$}}
  \put (87 , 185) {\fontsize{8}{9}\selectfont \textbf{Reconstructed asymmetric mode at $(\omega_0+\gamma, k_0) \text{ with } \gamma L_c /u_\infty =0.43$}}
  \put (91 , 80) {\fontsize{8}{9}\selectfont \textbf{Reference asymmetric mode at $(\omega_0+\gamma, k_0) \text{ with } \gamma L_c /u_\infty =0.43$}}
  \put (353, 198.5) {\fontsize{8}{9}\selectfont $\hat{u}$}
  \put (365, 190) {\fontsize{8}{9}\selectfont 0.03}
  \put (362, 174) {\fontsize{8}{9}\selectfont -0.03}
  \put (62 , 200) {\fontsize{8}{9}\selectfont Mode rank 1}
  \put (165 , 200) {\fontsize{8}{9}\selectfont Mode rank 2}
  \put (275 , 200) {\fontsize{8}{9}\selectfont Mode rank 3}

\end{overpic}}}
\caption{Energy of the 30 leading response modes that can be captured through the phase interference of symmetric and anti-symmetric response modes, where the asymmetric base flows are obtained by different asymmetric forcing: (a) volumetric forcing in the base flows is applied with different forcing strengths of $C_\mu = 0.01$, $0.02$, 0.03 and $0.1$ at the spanwise location of $z_0/b=1/2$, (b) volumetric forcing in the base flows is applied with the strength of $C_\mu = 0.1$ at different spanwise locations $z_0/b=1/2$, $3/4$, and $1$. (c) Averaged captured energy, $\bar{E}$, of the 30 leading response modes for different base flow asymmetry levels $\psi_b$. (d) Reconstruction of the first three leading modes at the fundamental frequency–wavenumber pair ($\omega_0+\gamma, k_0$) with $\gamma L_c/u_\infty = 0.43$, compared with the corresponding response modes obtained from asymmetric base flows with forcing applied at the spanwise location $z_0/b=1/2$, with $C_\mu=0.1$.} 
\label{fig.Forcing_Energy}
\end{figure}
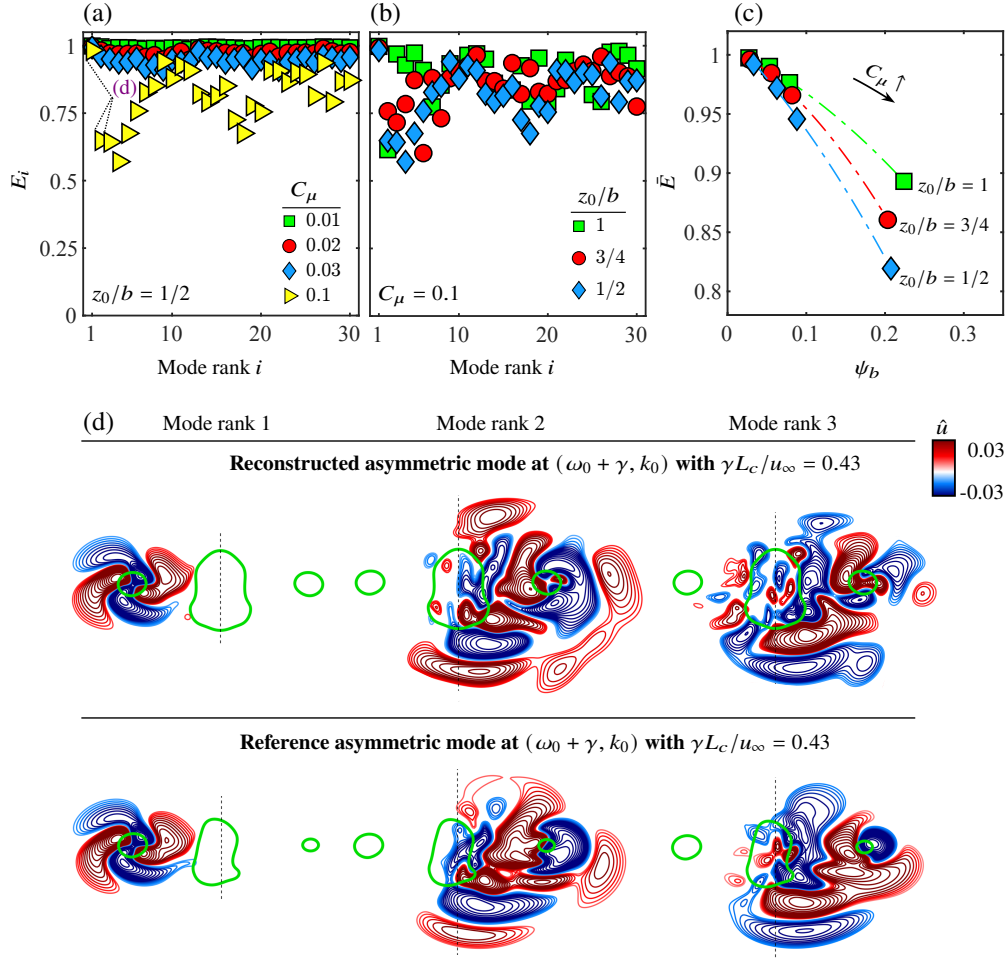

We next examine the influence of the location and strength of the asymmetric volumetric forcing introduced to the wake flow on the reconstruction accuracy of the asymmetric modes. In figure~\ref{fig.Forcing_Energy}(a), we examine the influence of forcing strength by varying $C_\mu$ from $0.01$ to $0.1$, with the forcing placed at a fixed spanwise location of $z_0/b=1/2$. As the forcing strength increases, the energy of the first 30 response modes captured by the symmetric-flow modal basis gradually decreases. This general trend is expected, as the increasing nonlinearity due to the finite-amplitude departure from the symmetric state would decrease the accuracy of the reconstruction that is linear by nature. Surprisingly, the results in figure~\ref{fig.Forcing_Energy}(a) show that more than $99\%$ of the energy of the rank-1 mode is still captured by the basis of symmetric-flow modes, even for the asymmetric base flow with $C_\mu = 0.1$. For moderate forcing strengths ($C_\mu=0.01$-0.03), the captured energy of higher-ranked modes remains above $90\%$. As the forcing strength increases to $C_\mu=0.1$, the captured energy of higher-ranked asymmetric modes decreases to approximately $57\%$, with an average captured energy of about $83\%$ across the first 30 leading modes.  In figure~\ref{fig.Forcing_Energy}(d), we reconstruct the first three leading asymmetric modes obtained from the asymmetric base flow with forcing applied at $z_0/b=1/2$ with $C_\mu=0.1$. The leading asymmetric mode, with approximately $99\%$ of its energy captured by the symmetric-flow modes, shows accurate reconstruction with respect to the modal structures, which is associated with the elliptic vortex instability. For the second and third modes, the captured energy decreases to $65\%$ and $64\%$, respectively.  Regardless of the lower reconstruction accuracy, the reconstructed modes still reasonably capture the asymmetric wake structures and the perturbations around the tip vortex. Overall, these analyses suggest that phase interference between the symmetric and anti-symmetric modes from the originally symmetric flow can be a potential mechanism for the emergence of asymmetric flow structures.

In figure~\ref{fig.Forcing_Energy}(b), we focus on the effect of forcing location by considering the largest forcing strength of $C_\mu=0.1$ and varying the forcing location from $z_0/b=1/2$ to the wing tip ($z_0/b=1$). The results show that the captured energy decreases slightly as the forcing location moves toward the wing root. This indicates the increased sensitivity of the wake to external disturbances, consistent with the non-normal dynamics of the wake evidenced by the spectra presented in figure~\ref{fig.LinStabilitySpectrum}. In figure~\ref{fig.Forcing_Energy}(c), we show that despite up to $22\%$ time-averaged asymmetry in the base flows, $\psi_b$, arising from different forcing configurations, on average, more than $83\%$ of the energy of the 30 leading asymmetric modes responsible for symmetry-breaking is captured through phase interference between symmetric and anti-symmetric modes obtained from the symmetric base flow. 

\subsection{Capturing the asymmetric base flow departure using the harmonic resolvent modes}
\label{BaseFlowShift}
The emergence of asymmetries not only manifests in the oscillatory fluctuations, but also appears as the base flow departure from the symmetric state. Following the analysis of asymmetric oscillatory fluctuations in~\S\ref{phase_interference}, in this section we investigate if such an asymmetric departure of the base flow can also be captured by the harmonic resolvent modes from the symmetric flow. To formalize the objective, let us consider the asymmetric base flow 
$\boldsymbol{q}_b^\text{asym}$ as a departure $\delta \boldsymbol{q}_b$ 
from the symmetric base flow $\boldsymbol{q}_b^\text{sym}$. Accordingly, this departure from the symmetric state can be expressed as
\begin{equation}
\delta \boldsymbol{q}_{b}(\boldsymbol{x},t)
=
\boldsymbol{q}_b^\text{asym}(\boldsymbol{x},t)
-
\boldsymbol{q}_b^\text{sym}(\boldsymbol{x},t).
\label{meanshift}
\end{equation}
Using the same Fourier mode representation of the base flow in equation \ref{eq.periodicBFFourier}, this base flow departure can also be expressed as
\begin{equation}
\delta \boldsymbol{q}_b(\boldsymbol{x},t)
=\sum_{n=-\infty}^{+\infty}
\underbrace{
    \left[
    \tilde{\boldsymbol{q}}_b^\text{asym}(y,z) - \tilde{\boldsymbol{q}}_b^\text{sym}(y,z)
    \right]_{n\omega_0,\,n k_0}
}_{\delta \tilde{\boldsymbol{q}}_{n\omega_0,\,n k_0}}
\,\mathrm{e}^{\mathrm{i}n(\omega_0 t - k_0 x)},
\quad n \in \mathbb{Z}.
\label{eq.BFshift}
\end{equation}
where $\tilde{\boldsymbol{q}}_b^\text{asym}$ and $\tilde{\boldsymbol{q}}_b^\text{sym}$ are respectively the Fourier modes of the asymmetric and symmetric base flows at the frequency–wavenumber pair $(n\omega_0,nk_0)$, and $\delta \tilde{\boldsymbol{q}}_{n\omega_0,\,n k_0}$ denotes the Fourier modes of the base-flow departure. Note that the departure mode $\delta \tilde{\boldsymbol{q}}$ is normalized such that it has a unit 2-norm in the following analyses.  From such a perspective, we can also view this base-flow departure as a perturbation mode and perform a similar analysis to that in~\S\ref{phase_interference}. Our goal is to quantify the energy of this base-flow departure that can be captured by the basis comprised of the symmetric and anti-symmetric modes ${\hat{\boldsymbol{\mathcal{Q}}}_j^\text{sym}}$ from the symmetric base flow.  Here, we focus on the harmonic resolvent modes at the harmonics of the fundamental frequency-wavenumber pair (so the frequency shift $\gamma=0$). The fraction of captured energy by projecting the base-flow departure, ${\delta \tilde{\boldsymbol{q}}}$, onto this basis can be found via
\begin{equation}
E=\sum_{j=1}^{n_s}
\left|\left\langle
 {\delta \tilde{\boldsymbol{q}}},~
{{\hat{\boldsymbol{\mathcal{Q}}}_j^\text{sym}}}
\right\rangle\right|^2,
\label{eq.BFE}
\end{equation}
where the total number of symmetric and anti-symmetric in the basis is $n_s=60$. 

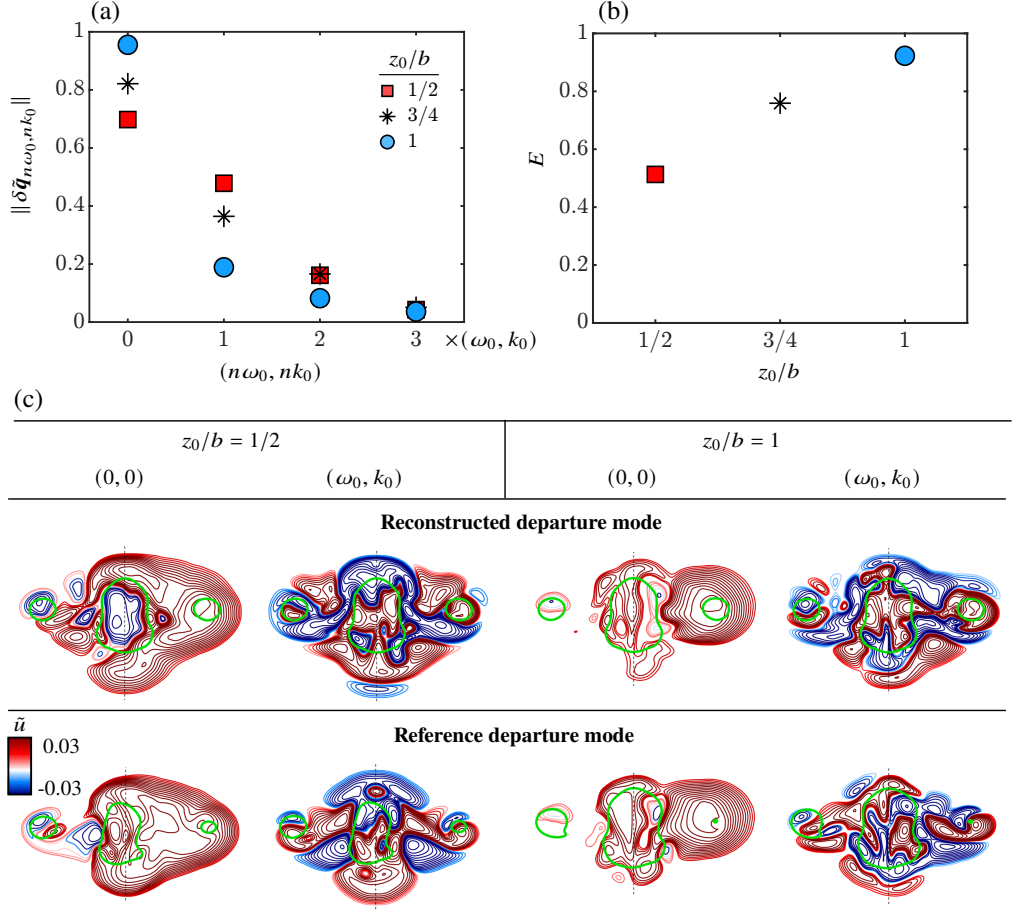
\begin{figure}
  \centerline{\begin{overpic}[width=5.32in]{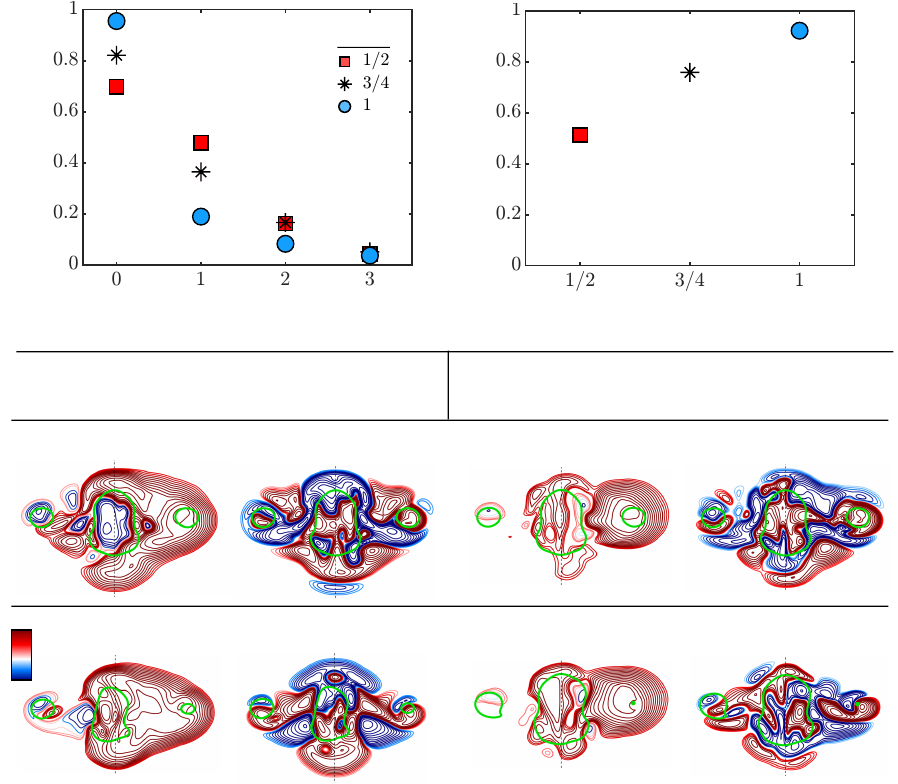} 
  \put (36 , 336) {(a)}
  \put (227 , 336) {(b)}
  \put (7 , 190) {(c)}
  \put (145 , 144) {\fontsize{8}{9}\selectfont \textbf{Reconstructed departure mode}}
 \put (150 , 64) {\fontsize{8}{9}\selectfont \textbf{Reference departure mode}}
 \put (37 , 161) {\fontsize{8}{9}\selectfont $(0,0)$}
 \put (124 , 161) {\fontsize{8}{9}\selectfont $(\omega_0,k_0)$}
 \put (230 , 161) {\fontsize{8}{9}\selectfont $(0,0)$}
 \put (319 , 161) {\fontsize{8}{9}\selectfont $(\omega_0,k_0)$}
 \put (7, 68) {\fontsize{8}{9}\selectfont $\tilde{u}$}
 \put (18 , 60) {\fontsize{8}{9}\selectfont 0.03}
 \put (16 , 44) {\fontsize{8}{9}\selectfont -0.03}
 \put (148 , 319) {\fontsize{8}{9}\selectfont $z_0/b$}
 \put (70 , 175) {\fontsize{8}{9}\selectfont $z_0/b=1/2$}
 \put (266 , 175) {\fontsize{8}{9}\selectfont $z_0/b=1$}
 \put (6, 262){\rotatebox{90} {\fontsize{8}{9}\selectfont $\| \delta \tilde{\boldsymbol{q}}_{n \omega_0,n k_0}  \|$}}
 \put (201 , 280){\rotatebox{90} {\fontsize{8}{9}\selectfont $E$}}
 \put (84 , 201){\rotatebox{0} {\fontsize{8}{9}\selectfont $(n \omega_0, nk_0)$}}
 \put (169 , 213){\rotatebox{0} {\fontsize{8}{9}\selectfont $\times (\omega_0, k_0)$}}
 \put (287 , 201){\rotatebox{0} {\fontsize{8}{9}\selectfont $z_0/b$}}
\end{overpic}}
\caption{(a) Energy distribution of the base flow departure harmonics, $\delta \tilde{\boldsymbol{q}}_{n\omega_0,\,n k_0}$, under different spanwise forcing locations with $C_\mu = 0.1$. (b) Energy of the base flow departure captured through the superposition of symmetric and anti-symmetric modal structures for different spanwise forcing locations. (c) Reconstructed zeroth and fundamental harmonics of $\delta \tilde{\boldsymbol{q}}_{n\omega_0,\,n k_0}$ obtained through phase interference of symmetric and anti-symmetric modes, are compared with the corresponding departure modes computed from the difference between the asymmetric and symmetric base flows.}  
\label{fig.BFShift}
\end{figure}

We consider three asymmetric base flows obtained by introducing the volumetric forcing of the highest strength ($C_\mu = 0.1$) at the spanwise locations of $z_0/b = 1/2$, $3/4$, and $1$. First, we examine the energy distribution over the harmonics of $(\omega_0, k_0)$ for the base-flow departure in figure~\ref{fig.BFShift}(a). Since the asymmetric volumetric forcing applied to the base flow is steady, the modification of the zero frequency-wavenumber component ($n=0$) exhibits the largest amplitude among all harmonics of $\tilde{\boldsymbol{q}}_{n\omega_0, n k_0}$. Despite the overall dominance of the zero frequency-wavenumber component, the energy distribution of the base flow departure strongly depends on the location of the asymmetric forcing. As shown in figure~\ref{fig.BFShift}(a), when the forcing is introduced near the wing-tip ($z_0/b=1$), the energy content of the base-flow departure concentrates at the zero frequency-wavenumber harmonic. However, when asymmetric disturbances are introduced near the wing root ($z_0/b=1/2$), the relative contribution of the nonzero harmonics increases, indicating that the unsteady structure in the wake is also responding to the time-invariant forcing. Such a sensitivity of wake modes can again be attributed to the high non-normality of the associated wake modes, as evidenced by the eigenmode characteristics discussed in~\S\ref{Eigenvalueanalysis}.

To examine how well the base-flow departure can be represented by the basis of symmetric-flow modes, we compute the captured energy defined in equation~\eqref{eq.BFE}. Figure~\ref{fig.BFShift}(b) shows the captured energy of the base-flow departure stacked over all frequency-wavenumber pairs $(m\omega_0,rk_0)$ with $m=r=0,\dots,\pm3$. We observe that the projection captures approximately $51\%$ of the base-flow departure when asymmetric forcing is applied near the wing root ($z_0/b=1/2$). This fraction exceeds $92\%$ when the asymmetric forcing is introduced near the wing tip $(z_0/L_c = 1)$. The reduced captured energy of the base-flow departure when forcing is applied near the wing root can be attributed to the increased contribution of nonzero harmonics to the base-flow departure, where the unsteady wake structures respond to the steady forcing in a highly nonlinear fashion. The difference in the captured energy is also reflected in the reconstructed streamwise velocity field in figure~\ref{fig.BFShift}(c). While the reconstruction error is noticeably larger for the fundamental frequency-wavenumber, closer agreement is observed for the zeroth harmonic component. The zeroth harmonic component reconstructed through the symmetric-flow modes captures an asymmetric region of increased streamwise velocity and is in close agreement with the departure field defined as the difference between the asymmetric and symmetric base flows. In summary, these findings suggest that the departure of the mean flow from the symmetric state can also be approximated through the superposition of the intrinsic modal structures of the symmetric flow.  Moreover, they imply that it is possible to modify the flow to a desirable asymmetric flow state by manipulating the phase dynamics of the underlying symmetric-flow modal structures, providing opportunities for physics-guided flow control to enhance the performance of aerodynamic vehicles. 

\section{Conclusion}
\label{Conclusion}

This study investigated the mechanisms underlying the emergence of long-time spanwise flow asymmetries in the unsteady wake of a rectangular finite-span wing at a chord-based Reynolds number of $1,000$ and angle of attack $\alpha = 10^\circ$.  To this end, we aimed to disentangle the asymmetric wake structures into the intrinsic symmetric and anti-symmetric structures that reside in the symmetric wake flow by performing global mode sensitivity and STHR analyses for both symmetry-enforced and asymmetrically disturbed wake flows. 

The global mode sensitivity analysis revealed that asymmetric structures preferentially emerge when highly non-normal symmetric and anti-symmetric eigenmodes in the nominally symmetric wake appear as pairs. The high levels of modal non-normality enhance the sensitivity of the flow to small asymmetric disturbances, facilitating modal interference between paired symmetric and anti-symmetric eigenmodes that unfolds them into asymmetric eigenmodes. In contrast, unpaired symmetric or anti-symmetric modes and mode pairs that exhibit low non-normality retain high levels of spanwise symmetry even when the wake flow is asymmetrically disturbed. These conditions provide implications for predicting the emergence of flow asymmetry in a nominally symmetric flow by performing structural sensitivity analysis about the symmetric flow.

Motivated by the modal interference as a potential mechanism for flow asymmetries to emerge, we developed the STHR framework to elucidate the manifestation of modal interference by combining the harmonic resolvent formulations of two prior studies. This formulation enables interactions between perturbations propagating with different phase velocities and explicitly embeds the phase information of the base flow into the linear operator.  The STHR analysis revealed that the long-time symmetry-breaking is associated with elliptic vortex instabilities that propagate with phase velocities slightly higher than that of the base flow, consistent with the convective instability characteristics identified through the linear stability analysis.  It also showed that the cross-frequency coupling serves as a crucial mechanism for modal asymmetries to emerge.  A perturbation component can be highly asymmetric even if it exhibits low amplification, as long as it is coupled with another highly asymmetric component of high amplification.  Moreover, even when the asymmetric departure of the base flow was increased to a large amplitude, we observed that asymmetric wake structures can still be understood as a phase interference between symmetric and anti-symmetric modes residing in the symmetric base flow, underscoring again the essential role of phase interference in the emergence of asymmetries in the unsteady wing wake. 

Lastly, we viewed the asymmetric departures of the base flow as a perturbation about the nominally symmetric base flow and showed that the underlying symmetric and anti-symmetric STHR modes can still provide insights into such mean-flow departures. This observation implies that a desirable asymmetric flow state can be achieved by manipulating the phase dynamics of the symmetric-flow modal structures, providing opportunities for physics-guided flow control to enhance the performance of aerodynamic vehicles. Overall, the present study establishes a phase-based perspective for understanding the emergence and manipulation of long-time flow asymmetries in the finite-wing wake. Beyond providing physical insight into the origin of asymmetric wakes, these results lay a foundation for predictive and control-oriented strategies that target modal phase relationships to suppress or promote flow asymmetries over finite wings.

\vspace{0.5cm}
\noindent\textbf{Declaration of interests}. The authors report no conflict of interest.
\appendix
\section{Global mode sensitivity against random asymmetric disturbances}
\label{apndx.randforcing}
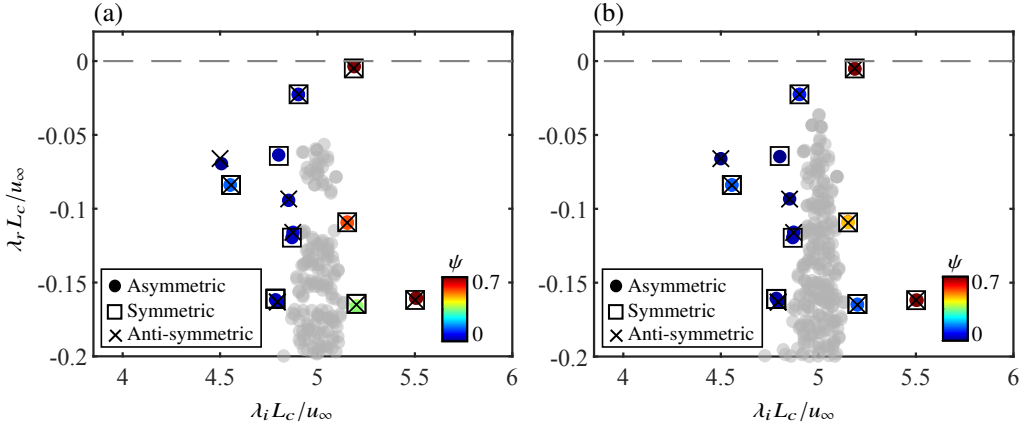
\begin{figure}
  \centerline{\begin{overpic}[width=5.32in]{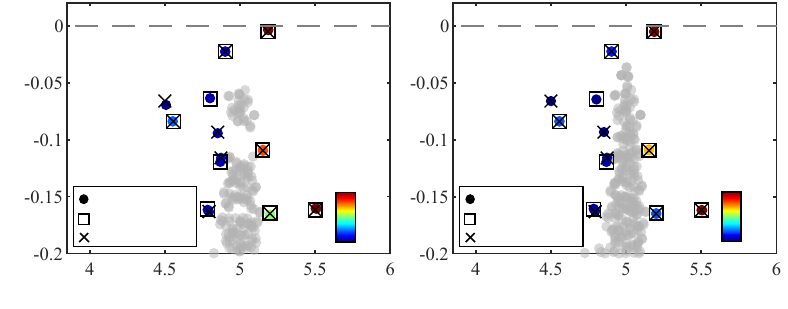} 
   \put (33 , 153) {(a)}
   \put (221 , 153) {(b)}
   
   \put (166 , 60) {\fontsize{8}{9}\selectfont $\psi$}
   \put (175.5 , 32.5) {\fontsize{8}{9}\selectfont $0$}
   \put (175.5 , 52) {\fontsize{8}{9}\selectfont $0.7$}
   
   \put (354 , 60) {\fontsize{8}{9}\selectfont $\psi$}
   \put (363.5 , 32.5) {\fontsize{8}{9}\selectfont $0$}
   \put (363.5 , 52) {\fontsize{8}{9}\selectfont $0.7$}   
   
   \put (45.5 , 51) {\fontsize{7.5}{9}\selectfont Asymmetric}
   \put (45.5 , 41.5) {\fontsize{7.5}{9}\selectfont Symmetric}
   \put (45.5 , 33) {\fontsize{7.5}{9}\selectfont Anti-symmetric}

   \put (234 , 51) {\fontsize{7.5}{9}\selectfont Asymmetric}
   \put (234 , 41.5) {\fontsize{7.5}{9}\selectfont Symmetric}
   \put (234 , 33) {\fontsize{7.5}{9}\selectfont Anti-symmetric}
   \put (0 , 66) {\rotatebox{90}{\fontsize{8}{9}\selectfont$\lambda_r L_c/u_\infty$}}
   \put (281 , 4) {\rotatebox{0}{\fontsize{8}{9}\selectfont$\lambda_i L_c/u_\infty$}}
   \put (92 , 4) {\rotatebox{0}{\fontsize{8}{9}\selectfont$\lambda_i L_c/u_\infty$}}
\end{overpic}}
\caption{Comparison of eigenvalue spectra of asymmetric base flows under two different asymmetric disturbances: (a) base flow obtained from Gaussian volumetric forcing applied in the wing upstream at $z_0/b=1/2$; (b) base flow disturbed by a random asymmetric disturbance of the same $C_\mu$.}  
\label{fig.SpectrumsCmpr}
\end{figure}
The asymmetric base flow considered in the global mode sensitivity analysis in~\S\ref{Eigenvalueanalysis} is obtained by introducing a spherical Gaussian forcing that is asymmetric about the root plane, as discussed in~\S\ref{asymforcing}. To ensure that our findings are robust against the shape of the asymmetric disturbances, we also consider another asymmetric base flow obtained by random asymmetric disturbances of equal strength. In figure~\ref{fig.SpectrumsCmpr}, we compare the eigenvalue spectra of two asymmetric base flows obtained from different shapes of disturbances. We observe that the eigenmodes of high levels of asymmetries appear in the same clusters for both asymmetric base flows obtained from different asymmetric forcings.  This indicates that our conclusive assessments for the emergence of asymmetric modes remain valid regardless of the spatial shapes of asymmetric disturbances.

\section{Spatial-temporal harmonic resolvent analysis}
Here, we provide further details for constructing the STHR operator. We start from recalling equation~\eqref{eq.harmonic}, which formulates the STHR analysis as a coupled system of equations for the flow response at different frequencies and wavenumbers, given by
\begin{equation*}
    \mathrm{i}(m \omega_0+\gamma) \mathsfbi{M} \hat{\boldsymbol{q}}_{m \omega_0+\gamma, r k_0} 
    = 
    \sum_{n + \check{m} = m} \sum_{n + \check{r} = r}
    \mathsfbi{\hat{A}}_{n\omega_0, nk_0} (\check{r}k_0) \hat{\boldsymbol{q}}_{\check{m}\omega_0+\gamma, \check{r}k_0} 
    +
    \mathsfbi{M}\hat{\boldsymbol{f}}_{m \omega_0+\gamma, r k_0}.
\end{equation*}
Also recall that $\mathsfbi{\hat{A}}_{n\omega_0, nk_0}$ are the Fourier components appear in equation~\eqref{eq.LinOptA}, given by
\begin{align*}
    \mathsfbi{A}_{\bar{\boldsymbol{q}}(\boldsymbol{x}, t)}(rk_0)
    = \sum_{n = -\infty}^{+\infty} \mathsfbi{\hat{A}}_{n\omega_0, n k_0} (rk_0) \mathrm{e}^{\mathrm{i}n( \omega_0 t -  k_0 x)}.
\end{align*}
For simplicity, in this section we adopt the following compact notations,
\[
	\mathsfbi{\hat{A}}_{n\omega_0, n k_0} (rk_0) \rightarrow \mathsfbi{\hat{A}}_n^{(r)}
	,~~~
	\tilde{\boldsymbol{q}}_{n\omega, nk_0} \rightarrow \tilde{\boldsymbol{q}}_{n}
	,~~~
	\hat{\boldsymbol{q}}_{m\omega+\gamma, rk_0} \rightarrow \hat{\boldsymbol{q}}_{m,\gamma}^{(r)}
	,~~~
	\hat{\boldsymbol{f}}_{m\omega+\gamma, rk_0} \rightarrow \hat{\boldsymbol{f}}_{m,\gamma}^{(r)},
\]
such that  equation~\eqref{eq.harmonic} becomes 
\begin{equation}
    \mathrm{i}\gamma \mathsfbi{M} \hat{\boldsymbol{q}}_{m,\gamma}^{(r)}
    +
    \mathrm{i}m \omega_0 \mathsfbi{M} \hat{\boldsymbol{q}}_{m,\gamma}^{(r)}
    - 
    \sum_{n + \check{m} = m} \sum_{n + \check{r} = r}
    \mathsfbi{\hat{A}}_n^{(\check{r})} \hat{\boldsymbol{q}}_{\check{m},\gamma}^{(\check{r})} 
    =
    \mathsfbi{M} \hat{\boldsymbol{f}}_{m,\gamma}^{(r)}.
    \label{eq.harmonic_compact}
\end{equation}
Below, we report explicitly the analytical expressions for $\mathsfbi{\hat{A}}_n^{(r)}$ and the formation of all finite-size matrices involved in the construction of the STHR operator. 

\subsection{Fourier components of the linear NS operator} 
\label{apndx.Ahats}
As mentioned, the Fourier component $\mathsfbi{\hat{A}}_{0}^{(r)}$ is equivalent to the linear NS operator about the time- and streamwise-averaged base flow, $\bar{\boldsymbol{q}}_0 = [\bar{u}, \bar{v}, \bar{w}, \bar{p}]^T$.  For an operator $\mathsfbi{\hat{A}}_{n}^{(r)}$ where $n \ne 0$, the operator is only related to the linearized convective terms about the $n$-th Fourier component of the base flow, $\tilde{\boldsymbol{q}}_{n} = [\tilde{u}_n, \tilde{v}_n, \tilde{w}_n, \tilde{p}_n]^T$.  For the cartesian coordinate system used in this study, these operators that act on the perturbation state vector $\hat{\boldsymbol{q}}_{m,\gamma}^{(r)} = [\hat{u}_{m, \gamma}^{(r)},~\hat{v}_{m, \gamma}^{(r)},~\hat{w}_{m, \gamma}^{(r)},~\hat{p}_{m, \gamma}^{(r)}]^{T}$ respectively write
{
\setlength{\arraycolsep}{3pt}
\begin{align*}
	\renewcommand\arraystretch{1.7}
	&\underbrace{
    	\begin{bmatrix}
		\bar{\nabla}^{(r)} &
		-\partial_y \bar{u}	&
		-\partial_z \bar{u} 	&
		\mathrm{i}rk_0 
        \\
        0 &
		\bar{\nabla}^{(r)} - \partial_y \bar{v}& 
		-\partial_z \bar{v}	&
		-\partial_y 
        \\
        0 &
		-\partial_y \bar{w}	&
		\bar{\nabla}^{(r)} - \partial_z \bar{w}& 
		-\partial_z 
        \\
        -\mathrm{i}rk_0 & 
        \partial_y & 
        \partial_z & 
        0
	\end{bmatrix}
	}_{\mathsfbi{\hat{A}}_{0}^{(r)}=~\mathsfbi{A}_{\bar{\boldsymbol{q}}_0}}
	~\text{and}~
	\underbrace{
    	\begin{bmatrix}
		\tilde{\nabla}_n^{(r)} + \mathrm{i}nk_0 \tilde{u}_n&
		-\partial_y \tilde{u}_n	&
		-\partial_z \tilde{u}_n 	&
		0 
        \\
		\mathrm{i}nk_0 \tilde{v}_n &
		\tilde{\nabla}_n^{(r)} - \partial_y \tilde{v}_n& 
		-\partial_z \tilde{v}_n	&
		0 
        \\
        \mathrm{i}nk_0 \tilde{w}_n &
		-\partial_y \tilde{w}	&
		\tilde{\nabla}_n^{(r)} - \partial_z \tilde{w}_n& 
		0 
        \\
        0 & 0 & 0 & 0 
	\end{bmatrix}
	}_{\mathsfbi{\hat{A}}_{n}^{(r)},~n \ne 0},
    \\
    &\hspace{0.5in}\text{where}~~
    \begin{cases}
    	\bar{\nabla}^{(r)} =
			\mathrm{i}r k_0 \bar{u} 
			- \bar{v} \partial_y - \bar{w} \partial_z 
			+ \nu \left(- r^2 k_0^2 + \partial_y^2 + \partial_z^2 \right) 
			&\text{in}~~ \mathsfbi{\hat{A}}_{0}^{(r)},\\
    	\tilde{\nabla}_n^{(r)} =	
			\mathrm{i} r k_0 \tilde{u}_n
			- \tilde{v}_n \partial_y - \tilde{w}_n \partial_z 
			&\text{in}~~ \mathsfbi{\hat{A}}_{n}^{(r)}.\\
    \end{cases}
\end{align*}
}
Once these operators are discretized in space, $\mathsfbi{\hat{A}}_{0}^{(r)}$ and $\mathsfbi{\hat{A}}_{n}^{(r)}$ of different values of $r$ and $n$ are used as the building blocks to form the finite-size matrices involved in the STHR analysis.

\subsection{Forming the finite-size matrices for STHR operator}
\label{apndx.HRMath}
Here, let us consider that the flow response vectors $\hat{\boldsymbol{q}}_{m,\gamma}^{(r)}$ are stacked into $\hat{\boldsymbol{\mathcal{Q}}}$ as
\[
	\hat{\boldsymbol{\mathcal{Q}}}^T = 
	\left[ 
		\dots,~
		\hat{\boldsymbol{q}}_{-1,\gamma}^{(-1)},~ 
		\hat{\boldsymbol{q}}_{0,\gamma}^{(-1)},~ 
		\hat{\boldsymbol{q}}_{1,\gamma}^{(-1)},~ 
		\dots,~
		\hat{\boldsymbol{q}}_{-1,\gamma}^{(0)},~ 
		\hat{\boldsymbol{q}}_{0,\gamma}^{(0)},~ 
		\hat{\boldsymbol{q}}_{1,\gamma}^{(0)},~ 
		\dots,~
		\hat{\boldsymbol{q}}_{-1,\gamma}^{(1)},~ 
		\hat{\boldsymbol{q}}_{0,\gamma}^{(1)},~ 
		\hat{\boldsymbol{q}}_{1,\gamma}^{(1)},~ 
		\dots
	\right],
\]
and similarly for the forcing ($\hat{\boldsymbol{\mathcal{F}}}$). With $\hat{\boldsymbol{\mathcal{F}}}$ and $\hat{\boldsymbol{\mathcal{Q}}}$, equation~\eqref{eq.harmonic_compact} can be re-written as
\begin{equation}
	\label{eq.harmonic_mat}
    \mathrm{i}\gamma \mathsfbi{I}_\mathsfbi{M} \hat{\boldsymbol{\mathcal{Q}}} 
    +
    \mathrm{i}\mathsfbi{D}_{\omega_0} \hat{\boldsymbol{\mathcal{Q}}} 
    - 
	\boldsymbol{\mathcal{A}}_{\tilde{\boldsymbol{q}}} \hat{\boldsymbol{\mathcal{Q}}} 
    =
    \mathsfbi{I}_\mathsfbi{M} \hat{\boldsymbol{\mathcal{F}}},
\end{equation} 
where $\mathsfbi{D}_{\omega_0}$ and $\boldsymbol{\mathcal{A}}_{\tilde{\boldsymbol{q}}}$ are related to the Hill matrix via
\begin{equation}
	\mathsfbi{T}_{\tilde{\boldsymbol{q}}} 
	= 
	- \mathrm{i}\mathsfbi{D}_{\omega_0}  
	+
	\boldsymbol{\mathcal{A}}_{\tilde{\boldsymbol{q}}},
	\label{eq.hill_DG}
\end{equation}
and hence equation~\eqref{eq.harmonic_mat} is equivalent to \eqref{eq.TQ}. Also, both $\mathsfbi{I}_\mathsfbi{M}$ and $\mathsfbi{D}_{\omega_0}$ are block-diagonal matrices. Following the stacking format of $\hat{\boldsymbol{\mathcal{Q}}}$, they can be straightforwardly formed via
\begin{align*}
	\mathsfbi{I}_\mathsfbi{M} 
	&=
	\text{diag}\left(
		\dots,~
		\mathsfbi{M},~ 
		\mathsfbi{M},~ 
		\mathsfbi{M},~ 
		\dots,~
		\mathsfbi{M},~ 
		\mathsfbi{M},~ 
		\mathsfbi{M},~ 
		\dots,~
		\mathsfbi{M},~ 
		\mathsfbi{M},~ 
		\mathsfbi{M},~ 
		\dots
	\right),
\end{align*}
and
\begin{align*}
	\mathsfbi{D}_{\omega_0}
	&=
	\text{diag}\left(
		\dots,~
		-\omega_0\mathsfbi{M},~ 
		\boldsymbol{0},~ 
		\omega_0\mathsfbi{M},~ 
		\dots,~
		-\omega_0\mathsfbi{M},~ 
		\boldsymbol{0},~ 
		\omega_0\mathsfbi{M},~ 
		\dots,~
		-\omega_0\mathsfbi{M},~ 
		\boldsymbol{0},~ 
		\omega_0\mathsfbi{M},~ 
		\dots
	\right). 
\end{align*}
Clearly, forming the matrix $\boldsymbol{\mathcal{A}}_{\tilde{\boldsymbol{q}}}$ is the most crucial part of the STHR analysis, as it accounts for the cross-frequency-wavenumber interactions between all components in $\hat{\boldsymbol{\mathcal{Q}}}$.  Once $\boldsymbol{\mathcal{A}}_{\tilde{\boldsymbol{q}}}$ is explicitly formed, the Hill matrix, $\mathsfbi{T}_{\tilde{\boldsymbol{q}}}$, can be readily formed via equation~\eqref{eq.hill_DG}. Similarly, the matrix $\left[\mathrm{i} \gamma \mathsfbi{I}_\mathsfbi{M}  - \mathsfbi{T}_{\tilde{\boldsymbol{q}}}\right]$ in equation~\eqref{eq.TQ} is also immediately formed for the STHR analysis.  Therefore, our discussion below will focus on how $\boldsymbol{\mathcal{A}}_{\tilde{\boldsymbol{q}}}$ is formed by placing $\mathsfbi{\hat{A}}_n^{(r)}$ in a designated block for the corresponding cross-frequency-wavenumber interactions.  

As a demonstrating example, let us consider that both the base flow and perturbations are represented by the zeroth and first harmonics only, i.e.~$\tilde{n} = \tilde{m} = \tilde{r} = 1$.  In such a case, we have nine frequency-wavenumber components in $\hat{\boldsymbol{\mathcal{Q}}}$, and the matrix $\boldsymbol{\mathcal{A}}_{\tilde{\boldsymbol{q}}}$ has a $9\times9$ block structure with different $\mathsfbi{\hat{A}}_n^{(r)}$ as the building blocks.  Now $\boldsymbol{\mathcal{A}}_{\tilde{\boldsymbol{q}}}$ can be explicitly form as
\[
\renewcommand\arraystretch{1.7}
	\boldsymbol{\mathcal{A}}_{\tilde{\boldsymbol{q}}} \hat{\boldsymbol{\mathcal{Q}}}
	=
	\underbrace{
	\begin{bmatrix}
		\mathsfbi{\hat{A}}_{0}^{(-1)}	&					
		0&	
		0&	\hspace{0.03in}
		0&	
		\mathsfbi{\hat{A}}_{-1}^{(0)} 	&					
		0&	\hspace{0.03in}
		0&	
		0&	
		0\\	
		0&	
		\mathsfbi{\hat{A}}_{0}^{(-1)} 	&					
		0&	\hspace{0.03in}
		0&	
		0&	
		\mathsfbi{\hat{A}}_{-1}^{(0)} 	&	\hspace{0.03in}	
		0&	
		0&	
		0\\	
		0&	
		0&	
		\mathsfbi{\hat{A}}_{0}^{(-1)} 	& 	\hspace{0.03in}	
		0&	
		0&	
		0&	\hspace{0.03in}
		0&	
		0&	
		0\\[3pt]	
		0&	
		0&	
		0&\hspace{0.03in}	
		\mathsfbi{\hat{A}}_{0}^{(0)} 	&					
		0&	
		0&\hspace{0.03in}	
		0&	
		\mathsfbi{\hat{A}}_{-1}^{(1)} 	&					
		0\\	
		\mathsfbi{\hat{A}}_{1}^{(-1)} 	&					
		0&	
		0&\hspace{0.03in}	
		0&	
		\mathsfbi{\hat{A}}_{0}^{(0)} 	&					
		0&\hspace{0.03in}	
		0&	
		0&	
		\mathsfbi{\hat{A}}_{-1}^{(1)}	\\					
		0&	
		\mathsfbi{\hat{A}}_{1}^{(-1)} 	&					
		0&\hspace{0.03in}	
		0&	
		0&	
		\mathsfbi{\hat{A}}_{0}^{(0)} 	&	\hspace{0.03in}	
		0&	
		0&	
		0\\[3pt]	
		0&	
		0&	
		0&	\hspace{0.03in}
		0&	
		0&	
		0&	\hspace{0.03in}
		\mathsfbi{\hat{A}}_{0}^{(1)} 	&					
		0&	
		0\\	
		0&	
		0&	
		0&	\hspace{0.03in}
		\mathsfbi{\hat{A}}_{1}^{(0)} 	&					
		0&	
		0&	\hspace{0.03in}
		0&	
		\mathsfbi{\hat{A}}_{0}^{(1)} 	&					
		0\\	
		0&	
		0&	
		0&	\hspace{0.03in}
		0&	
		\mathsfbi{\hat{A}}_{1}^{(0)} 	&					
		0&	\hspace{0.03in}
		0&	
		0&	
		\mathsfbi{\hat{A}}_{0}^{(1)}						
	\end{bmatrix}
	}_{\boldsymbol{\mathcal{A}}_{\tilde{\boldsymbol{q}}}}
	\underbrace{
	\begin{bmatrix}
		\hat{\boldsymbol{q}}_{-1,\gamma 	}^{(-1)} 	\\			
		\hat{\boldsymbol{q}}_{0,\gamma 	}^{(-1)} 	\\			
		\hat{\boldsymbol{q}}_{1,\gamma 	}^{(-1)} 	\\[3pt]	
		\hat{\boldsymbol{q}}_{-1,\gamma 	}^{(0)} 	\\			
		\hat{\boldsymbol{q}}_{0,\gamma 	}^{(0)} 	\\			
		\hat{\boldsymbol{q}}_{1,\gamma 	}^{(0)} 	\\[3pt]	
		\hat{\boldsymbol{q}}_{-1,\gamma 	}^{(1)} 	\\			
		\hat{\boldsymbol{q}}_{0,\gamma 	}^{(1)} 	\\			
		\hat{\boldsymbol{q}}_{1,\gamma	}^{(1)} 				
	\end{bmatrix}
	}_{\hat{\boldsymbol{\mathcal{Q}}}}.
\]
A few comments can be made on the structure of $\boldsymbol{\mathcal{A}}_{\tilde{\boldsymbol{q}}}$.  For this example where $\tilde{n} = \tilde{m} = \tilde{r} = 1$, $\boldsymbol{\mathcal{A}}_{\tilde{\boldsymbol{q}}}$ has a tri-block-diagonal structure with a block-wise size of $9 \times 9$.   Moreover, we notice that not all frequency-wavenumber components in $\hat{\boldsymbol{\mathcal{Q}}}$ are fully coupled.  For this demonstrating example, interactions are only taking place within three modal clusters, namely, 
$$
	\left[
        \hat{\boldsymbol{q}}_{1,\gamma}^{(1)},~
        \hat{\boldsymbol{q}}_{0,\gamma}^{(0)},~
        \hat{\boldsymbol{q}}_{-1, \gamma}^{(-1)}
    \right]
	,\quad
	\left[
        \hat{\boldsymbol{q}}_{0,\gamma}^{(1)},~
        \hat{\boldsymbol{q}}_{-1,\gamma}^{(0)}
    \right]
	,\quad
	\left[
        \hat{\boldsymbol{q}}_{1,\gamma}^{(0)},~
        \hat{\boldsymbol{q}}_{0,\gamma}^{(-1)}
    \right]
$$
and two decoupled modes, $\hat{\boldsymbol{q}}_{-1,\gamma}^{(1)}$ and $\hat{\boldsymbol{q}}_{1,\gamma}^{(-1)}$.  This suggests that, with appropriate pivoting, the non-zero off-diagonal blocks in $\boldsymbol{\mathcal{A}}_{\tilde{\boldsymbol{q}}}$ can be relocated to places immediately adjacent to the main diagonal, resulting in three decoupled clusters of diagonal blocks and two additional single diagonal blocks for the fully decoupled modes.  Such a structure is of great advantage to the STHR analysis, where a matrix inverse is involved in $\mathsfbi{H}_{\tilde{\boldsymbol{q}}} = \left[\mathrm{i}\gamma\mathsfbi{I}_\mathsfbi{M} - \mathsfbi{T}_{\tilde{\boldsymbol{q}}}\right]^{-1}$. Since the structure of $\boldsymbol{\mathcal{A}}_{\tilde{\boldsymbol{q}}}$ will be inherited by $\mathsfbi{T}_{\tilde{\boldsymbol{q}}}$ and, consequently, by $\left[\mathrm{i}\gamma\mathsfbi{I}_\mathsfbi{M} - \mathsfbi{T}_{\tilde{\boldsymbol{q}}}\right]$, we can invert the decoupled clusters of diagonal blocks one cluster at a time, instead of inverting the entire $\left[\mathrm{i}\gamma\mathsfbi{I}_\mathsfbi{M} - \mathsfbi{T}_{\tilde{\boldsymbol{q}}}\right]$.  In cases where the inversion is addressed by direct matrix solves (e.g., LU decomposition), taking advantage of such a structure can significantly reduce the memory usage.  Moreover, the process of inverting each diagonal cluster is embarrassingly parallelizable.  A similar structure in $\boldsymbol{\mathcal{A}}_{\tilde{\boldsymbol{q}}}$ is observed in our STHR analysis, where $\tilde{n} = 2$ and $\tilde{m} = \tilde{r} = 3$ are considered.  The matrix $\boldsymbol{\mathcal{A}}_{\tilde{\boldsymbol{q}}}$ has a penta-block-diagonal structure with $49 \times 49$ blocks of $\mathsfbi{\hat{A}}_n^{(r)}$ and multiple decoupled clusters of diagonal blocks. 

\subsection{Convergence study of frequency-wavenumber truncation}
\label{apndx.HRTruncation}
We seek the sets of truncated frequency and wavenumber for both the perturbations and the base flow such that convergence in the leading singular values of the STHR operator, $\mathsfbi{H}_{\tilde{\boldsymbol{q}}}(\gamma,\boldsymbol{\Omega}, \boldsymbol{K};~ \boldsymbol{\Omega}_b, \boldsymbol{K}_b)$, is achieved against the number of harmonics contained in these sets. Following equation~\eqref{eq.fwsets}, these frequency and wavenumber sets are denoted by 
\begin{align*}
    \begin{cases}
        \boldsymbol{\Omega} 
        &=\{ -\tilde{m}, \ldots, 0, \ldots, \tilde{m} \} \omega_0 + \gamma\\ 
    \boldsymbol{K}
        &=\{ -\tilde{r}, \ldots, 0, \ldots, \tilde{r} \} k_0
    \end{cases}
    ~~\text{and}\quad
    \begin{cases}
        \boldsymbol{\Omega}_b
            &=\{-\tilde{n}, \dots, 0, \dots, \tilde{n} \}\omega_0 \\
        \boldsymbol{K}_b
            &=\{-\tilde{n}, \dots, 0, \dots, \tilde{n} \}k_0
    \end{cases}
\end{align*}
for the perturbations and the base flow, respectively. Our goal is to incrementally increase the integer values of $\tilde{m}$, $\tilde{r}$, and $\tilde{n}$ until convergence in the leading singular values of the harmonic resolvent operator, $\mathsfbi{H}_{\tilde{\boldsymbol{q}}}$, is observed. 

\begin{figure}
\centering{\begin{overpic}[width=5.32in]{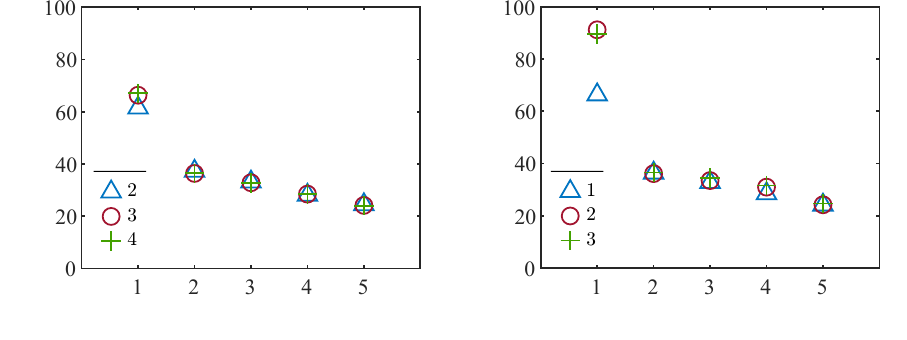} 
\put (34 , 147) {(a)}
\put (229 , 147) {(b)}
 \put (9 , 85){\rotatebox{90} {\fontsize{8}{9}\selectfont $\sigma_i$}}
 \put (202 , 85){\rotatebox{90} {\fontsize{8}{9}\selectfont $\sigma_i$}}
 \put (88 , 7){\rotatebox{0} {\fontsize{8}{9}\selectfont Mode rank $i$}}
 \put (281 , 7){\rotatebox{0} {\fontsize{8}{9}\selectfont Mode rank $i$}}
  \put (43 , 75){\rotatebox{0} {\fontsize{8}{9}\selectfont  $\tilde{m},~\tilde{r}$}}
  \put (241 , 75){\rotatebox{0} {\fontsize{8}{9}\selectfont  $\tilde{n}$}}

\end{overpic}}
\caption{Leading singular values of the STHR operator for symmetric base flow. In (a), we increase the number of harmonics for the perturbations from $\tilde{m} = \tilde{r} = 1$ to $4$ while fixing $\tilde{n} = 1$. In (b), we use $\tilde{m} = \tilde{r} = 3$ and increase the number of harmonics for the base flow from $\tilde{n} = 1$ to $3$.  Convergence in the singular values is observed for $\tilde{m} = \tilde{r} \ge 3$ and $\tilde{n} \ge 2$.}
\label{fig.trunc}
\end{figure}
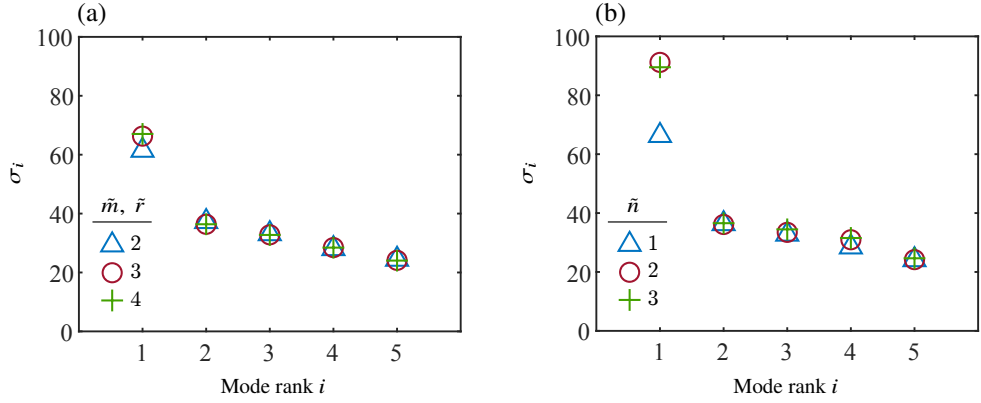

In figure~\ref{fig.trunc}, we consider different values of $\tilde{m} = \tilde{r} \in \{2,~3,~4\}$ and $\tilde{n} \in \{1,~2,~3\}$ used for the construction of the harmonic resolvent operator, $\mathsfbi{H}_{\tilde{\boldsymbol{q}}}$.  The convergence in the leading five singular values for different values of $\tilde{m} = \tilde{r}$ is shown in figure~\ref{fig.trunc}(a), where the value of $\tilde{n}$ is fixed to 1.  We observe convergence in these leading singular values for $(\tilde{m},\tilde{r}) \ge 3$ for $\tilde{n} = 1$.  Next, we set $\tilde{m} = \tilde{r} = 3$ and increase the value of $\tilde{n}$ from $1$ to $3$, as shown in figure~\ref{fig.trunc}(b). Here, convergence in the singular values is observed for $\tilde{n} \ge 2$.  According to this convergence analysis, we choose $\tilde{m} = \tilde{r} = 3$ and $\tilde{n} = 2$ throughout this study.

\subsection{Floquet--Hill analysis and discounting parameter for STHR}
\label{apndx.Floquet}
We employ the Floquet--Hill theory to determine the need for a discounting parameter in the present STHR analysis. We start with an unforced version of equation~\eqref{eq.PerdclinearNS},
\begin{equation}
    \mathsfbi{M}\frac{\mathrm{d}\boldsymbol{q}^\prime}{\mathrm{d} t} 
    = 
    \mathsfbi{A}_{\bar{\boldsymbol{q}}(\boldsymbol{x}, t)} \boldsymbol{q}^\prime,
    \label{eq.nonforcedLNS}
\end{equation}
and seek eigensolutions for the perturbations in the form of an exponentially modulated periodic signal \citep{wereley1990frequency}, given by 
\begin{equation}
\label{eq.EMP}
    \boldsymbol{q}^\prime(\boldsymbol{x}, t)
    = \mathrm{e}^{s t}\sum_{m,r=-\infty}^{+\infty}
        \hat{\boldsymbol{q}}_{m \omega_0+\gamma,\,r k_0}(y,z)
        \mathrm{e}^{\mathrm{i}(m \omega_0 t - r k_0 x)},
\end{equation}
where $s \in \mathbb{C}$ is the Floquet exponent. Note the resemblance between the expressions in \eqref{eq.EMP} and \eqref{eq.FourierRep}, except that the frequency shift introduced by $\gamma \in \mathbb{R}$ is replaced by the Floquet exponent $s \in \mathbb{C}$. Substituting the expression in equation~\eqref{eq.EMP} and the time-periodic linear NS operator from equation~\eqref{eq.LinOptA} into equation~\eqref{eq.nonforcedLNS}, we arrive at a generalized eigenvalue problem for the Hill matrix $\mathsfbi{T}_{\tilde{\boldsymbol{q}}}$, given by
\begin{equation}
    \mathsfbi{T}_{\tilde{\boldsymbol{q}}}\hat{\boldsymbol{\mathcal{Q}}}=s{\mathsfbi{I}_\mathsfbi{M}}\hat{\boldsymbol{\mathcal{Q}}}.
\end{equation}
If any of the Floquet exponents has a positive real part, $\Re(s) > 0$, the time-periodic linear system is unstable. In such a case, we consider the use of a discounting parameter, $\gamma_d$, for the present STHR formulation \citep{jovanovic2004modeling,yeh2020resolvent,leclercq2023mean}.  Following the convergence analysis in section \ref{apndx.HRTruncation}, the Hill matrix, $\mathsfbi{T}_{\tilde{\boldsymbol{q}}}$, is constructed with $\tilde{m} = \tilde{r} = 3$ and $\tilde{n} = 2$ for the Floquet--Hill analysis here. The resulting frequency-wavenumber sets are 
$\boldsymbol{\Omega}_b = \{-2,\ldots,0,\ldots,2\}\,\omega_0$
and
$\boldsymbol{K}_b = \{-2,\ldots,0,\ldots,2\}\,k_0$ 
for the base flow, and 
$\boldsymbol{\Omega} = \{-3,\ldots,0,\ldots,3\}\,\omega_0$ and
$\boldsymbol{K} = \{-3,\ldots,0,\ldots,3\}\,k_0$ for the perturbations. 

\begin{figure}
\centerline{\begin{overpic}[width=5.32in]{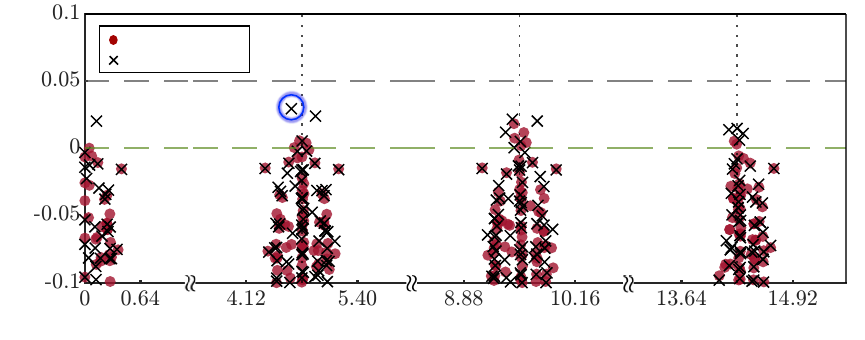} 
   \put (55, 131) {\fontsize{8}{9}\selectfont Symmetric}
   \put (55, 122) {\fontsize{8}{9}\selectfont Anti-symmetric}
   \put (71 , 105) {{\color{blue}{\fontsize{8}{9}\selectfont Most unstable}}}
   \put (66 , 96) {{\color{blue}{\fontsize{8}{9}\selectfont Floquet exponent}}}
   \put (186 , 2){\rotatebox{0} {\fontsize{9}{9}\selectfont $s_i L_c/u_\infty$}}
   \put (3 , 66){\rotatebox{90} {\fontsize{9}{9}\selectfont $s_r L_c/u_\infty$}}
   \put (150 , 119){\rotatebox{0} {\fontsize{9}{9}\selectfont $\gamma_d L_c/u_\infty=0.05$}}
   \put (126 , 125){\rotatebox{90} {\fontsize{9}{9}\selectfont $\omega_0$}}
   \put (221 , 123){\rotatebox{90} {\fontsize{9}{9}\selectfont $2\omega_0$}}
   \put (317 , 123){\rotatebox{90} {\fontsize{9}{9}\selectfont $3\omega_0$}}
   \end{overpic}}
\caption{Normalized Floquet exponents of the spatio-temporal linear operator $\mathsfbi{T}_{\tilde{\boldsymbol{q}}}$ for the symmetric flow in the complex plane.
}
\label{fig.Floquet}
\end{figure}
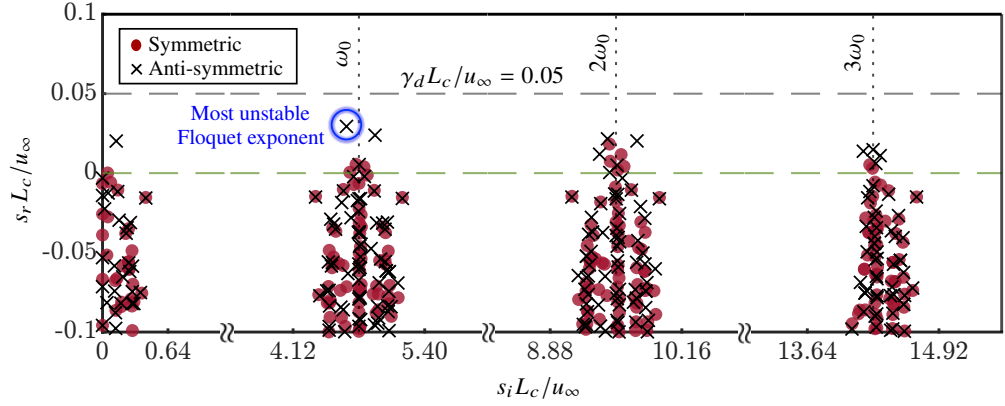

Figure~\ref{fig.Floquet} shows the complex-valued Floquet exponents, equivalently the eigenvalue spectrum of the Hill matrix $\mathsfbi{T}_{\tilde{\boldsymbol{q}}}$, obtained for symmetric ($\bullet$) and anti-symmetric ($\cross$) perturbations of the symmetric flow. We observe unstable Floquet exponents of positive real parts near the harmonics of the base flow frequency.  Notably, the most unstable Floquet exponents are predominantly associated with anti-symmetric structures. Excitation of these anti-symmetric modes that exhibit higher growth rates than the symmetric ones can potentially promote the emergence of flow asymmetries.  Due to the presence of these unstable Floquet exponents, we introduce a temporal discounting $\mathrm{e^{-\gamma_d t}}$, where $\gamma_d \in \mathbb{R}$, to the present STHR analysis.  This results in a modified harmonic resolvent operator given by
\[
    \mathsfbi{H}_{\tilde{\boldsymbol{q}}} 
    = [(\mathrm{i} \gamma + \gamma_d) \mathsfbi{I}_\mathsfbi{M} - \mathsfbi{T}_{\tilde{\boldsymbol{q}}}]^{-1}.
\]
This discounting parameter is chosen to be $\gamma_d L_c/u_{\infty} = 0.05$, which exceeds the growth rates of all unstable Floquet exponents found in figure~\ref{fig.Floquet}.

\bibliographystyle{jfm}
\bibliography{jfm}

@article{mckeon2010critical,
  title={A critical-layer framework for turbulent pipe flow},
  author={McKeon, B. J. and Sharma, A. S.},
  journal={Journal of Fluid Mechanics},
  volume={658},
  pages={336--382},
  year={2010},
  publisher={Cambridge University Press}
}

@article{yeh2019resolvent,
  title={Resolvent-analysis-based design of airfoil separation control},
  author={Yeh, C.-A. and Taira, K.},
  journal={Journal of Fluid Mechanics},
  volume={867},
  pages={572--610},
  year={2019},
  publisher={Cambridge University Press}
}

@article{jovanovic2005componentwise,
  title={Componentwise energy amplification in channel flows},
  author={Jovanovi{\'c}, M. R. and Bamieh, B.},
  journal={Journal of Fluid Mechanics},
  volume={534},
  pages={145--183},
  year={2005},
  publisher={Cambridge University Press}
}

@article{leweke2016dynamics,
  title={Dynamics and instabilities of vortex pairs},
  author={Leweke, T. and Le Diz{\`e}s, S. and Williamson, C. H. K.},
  journal={Annual Review of Fluid Mechanics},
  volume={48},
  pages={507--541},
  year={2016},
  publisher={Annual Reviews}
}

@article{padovan2020analysis,
  title={Analysis of amplification mechanisms and cross-frequency interactions in nonlinear flows via the harmonic resolvent},
  author={Padovan, A. and Otto, S. E. and Rowley, C. W.},
  journal={Journal of Fluid Mechanics},
  volume={900},
  pages={A14},
  year={2020},
  publisher={Cambridge University Press}
}

@article{padovan2022analysis,
  title={Analysis of the dynamics of subharmonic flow structures via the harmonic resolvent: application to vortex pairing in an axisymmetric jet},
  author={Padovan, A. and Rowley, C. W.},
  journal={Physical Review Fluids},
  volume={7},
  number={7},
  pages={073903},
  year={2022},
  publisher={American Physical Society}
}

@article{chavarin2020resolvent,
  title={Resolvent analysis for turbulent channel flow with riblets},
  author={Chavarin, A. and Luhar, M.},
  journal={AIAA Journal},
  volume={58},
  number={2},
  pages={589--599},
  year={2020},
  publisher={American Institute of Aeronautics and Astronautics}
}

@article{ham2006accurate,
  title={Accurate and stable finite volume operators for unstructured flow solvers},
  author={Ham, F. and Mattsson, K. and Iaccarino, G.},
  journal={Annual Research Briefs},
  pages = {243--261},
  year={2006},
  publisher={Center for Turbulence Research, NASA Ames/Stanford University}
}

@article{benton2021crow,
  title={{Crow} and elliptic instabilities in a streamwise vortex--wall interaction},
  author={Benton, S. I. and Bons, J. P.},
  journal={AIAA Journal},
  volume={59},
  number={3},
  pages={1109--1113},
  year={2021},
  publisher={American Institute of Aeronautics and Astronautics}
}

@article{edstrand2018parallel,
  title={A parallel stability analysis of a trailing vortex wake},
  author={Edstrand, A. M. and Schmid, P. J. and Taira, K. and Cattafesta III, L. N.},
  journal={Journal of Fluid Mechanics},
  volume={837},
  pages={858--895},
  year={2018},
  publisher={Cambridge University Press}
}

@article{sipp2003widnall,
  title={{Widnall} instabilities in vortex pairs},
  author={Sipp, D. and Jacquin, L.},
  journal={Physics of Fluids},
  volume={15},
  number={7},
  pages={1861--1874},
  year={2003},
  publisher={American Institute of Physics}
}

@article{luo2018time,
  title={Time-averaged asymmetries and oscillatory global modes of vortex flows over a slender wing},
  author={Luo, H. and Ma, B.-F.},
  journal={Physics of Fluids},
  volume={30},
  number={9},
  year={2018},
  publisher={AIP Publishing}
}

@article{ma2017symmetry,
  title={Symmetry breaking and instabilities of conical vortex pairs over slender delta wings},
  author={Ma, B.-F. and Wang, Z. and Gursul, I.},
  journal={Journal of Fluid Mechanics},
  volume={832},
  pages={41--72},
  year={2017},
  publisher={Cambridge University Press}
}

@article{gursul2005review,
  title={Review of unsteady vortex flows over slender delta wings},
  author={Gursul, I.},
  journal={Journal of Aircraft},
  volume={42},
  number={2},
  pages={299--319},
  year={2005}
}

@article{shen2016asymmetric,
  title={Asymmetric flow control on a delta wing with dielectric barrier discharge actuators},
  author={Shen, L. and Wen, C.-y. and Chen, H.-A.},
  journal={AIAA Journal},
  volume={54},
  number={2},
  pages={652--658},
  year={2016},
  publisher={American Institute of Aeronautics and Astronautics}
}

@inproceedings{hunt1982asymmetric,
  title={Asymmetric vortex forces and wakes on slender bodies},
  author={Hunt, B.},
  booktitle={9th Atmospheric Flight Mechanics Conference},
  pages={1336},
  year={1982}
}

@article{zong2021structural,
  title={Structural control of asymmetric forebody vortices over a slender body},
  author={Zong, S. and Wang, Y. and Qi, Z.},
  journal={Physics of Fluids},
  volume={33},
  number={11},
  year={2021},
  publisher={AIP Publishing}
}

@article{theofilis2003advances,
  title={Advances in global linear instability analysis of nonparallel and three-dimensional flows},
  author={Theofilis, V.},
  journal={Progress in Aerospace Sciences},
  volume={39},
  number={4},
  pages={249--315},
  year={2003},
  publisher={Elsevier}
}

@techreport{mckeon2019applications,
  title={Applications of resolvent analysis in fluid mechanics},
  author={McKeon, B.},
  year={2019},
  institution={EPSRC Summer School on Modal Decompositions in Fluid Mechanics}
}

@inproceedings{ericsson1990control,
  title={Control of forebody flow asymmetry-A critical review},
  author={Ericsson, L. E.},
  booktitle={17th Atmospheric Flight Mechanics Conference},
  pages={2833},
  year={1990}
}

@article{degani1991effect,
  title={Effect of geometrical disturbance on vortex asymmetry},
  author={Degani, D.},
  journal={AIAA Journal},
  volume={29},
  number={4},
  pages={560--566},
  year={1991}
}

@article{chen2023effect,
  title={Effect of the free-stream turbulence on the bi-modal wake dynamics of square-back bluff body},
  author={Chen, G. and Li, X.-B. and He, K. and Cheng, Z. and Zhou, D. and Liang, X.-F.},
  journal={Physics of Fluids},
  volume={35},
  number={1},
  year={2023},
  publisher={AIP Publishing}
}

@article{degani2022development,
  title={Development of nonstationary side forces along a slender body of revolution at incidence},
  author={Degani, D.},
  journal={Physical Review Fluids},
  volume={7},
  number={12},
  pages={124101},
  year={2022},
  publisher={American Physical Society}
}

@inproceedings{degani1992effect,
  title={Effect of upstream disturbance on flow asymmetry},
  author={Degani, D. and Tobak, M.},
  booktitle={30th Aerospace Sciences Meeting and Exhibit},
  pages={408},
  year={1992}
}

@article{levy1996systematic,
  title={Systematic study of the correlation between geometrical disturbances and flow asymmetries},
  author={Levy, Y. and Hesselnik, L. and Degani, D.},
  journal={AIAA Journal},
  volume={34},
  number={4},
  pages={772--777},
  year={1996}
}

@article{williams2025asymmetries,
  title={Asymmetries in nominally symmetric flows},
  author={Williams, O. J. H. and Smits, A. J.},
  journal={Annual Review of Fluid Mechanics},
  volume={57},
  year={2025},
  publisher={Annual Reviews}
}

@article{grandemange2012reflectional,
  title={Reflectional symmetry breaking of the separated flow over three-dimensional bluff bodies},
  author={Grandemange, M. and Cadot, O. and Gohlke, M.},
  journal={Physical Review E},
  volume={86},
  number={3},
  pages={035302},
  year={2012},
  publisher={American Physical Society}
}

@article{bury2012transitions,
  title={Transitions to chaos in the wake of an axisymmetric bluff body},
  author={Bury, Y. and Jardin, T.},
  journal={Physics Letters A},
  volume={376},
  number={45},
  pages={3219--3222},
  year={2012},
  publisher={Elsevier}
}

@article{prasad1997instability,
  title={The instability of the shear layer separating from a bluff body},
  author={Prasad, A. and Williamson, C. H. K.},
  journal={Journal of Fluid Mechanics},
  volume={333},
  pages={375--402},
  year={1997},
  publisher={Cambridge University Press}
}

@article{haffner2020mechanics,
  title={Mechanics of bluff body drag reduction during transient near-wake reversals},
  author={Haffner, Y. and Bor{\'e}e, J. and Spohn, A. and Castelain, T.},
  journal={Journal of Fluid Mechanics},
  volume={894},
  pages={A14},
  year={2020},
  publisher={Cambridge University Press}
}

@article{brackston2016stochastic,
  title={Stochastic modelling and feedback control of bistability in a turbulent bluff body wake},
  author={Brackston, R. D. and De La Cruz, J. M. G. and Wynn, A. and Rigas, G. and Morrison, J. F.},
  journal={Journal of Fluid Mechanics},
  volume={802},
  pages={726--749},
  year={2016},
  publisher={Cambridge University Press}
}

@article{grandemange2013turbulent,
  title={Turbulent wake past a three-dimensional blunt body. {Part 1}. {Global} modes and bi-stability},
  author={Grandemange, M. and Gohlke, M. and Cadot, O.},
  journal={Journal of Fluid Mechanics},
  volume={722},
  pages={51--84},
  year={2013},
  publisher={Cambridge University Press}
}

@article{ribeiro2023triglobal,
  title={Triglobal resolvent analysis of swept-wing wakes},
  author={Ribeiro, J. H. M. and Yeh, C.-A. and Taira, K.},
  journal={Journal of Fluid Mechanics},
  volume={954},
  pages={A42},
  year={2023},
  publisher={Cambridge University Press}
}

@article{gursul2004recent,
  title={Recent developments in delta wing aerodynamics},
  author={Gursul, I.},
  journal={The Aeronautical Journal},
  volume={108},
  number={1087},
  pages={437--452},
  year={2004},
  publisher={Cambridge University Press}
}

@book{trefethen2005spectra,
  title={Spectra and pseudospectra: the behavior of nonnormal matrices and operators},
  author={Trefethen, L. N. and Embree, M.},
  year={2005},
  publisher={Princeton University Press}
}

@article{schmid2007nonmodal,
  title={Nonmodal stability theory},
  author={Schmid, P. J.},
  journal={Annual Review of Fluid Mechanics},
  volume={39},
  number={1},
  pages={129--162},
  year={2007},
  publisher={Annual Reviews}
}

@book{schmid2001stability,
  title    ={Stability and transition in shear flows},
  author   ={Schmid, P. J. and Henningson, D. S.},
  series={Applied Mathematical Sciences},
  volume={142},
  year     ={2001},
  publisher={Springer}
}

@article{magri2023linear,
  title={Linear flow analysis inspired by mathematical methods from quantum mechanics},
  author={Magri, L. and Schmid, P. J. and Moeck, J. P.},
  journal={Annual Review of Fluid Mechanics},
  volume={55},
  number={1},
  pages={541--574},
  year={2023},
  publisher={Annual Reviews}
}

@incollection{andersson2018instabilities,
  title={Instabilities in the wake of an inclined prolate spheroid},
  author={Andersson, H. I. and Jiang, F. and Okulov, V. L.},
  booktitle={Computational Modelling of Bifurcations and Instabilities in Fluid Dynamics},
  pages={311--352},
  year={2018},
  publisher={Springer}
}

@inproceedings{rigas2016symmetry,
  title={Symmetry breaking in a {3D} bluff-body wake},
  author={Rigas, G. and Esclapez, L. and Magri, L.},
  pages = {193-202},
  year={2016},
  organization={Center for Turbulence Research, Stanford University}
}

@article{regan2019adjoint,
  title={Adjoint sensitivity and optimal perturbations of the low-speed jet in cross-flow},
  author={Regan, M. A. and Mahesh, K.},
  journal={Journal of Fluid Mechanics},
  volume={877},
  pages={330--372},
  year={2019},
  publisher={Cambridge University Press}
}

@article{degani1992asymmetric,
  title={Asymmetric turbulent vortical flows over slender bodies},
  author={Degani, D. and Levy, Y.},
  journal={AIAA Journal},
  volume={30},
  number={9},
  pages={2267--2273},
  year={1992}
}

@article{giannetti2007structural,
  title={Structural sensitivity of the first instability of the cylinder wake},
  author={Giannetti, F. and Luchini, P.},
  journal={Journal of Fluid Mechanics},
  volume={581},
  pages={167--197},
  year={2007},
  publisher={Cambridge University Press}
}

@article{marquet2008sensitivity,
  title={Sensitivity analysis and passive control of cylinder flow},
  author={Marquet, O. and Sipp, D. and Jacquin, L.},
  journal={Journal of Fluid Mechanics},
  volume={615},
  pages={221--252},
  year={2008},
  publisher={Cambridge University Press}
}

@article{schmid2014analysis,
  title={Analysis of fluid systems: stability, receptivity, sensitivity},
  author={Schmid, P. J. and Brandt, L.},
  journal={Applied Mechanics Reviews},
  volume={66},
  number={2},
  pages={024803},
  year={2014},
  publisher={American Society of Mechanical Engineers}
}

@inproceedings{wereley1990frequency,
  title={Frequency response of linear time periodic systems},
  author={Wereley, N. M. and Hall, S. R.},
  booktitle={29th IEEE Conference on Decision and Control},
  pages={3650--3655},
  year={1990},
  organization={IEEE}
}

@article{keener1977similarity,
  title={Similarity in vortex asymmetries over slender bodies and wings},
  author={Keener, E. R. and Chapman, G. T.},
  journal={AIAA Journal},
  volume={15},
  number={9},
  pages={1370--1372},
  year={1977}
}

@article{pavia2020salient,
  title={Salient three-dimensional features of the turbulent wake of a simplified square-back vehicle},
  author={Pavia, G. and Passmore, M. A. and Varney, M. and Hodgson, G.},
  journal={Journal of Fluid Mechanics},
  volume={888},
  pages={A33},
  year={2020},
  publisher={Cambridge University Press}
}

@article{wang2019enhanced,
  title={Enhanced stability of flows through contraction channels: Combining shape optimization and linear stability analysis},
  author={Wang, Y. and Ferrer, E. and Mart{\'\i}nez-Cava, A. and Zheng, Y. and Valero, E.},
  journal={Physics of Fluids},
  volume={31},
  number={7},
  year={2019},
  publisher={AIP Publishing}
}

@article{theofilis2011global,
  title={Global linear instability},
  author={Theofilis, V.},
  journal={Annual Review of Fluid Mechanics},
  volume={43},
  number={1},
  pages={319--352},
  year={2011},
  publisher={Annual Reviews}
}

@article{goswami2022mechanisms,
  title={Mechanisms of wake asymmetry and secondary structures behind low aspect-ratio wall-mounted prisms},
  author={Goswami, S. and Hemmati, A.},
  journal={Journal of Fluid Mechanics},
  volume={950},
  pages={A31},
  year={2022},
  publisher={Cambridge University Press}
}

@article{lashgari2014planar,
  title={The planar X-junction flow: stability analysis and control},
  author={Lashgari, I. and Tammisola, O. and Citro, V. and Juniper, M. P. and Brandt, L.},
  journal={Journal of Fluid Mechanics},
  volume={753},
  pages={1--28},
  year={2014},
  publisher={Cambridge University Press}
}

@article{porter2014closed,
  title={Closed-loop flow control of a forebody at a high incidence angle},
  author={Porter, C. and Fagley, C. and Farnsworth, J. and Seidel, J. and McLaughlin, T.},
  journal={AIAA Journal},
  volume={52},
  number={7},
  pages={1430--1440},
  year={2014},
  publisher={American Institute of Aeronautics and Astronautics}
}

@article{hitzel2018enhanced,
  title={Enhanced maneuverability of a delta-canard combat aircraft by vortex flow control},
  author={Hitzel, S. M. and Osterhuber, R.},
  journal={Journal of Aircraft},
  volume={55},
  number={3},
  pages={1090--1102},
  year={2018},
  publisher={American Institute of Aeronautics and Astronautics}
}

@article{leclercq2023mean,
  title={Mean resolvent operator of a statistically steady flow},
  author={Leclercq, C. and Sipp, D.},
  journal={Journal of Fluid Mechanics},
  volume={968},
  pages={A13},
  year={2023},
  publisher={Cambridge University Press}
}

@article{song2025symmetry,
  title={Symmetry-breaking bifurcations and subharmonic lock-in of a flexible splitter plate in cylinder wake flow},
  author={Song, B. and Ping, H. and Chen, W.-L. and Cao, Y. and Zhou, D.},
  journal={Journal of Fluid Mechanics},
  volume={1025},
  pages={A47},
  year={2025},
  publisher={Cambridge University Press}
}

@article{tang1997symmetry,
  title={On the symmetry breaking instability leading to vortex shedding},
  author={Tang, S. and Aubry, N.},
  journal={Physics of Fluids},
  volume={9},
  number={9},
  pages={2550--2561},
  year={1997},
  publisher={American Institute of Physics}
}

@article{matsumoto1999vortex,
  title={Vortex shedding of bluff bodies: a review},
  author={Matsumoto, M.},
  journal={Journal of Fluids and Structures},
  volume={13},
  number={7-8},
  pages={791--811},
  year={1999},
  publisher={Elsevier}
}

@article{agrawal1992numerical,
  title={Numerical investigation of vortex breakdown on a delta wing},
  author={Agrawal, S. and Barnett, R. M. and Robinson, B. A.},
  journal={AIAA Journal},
  volume={30},
  number={3},
  pages={584--591},
  year={1992}
}

@article{visbal1994onset,
  title={Onset of vortex breakdown above a pitching delta wing},
  author={Visbal, M. R.},
  journal={AIAA Journal},
  volume={32},
  number={8},
  pages={1568--1575},
  year={1994}
}

@article{mao2014structure,
  title={The structure of primary instability modes in the steady wake and separation bubble of a square cylinder},
  author={Mao, X. and Blackburn, H. M.},
  journal={Physics of Fluids},
  volume={26},
  number={7},
  year={2014},
  publisher={AIP Publishing}
}

@article{williams1992response,
  title={The response and symmetry properties of a cylinder wake subjected to localized surface excitation},
  author={Williams, D. R. and Mansy, H. and Amato, C.},
  journal={Journal of Fluid Mechanics},
  volume={234},
  pages={71--96},
  year={1992},
  publisher={Cambridge University Press}
}

@article{marasli1989modal,
  title={Modal decomposition of velocity signals in a plane, turbulent wake},
  author={Marasli, B. and Champagne, F. H. and Wygnanski, I. J.},
  journal={Journal of Fluid Mechanics},
  volume={198},
  pages={255--273},
  year={1989},
  publisher={Cambridge University Press}
}

@book{griffiths2018introduction,
  title={Introduction to quantum mechanics},
  author={Griffiths, D. J. and Schroeter, D. F.},
  year={2018},
  publisher={Cambridge University Press}
}

@article{johnson2026flow,
  title={Flow asymmetry over a body of revolution},
  author={Johnson, S. M. and Schmitz, S. and Yang, X. I. A. and Chyczewski, T. S.},
  journal={Journal of Fluid Mechanics},
  volume={1032},
  pages={A2},
  year={2026},
  publisher={Cambridge University Press}
}

@article{tsai1976stability,
  title={The stability of short waves on a straight vortex filament in a weak externally imposed strain field},
  author={Tsai, C.-Y. and Widnall, S. E.},
  journal={Journal of Fluid Mechanics},
  volume={73},
  number={4},
  pages={721--733},
  year={1976},
  publisher={Cambridge University Press}
}

@article{huerre1985absolute,
  title={Absolute and convective instabilities in free shear layers},
  author={Huerre, P. and Monkewitz, P. A.},
  journal={Journal of Fluid Mechanics},
  volume={159},
  pages={151--168},
  year={1985},
  publisher={Cambridge University Press}
}

@article{monkewitz1988absolute,
  title={The absolute and convective nature of instability in two-dimensional wakes at low {Reynolds} numbers},
  author={Monkewitz, P. A.},
  journal={Physics of Fluids},
  volume={31},
  number={5},
  pages={999--1006},
  year={1988},
  publisher={American Institute of Physics}
}

@article{ghoreyshi2023vortex,
  title={Vortex interaction characteristics of multiswept wings at subsonic speeds},
  author={Ghoreyshi, M. and Fagley, C. and Seidel, J.},
  journal={AIAA Journal},
  volume={61},
  number={7},
  pages={2932--2947},
  year={2023},
  publisher={American Institute of Aeronautics and Astronautics}
}

@article{rojas2025flow,
  title={Flow control over a generic tailless chined forebody delta wing},
  author={Rojas Carvajal, T. E. and Amitay, M.},
  journal={AIAA Journal},
  volume={63},
  number={7},
  pages={2750--2763},
  year={2025},
  publisher={American Institute of Aeronautics and Astronautics}
}

@article{shah1999turbulent,
  title={Turbulent transport in the core of a trailing half-delta-wing vortex},
  author={Shah, P. N. and Atsavapranee, P. and Hsu, T. Y. and Wei, T. and McHugh, J.},
  journal={Journal of Fluid Mechanics},
  volume={387},
  pages={151--175},
  year={1999},
  publisher={Cambridge University Press}
}

@article{williams2008active,
  title={Active flow control on a nonslender delta wing},
  author={Williams, N. M. and Wang, Z. and Gursul, I.},
  journal={Journal of Aircraft},
  volume={45},
  number={6},
  pages={2100--2110},
  year={2008}
}

@article{rolandi2024invitation,
  title={An invitation to resolvent analysis},
  author={Rolandi, L. V. and Ribeiro, J. H. M. and Yeh, C.-A. and Taira, K.},
  journal={Theoretical and Computational Fluid Dynamics},
  volume={38},
  number={5},
  pages={603--639},
  year={2024},
  publisher={Springer}
}

@article{trefethen1993hydrodynamic,
  title={Hydrodynamic stability without eigenvalues},
  author={Trefethen, L. N. and Trefethen, A. E. and Reddy, S. C. and Driscoll, T. A.},
  journal={Science},
  volume={261},
  number={5121},
  pages={578--584},
  year={1993},
  publisher={American Association for the Advancement of Science}
}

@article{yeh2020resolvent,
  title={Resolvent analysis of an airfoil laminar separation bubble at {Re= 500,000}},
  author={Yeh, C.-A. and Benton, S. I. and Taira, K. and Garmann, D. J.},
  journal={Physical Review Fluids},
  volume={5},
  number={8},
  pages={083906},
  year={2020},
  publisher={American Physical Society}
}

@phdthesis{jovanovic2004modeling,
  title={Modeling, analysis, and control of spatially distributed systems},
  author={Jovanovic, M. R.},
  year={2004},
  school={University of California, Santa Barbara}
}

\end{document}